\documentclass[twocolumn,trackchanges]{aastex631}
\usepackage{graphicx} 

\usepackage{graphicx}
\usepackage[suffix=]{epstopdf}
\usepackage{amsmath}
\usepackage{url}
\usepackage{xspace}
\usepackage{mathrsfs}
\usepackage{enumitem}
\usepackage[caption=false]{subfig}
\usepackage{float}
\usepackage{amssymb}
\usepackage{comment}

\makeatletter
\DeclareRobustCommand{\HII}{%
  \mbox{H\check@mathfonts\fontsize\sf@size\z@\selectfont II}%
}
\makeatother

\begin{document}

\newcommand{\yale}{Department of Astronomy, Yale University, 219 Prospect Street, New Haven, CT 06511}
\newcommand{\nrao}{National Radio Astronomy Observatory, PO Box O, 1011 Lopezville Road, Socorro, NM 87801}
\newcommand{\nmt}{New Mexico Institute of Mining and Technology, 801 Leroy Place, Socorro, NM 87801}

\author[0000-0002-5956-5167]{Andy Nilipour}
\affiliation{\yale}
\affiliation{\nrao}

\author[0000-0001-8224-1956]{Juergen Ott}
\affiliation{\nrao}

\author[0000-0001-9436-9471]{David S. Meier}
\affiliation{\nmt}
\affiliation{\nrao}

\author[0000-0002-8502-6431]{Brian Svoboda}
\affiliation{\nrao}

\author[0000-0001-6113-6241]{Mattia C. Sormani}
\affiliation{Universit{\`a} dell’Insubria, via Valleggio 11, 22100 Como, Italy}

\author[0000-0001-6431-9633]{Adam Ginsburg}
\affiliation{Department of Astronomy, University of Florida, P.O. Box 112055, Gainesville, FL 32611}

\author[0000-0002-1313-429X]{Savannah R. Gramze}
\affiliation{Department of Astronomy, University of Florida, P.O. Box 112055, Gainesville, FL 32611}

\author[0000-0002-4013-6469]{Natalie O. Butterfield}
\affiliation{National Radio Astronomy Observatory, 520 Edgemont Road, Charlottesville, VA 22903, USA}

\author[0000-0002-0560-3172]{Ralf S.\ Klessen}
\affiliation{Universit\"at Heidelberg, Zentrum f\"ur Astronomie, Institut f\"ur Theoretische Astrophysik, Albert-Ueberle-Str. 2, 69120 Heidelberg, Germany}
\affiliation{Universit\"{a}t Heidelberg, Interdisziplin\"{a}res Zentrum f\"{u}r Wissenschaftliches Rechnen, Im Neuenheimer Feld 205, 69120 Heidelberg, Germany}
\affiliation{Harvard-Smithsonian Center for Astrophysics, 60 Garden Street, Cambridge, MA 02138, U.S.A.}
\affiliation{Elizabeth S. and Richard M. Cashin Fellow at the Radcliffe Institute for Advanced Studies at Harvard University, 10 Garden Street, Cambridge, MA 02138, U.S.A.}

\correspondingauthor{Andy Nilipour}
\email{andy.nilipour@yale.edu}

\title{Turbulent Pressure Heats Gas and Suppresses Star Formation in Galactic Bar Molecular Clouds}
\shortauthors{Nilipour et al.}


\begin{abstract}
The Central Molecular Zone (CMZ) of the Milky Way is fed by gas inflows from the Galactic disk along almost radial trajectories aligned with the major axis of the Galactic bar. However, despite being fundamental to all processes in the nucleus of the Galaxy, these inflows have been studied significantly less than the CMZ itself. We present observations of various molecular lines between 215 and 230 GHz for 20 clouds with $\lvert \ell \rvert < 10^\circ$, which are candidates for clouds in the Galactic bar due to their warm temperatures and broad lines relative to typical Galactic disk clouds, using the Atacama Large Millimeter/submillimeter Array (ALMA) Atacama Compact Array (ACA). We measure gas temperatures, shocks, star formation rates, turbulent Mach numbers, and masses for these clouds. Although some clouds may be in the Galactic disk despite their atypical properties, nine clouds are likely associated with regions in the Galactic bar, and in these clouds, turbulent pressure is suppressing star formation. In clouds with no detected star formation, turbulence is the dominant heating mechanism, whereas photoelectric processes heat the star-forming clouds. We find that the ammonia (NH\textsubscript{3}) and formaldehyde (H\textsubscript{2}CO) temperatures probe different gas components, and in general, each transition appears to trace different molecular gas phases within the clouds. We also measure the CO-to-H\textsubscript{2} X-factor in the bar to be an order of magnitude lower than the typical Galactic value. These observations provide evidence that molecular clouds achieve CMZ-like properties before reaching the CMZ.
\end{abstract}



\section{Introduction}
\label{sec:intro}

The Central Molecular Zone (CMZ) is a region of molecular gas at high density and pressure within the inner $\sim$250 pc of the Milky Way \citep{morris_galactic_1996, henshaw_star_2023}. The CMZ is contained within the Galactic bar, which has a semimajor axis length of about 3.5\,kpc and a semiminor axis length of around 1\,kpc (e.g.,\ \citealt{queiroz_milky_2021, lucey_dynamically_2023}), although some have proposed that it may be longer \citep{wegg_structure_2015, bland-hawthorn_galaxy_2016, sormani_stellar_2022}. Two important families of closed orbits in a bar potential are x\textsubscript{1} and x\textsubscript{2} orbits (e.g.,\ \citealt{contopoulos_orbits_1989}). x\textsubscript{1} orbits are elongated roughly parallel to the major axis of the bar and can form cusps and self-intersecting loops, whereas x\textsubscript{2} orbits are more circular, closer to the bar center, and elongated roughly parallel to the minor axis of the bar. Shocks mainly form at the self-intersections of material on x\textsubscript{1} orbits or at the x\textsubscript{1}-x\textsubscript{2} orbit intersections \citep{athanassoula_existence_1992}. The CMZ is suggested to be associated with material on x\textsubscript{2} orbits \citep{binney_understanding_1991, fux_3d_1999, sormani_gas_2015, busch_living_2022}, and the gas transitioning from x\textsubscript{1} to x\textsubscript{2} orbits is believed to form bar lanes that feed the CMZ.

While the CMZ has been extensively studied at many wavelengths (e.g.,\ \citealt{baganoff_chandra_2003, oka_hot_2005, goto_absorption_2008, jones_spectral_2012, kruijssen_what_2014, yang_fermi-lat_2015, ginsburg_dense_2016, oka_central_2019}), the bar dust lanes and inflows have been relatively neglected. \citet{sormani_mass_2019} calculate the gas inflow rate along the Milky Way bar lanes to be $2.7^{+1.5}_{-1.7} \text{ M}_\odot \text{ yr}^{-1}$ using previous observations of \textsuperscript{12}CO and a simple geometrical model of the inner Galaxy, but simulations show more complex dynamics, such as inflowing gas overshooting the CMZ and subsequently colliding with the bar lane on the opposite side \citep{sormani_geometry_2019}, which reduces the CMZ gas accretion rate to $0.8 \pm 0.6 \text{ M}_\odot \text{ yr}^{-1}$ \citep{hatchfield_dynamically_2021}. \citet{gramze_evidence_2023} suggest that this is an overestimate based on evidence that the CO-to-H\textsubscript{2} X-factor in the bar region may be significantly lower than the standard Galactic value.

Observations and simulations have shown that the bar inflows induce strong shocks, large shears, and collisions with gas clouds \citep{fux_3d_1999, li_gas_2016}. \citet{marshall_large_2008} find that the gas and dust distributions of the inflows overlap but are offset, which may be due to shock destruction of the dust material. \citet{liszt_molecular_2006} conclude that the presence of localized wide lines, indicating extremely large velocity gradients, in this region results from material crossing standing shocks along the inner edge of the inflows. It has also been suggested that clouds such as Bania 2 (B2; \citealt{stark_clump_1986}) and G5, the cloud at $(\ell, b) = (+5.4, -0.4)$ \citep{gramze_evidence_2023}, which have broad lines, warm temperatures, and shocked gas, are sites of high-velocity cloud-cloud collisions along the bar lanes \citep{fux_3d_1999, sormani_geometry_2019}. 

Star formation may be suppressed by the aforementioned extreme processes in the Galactic bar, but also more generally in the bar regions of barred galaxies \citep{athanassoula_existence_1992}. However, observations of star formation in the bar region have been largely limited to nearby galaxies, and the effect of these dynamics on star formation efficiency is still poorly understand \citep{diaz-garcia_molecular_2021, maeda_statistical_2023}. For example, \citet{meier_molecular_2001} find that star formation in IC 342 may be triggered by shocks at the intersection of x\textsubscript{1} and x\textsubscript{2} orbits, where molecular gas driven toward the nucleus by the bar potential collides with gas on the inner orbits. \citet{meier_nuclear_2008} observe a similar presence of dynamically induced star formation at the intersection of the nuclear bar orbits of the galaxy Maffei 2. Meanwhile, \citet{reynaud_kinematics_1998} find that star formation in NGC 1530 is strong at the center of the galaxy and at the ends of the bar but weak along the length of the bar, and they conclude that it is likely inhibited by shocks and shears in the bar lanes. Similarly, observations of four nearby barred galaxies by \citet{james_stellar_2016} show the existence of a central peak and outer ring of strong star formation, but also a region between the two where the star formation is strongly suppressed. And, with a sample of over 1000 galaxies, \citet{vera_effect_2016} show that strong-barred galaxies have lower star formation efficiencies than galaxies with weak or no bars.

In this paper, we have selected 20 clouds that are candidates for being in the Milky Way Galactic bar region outside the CMZ, including several located within B2 and G5. These clouds show bright NH\textsubscript{3} (3,3) emission in the Mopra H\textsubscript{2}O Southern Galactic Plane Survey \cite[HOPS; see][]{walsh_h2o_2011,purcell_h2o_2012, longmore_h2o_2017}, have line widths greater than a few kilometers per second, and exhibit gas temperatures warmer than typical clouds in the Galactic disk ($\gtrsim 10 - 20$ K). We use the Total Power (TP) antennas of the Atacama Large Millimeter/submillimeter Array (ALMA) Atacama Compact Array (ACA) to observe several molecular lines from these 20 clouds in order to probe physical parameters including temperature, shocks, ionization, star formation, and turbulence, and to investigate where these clouds are located. In Section \ref{sec:obs} we describe the observations, ancillary data, and data reduction procedure. In Section \ref{sec:loc} we discuss the locations of the clouds. In Section \ref{sec:results} we calculate and compare the temperature, turbulence, star formation rates (SFRs), shock properties, and masses of the clouds. In Section \ref{sec:discussion} we discuss the relationships between the cloud properties, and we conclude in Section \ref{sec:conclusion}.

\begin{figure*}[t]
\centering
\includegraphics[width=7in]{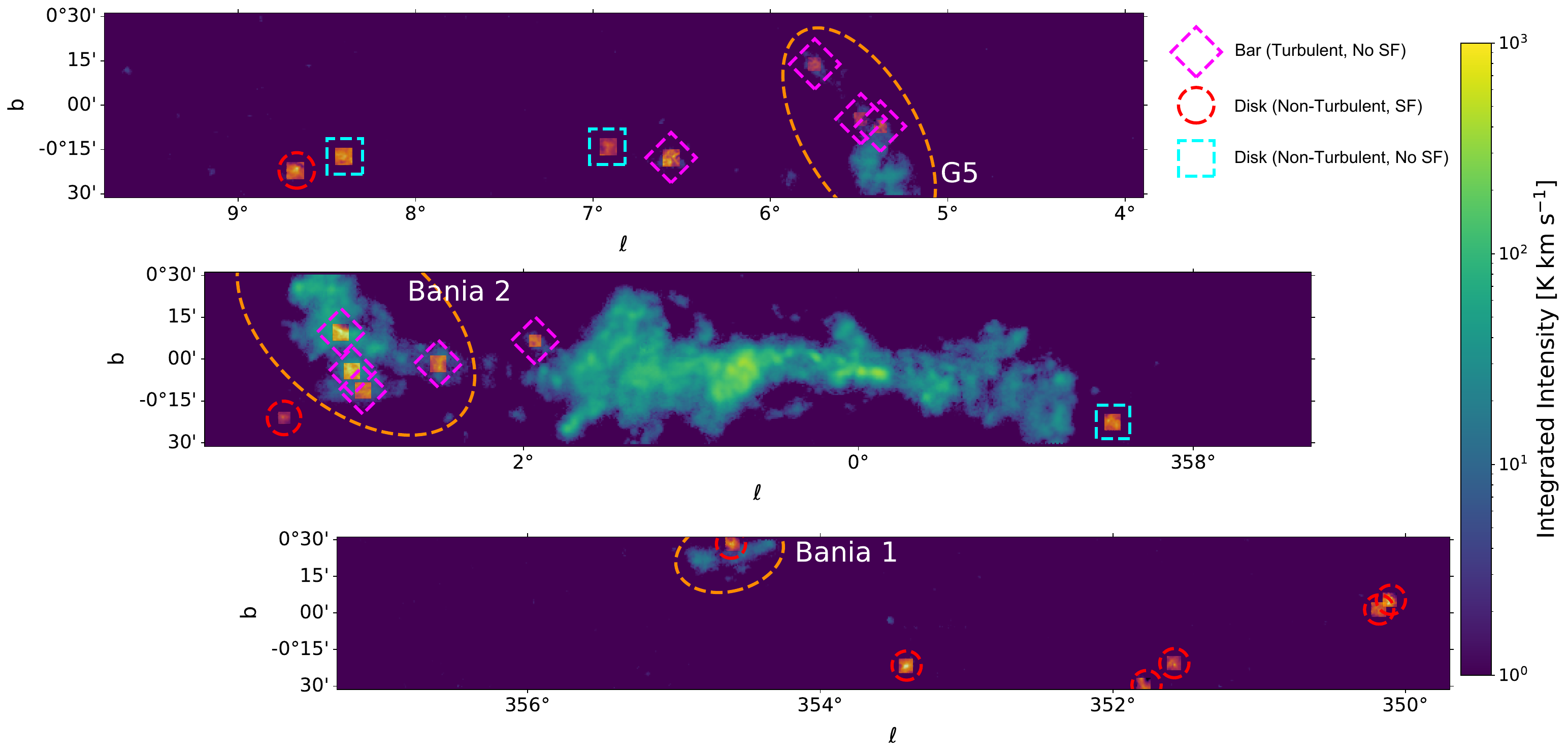}
\caption{Spatial distribution of the 20 molecular clouds in our sample. The background and associated color bar is the integrated intensity of NH\textsubscript{3} (3,3) from HOPS \citep{walsh_h2o_2011,purcell_h2o_2012, longmore_h2o_2017}. The overlays are our \textsuperscript{13}CO $J=2 \to 1$ moment 0 maps from ALMA. The clouds G5 at $(\ell, b) = (+5.4, -0.4)$, Bania 1 (B1; \citealt{bania_clump_1986}) at $(\ell, b) = (-5.4, +0.4)$, and Bania 2 (B2; \citealt{stark_clump_1986}) at $(\ell, b) = (+3, +0.2)$ are marked with large orange ellipses. The colored shapes around each of the clouds in our sample correspond to our categories, first mentioned in Section \ref{sec:loc} and refined in Section \ref{subsec:loc2}. The pink diamonds indicate the clouds that are highly turbulent and have no detected star formation; these are the most likely to be in the bar region. The red circles indicate the clouds with detected star formation but low turbulence; these are likely to be typical star-forming regions in the Galactic disk. Lastly, the blue squares indicate the clouds with low turbulence and no detected star formation; these are likely to be typical non-star-forming clouds in the Galactic disk.}
\label{fig:hopsOverlay}
\end{figure*}

\section{Observations and Data Reduction}
\label{sec:obs}

\subsection{Sample Selection}
\label{subsec:sampselec}
We selected 20 clouds from HOPS that may be in the Milky Way Galactic bar region. The clouds were selected on the basis of bright, large-scale NH\textsubscript{3} (3,3) emission,  which is only observed toward a few objects, and which indicates that the gas temperatures may be warmer than typical clouds in the Galactic disk. In addition, several of the clouds have ammonia line widths much broader than typical clouds in the Galactic disk. The combination of these two properties suggests that the clouds may not be located in the Milky Way disk but rather be part of the central Galactic bar.

The spatial distribution of the 20 selected clouds is shown in Figure \ref{fig:hopsOverlay}. We discuss the location of these clouds in the Galaxy in Section \ref{sec:loc}. 

\subsection{ALMA Observations and Data Reduction}
\label{subsec:alma}

The clouds were observed with the ALMA-ACA between 2021 May and 2023 May (project codes: 2019.2.00068.S, 2021.2.00001.S, 2022.1.00591.S; PI: Ott, J.) using both the 7 m array and the TP antennas. We analyze the TP data in this work. The spectral windows used cover lines of several important molecules: the carbon monoxide isotopologues \textsuperscript{12}CO, \textsuperscript{13}CO, and C\textsuperscript{18}O; the shock tracers SiO (e.g.,\ \citealt{schilke_sio_1997}) and methanol (CH\textsubscript{3}OH; e.g.,\ \citealt{meier_spatially_2005}); formaldehyde (H\textsubscript{2}CO), from which the gas temperature can be determined (e.g.,\ \citealt{mangum_formaldehyde_1993, ginsburg_dense_2016}); the dense molecular gas tracer HC\textsubscript{3}N (e.g.,\ \citealt{meier_cyanoacetylene_2011, mills_dense_2018}); and the radio recombination line H30$\alpha$. The observing parameters of the spectral windows and the specific lines are shown in Table \ref{table:obs_params}.

\begin{table*}[t]
\centering
\setlength{\tabcolsep}{2em}
\begin{tabular}{c  c  c  c} 
 \hline
 Spectral Line & Rest Frequency (GHz) & Bandwidth (GHz) & No. of Channels \\ [0.5ex] 
 \hline\hline
 SiO $J=5\to 4$ & 217.1050 & 0.25 & 512 \\
 \hline
 H\textsubscript{2}CO $J=3_{21} \to 2_{20}$ & 218.7601 & 0.25 & 512 \\
 \hline
 H\textsubscript{2}CO $J=3_{03} \to 2_{02}$ & 218.2222 & 0.25 & 512 \\
 \hline
 HC\textsubscript{3}N $J=24 \to 23$ & 218.3247 & 0.25 & 512 \\ 
 CH\textsubscript{3}OH $J = 4_{22} \to 3_{12}$ & 218.4401 & & \\
 \hline
 C\textsuperscript{18}O $J=2 \to 1$ & 219.5604 & 0.25 & 1024 \\
 \hline
 \textsuperscript{13}CO $J=2 \to 1$ & 220.3987 & 0.25 & 1024 \\
 \hline
 \textsuperscript{12}CO $J=2 \to 1$ & 230.5380 & 0.25 & 2048 \\
 \hline
 H30$\alpha$ & 231.9009 & 2 & 2048 \\
 \hline
\end{tabular}
\caption{Observing parameters for each spectral window.}
\label{table:obs_params}
\end{table*}

\begin{table*}[t]
\centering
\setlength{\tabcolsep}{0.7em}
\begin{tabular}{c  c  c  c  c  c  c  c  c} 
 \hline
 Cloud & $\ell$ [$^\circ$] & $b$ [$^\circ$] & NH\textsubscript{3} (1,1) $\Delta$v [km\ s$^{-1}$] & v$_\mathrm{lsr}$ [km\ s$^{-1}$] & v$_\mathrm{int}$ [km\ s$^{-1}$] & $D$ [kpc] & $R_\text{Gal}$ [kpc] & $D_\text{ref}$  \\ [0.5ex]
 \hline\hline
G008.67 & 8.67 & -0.37 & $3.31 \pm 0.13$ & $35.10 \pm 0.19$ & $24\textup{--}52$ & $3.4 \pm 0.3$ & $4.8 \pm 1.0$ & \textbf{1} \\ 
 
G008.40 & 8.40 & -0.29 & $2.38 \pm 0.08$ & $36.79 \pm 0.07$ & $31\textup{--}42$ & $4.5 \pm 1.0$ & $3.8 \pm 2.1$ & \textbf{2} \\ 
 
G006.92 & 6.92 & -0.23 & $2.73 \pm 0.23$ & $19.21 \pm 0.26$ & $5\textup{--}35$ & $3.5 \pm 1.0$ & $4.7 \pm 3.3$ & \textbf{2} \\ 
 
\textbf{\underline{G006.56}} & 6.56 & -0.30 & $24.34 \pm 1.98$ & $7.41 \pm 0.04$ & $-10\textup{--}45$ & $6.9 \pm 1.0$ & $1.6 \pm 0.6$ & \textbf{3}  \\ 
 
\textbf{\underline{G005.75}} & 5.75 & 0.23 & $27.43 \pm 1.93$ & $70.11 \pm 0.27$ & $31\textup{--}100$ & $7.0 \pm 1.0$ & $1.4 \pm 0.5$ & \textbf{3} \\ 
 
\textbf{\underline{G005.49}} & 5.49 & -0.08 & $20.35 \pm 2.48$ & $53.57 \pm 0.16$ & $25\textup{--}95$ & $7.0 \pm 1.0$ & $1.3 \pm 0.5$ & \textbf{3} \\ 
 
\textbf{\underline{G005.38}} & 5.38 & -0.12 & $27.36 \pm 2.06$ & $38.16 \pm 0.18$ & $0\textup{--}75$ & $7.1 \pm 1.0$ & $1.3 \pm 0.5$ & \textbf{3} \\ 
 
G003.43 & 3.43 & -0.35 & $3.54 \pm 0.22$ & $-25.36 \pm 0.26$ & $-50\textup{--}5$ & $2.5 \pm 1.0$ & $5.7 \pm 5.0$ & \textbf{3}, \textbf{4} \\ 
 
\textbf{\underline{G003.09}} & 3.09 & 0.16 & $42.17 \pm 0.64$ & $151.06 \pm 0.41$ & $90\textup{--}200$ & $7.5 \pm 1.0$ & $0.8 \pm 0.3$ & \textbf{3} \\ 
 
\textbf{\underline{G003.02}} & 3.02 & -0.07 & $38.43 \pm 0.62$ & $42.17 \pm 1.00$ & $-48\textup{--}145$ & $7.5 \pm 1.0$ & $0.8 \pm 0.3$ & \textbf{3} \\ 
 
\textbf{\underline{G002.96}} & 2.96 & -0.19 & $34.09 \pm 0.63$ & $97.16 \pm 0.59$ & $30\textup{--}130$ & $7.5 \pm 1.0$ & $0.8 \pm 0.3$ & \textbf{3} \\ 
 
\textbf{\underline{G002.51}} & 2.51 & -0.03 & $37.74 \pm 0.96$ & $8.21 \pm 0.53$ & $-35\textup{--}20$ & $7.6 \pm 1.0$ & $0.7 \pm 0.2$ & \textbf{3} \\ 
 
\textbf{\underline{G001.93}} & 1.93 & 0.11 & $30.61 \pm 1.59$ & $23.64 \pm 0.63$ & $2\textup{--}76$ & $7.7 \pm 1.0$ & $0.5 \pm 0.2$ & \textbf{3} \\ 
 
G358.48 & 358.48 & -0.38 & $2.80 \pm 0.17$ & $-1.67 \pm 0.06$ & $-17\textup{--}30$ &  $2.5 \pm 1.0$ & $5.7 \pm 5.0$ & \textbf{3} \\ 
 
G354.61 & 354.61 & 0.47 & $2.30 \pm 0.20$ & $-20.57 \pm 0.09$ & $-35\textup{--}-9$ & $12.3 \pm 2.2$ & $4.2 \pm 1.8$ & \textbf{5} \\ 
 
G353.41 & 353.41 & -0.36 & $3.73 \pm 0.15$ & $-15.44 \pm 0.20$ & $-25\textup{--}-4$ &$2.0 \pm 0.7$ & $6.2 \pm 5.0$ & \textbf{1} \\ 
 
G351.77 & 351.77 & -0.49 & $3.05 \pm 0.23$ & $-3.23 \pm 0.25$ & $-15\textup{--}20$ & $2.0 \pm 0.1$ & $6.2 \pm 5.0$ & \textbf{6} \\ 
 
G351.58 & 351.58 & -0.34 & $4.22 \pm 0.28$ & $-92.18 \pm 0.28$ & $-105\textup{--}-80$ &$5.2 \pm 1.0$ & $3.1 \pm 1.4$ & \textbf{7} \\ 
 
G350.18 & 350.18 & 0.02 & $24.32 \pm 1.37$ & $-66.50 \pm 0.13$ & $-90\textup{--}-49$ & $6.2 \pm 0.8$ & $2.3 \pm 0.7$ & \textbf{4}, \textbf{8} \\ 
 
G350.10 & 350.10 & 0.09 & $21.10 \pm 0.94$ & $-69.25 \pm 0.16$ & $-85\textup{--}-45$ & $6.2 \pm 0.8$ & $2.3 \pm 0.7$ & \textbf{8} \\ 
\hline

\end{tabular}
\caption{Locations of all clouds. The clouds we take to be inside the bar have their names underlined and in bold. The NH\textsubscript{3} (1,1) line width and its associated error are calculated as the FWHM of a Gaussian fit to the spectrum of a circular area around the peak integrated emission pixel in the Cube Analysis and Rendering Tool for Astronomy (CARTA; \citealt{comrie_carta_2021}). The central velocity of each cloud, v$_\mathrm{lsr}$, and its associated error are calculated from the H\textsubscript{2}CO $J=3_{03} \to 2_{02}$ data cubes as the mean of a Gaussian fit to the spectrum of the 1' box centered on the pixel with maximum integrated intensity. v$_\mathrm{int}$ indicates the velocities over which the emission is integrated for the maps detailed in Section \ref{subsec:maps}, as determined by the extent of the cloud in \textsuperscript{13}CO $J = 2\to 1$ emission. For clouds with no kinematic distance error reported, we take the error to be 1.0 kpc and note that the actual error may be larger. The distance references are as follows: \textbf{1)} \citet{motte_alma-imf_2022}; \textbf{2)} \citet{yang_sedigism_2022}; \textbf{3)} This work; \textbf{4)} \citet{longmore_h2o_2017}; \textbf{5)} \citet{jones_h_2013}; \textbf{6)} \citet{reyes-reyes_benchmarking_2024}; \textbf{7)} \citet{green_distances_2011}; \textbf{8)} \citet{nandakumar_star-forming_2016}}
\label{table:cloudLocations}
\end{table*}

\begin{table*}[t]
\centering
\begin{tabular}{c  c  c  c  c  c  c  c  c  c  c  c} 
 \hline
 Cloud & SiO & H\textsubscript{2}CO $(3_{21}-2_{20})$ & H\textsubscript{2}CO $(3_{03}-2_{02})$ & HC\textsubscript{3}N & CH\textsubscript{3}OH & H30$\alpha$ & NH\textsubscript{3} $(2,2)$ & NH\textsubscript{3} $(6,6)$\\ [0.5ex] 
 \hline\hline
 G008.67 & \checkmark & \checkmark & \checkmark & \checkmark & \checkmark & \checkmark & \checkmark & $\hdots$ \\
 
 G008.40 & \checkmark & \checkmark & \checkmark & $\hdots$  & \checkmark & $\hdots$ & \checkmark & $\hdots$ \\
 
 G006.92 & \checkmark & \checkmark & \checkmark & $\hdots$ & \checkmark & $\hdots$ & \checkmark & $\hdots$ \\
 
 G006.56 & \checkmark & \checkmark & \checkmark & \checkmark & \checkmark & $\hdots$ & \checkmark &  $\hdots$\\
 
 G005.75 & $\hdots$ & $\hdots$ & \checkmark & $\hdots$ & \checkmark & $\hdots$ & \checkmark & $\hdots$\\
 
 G005.49 &  $\hdots$ & $\hdots$ & \checkmark & $\hdots$ & $\hdots$ & $\hdots$ & $\hdots$ & $\hdots$ \\
 
 G005.38 & \checkmark & \checkmark & \checkmark & $\hdots$ & \checkmark & $\hdots$ & \checkmark & $\hdots$ \\
 
 G003.43 & \checkmark & \checkmark & \checkmark & \checkmark & \checkmark & $\hdots$ & \checkmark &  $\hdots$ \\
 
 G003.09 & \checkmark & \checkmark & \checkmark & \checkmark & \checkmark & $\hdots$ & \checkmark & \checkmark \\
 
 G003.02 & \checkmark & \checkmark & \checkmark & \checkmark & \checkmark & $\hdots$ & \checkmark & \checkmark \\
 
 G002.96 & \checkmark & \checkmark & \checkmark & $\hdots$ & \checkmark & $\hdots$ & \checkmark & \checkmark \\
 
 G002.51 & \checkmark & \checkmark & \checkmark & $\hdots$ & \checkmark & $\hdots$ & \checkmark & \checkmark \\
 
 G001.93 & \checkmark & \checkmark & \checkmark & $\hdots$ & \checkmark & $\hdots$ & \checkmark & $\hdots$ \\
 
 G358.48 & \checkmark & \checkmark & \checkmark & \checkmark & \checkmark & $\hdots$ & \checkmark & $\hdots$ \\
 
 G354.61 & \checkmark & \checkmark & \checkmark & \checkmark & \checkmark & \checkmark & \checkmark &  $\hdots$ \\
 
 G353.41 & \checkmark & \checkmark & \checkmark & \checkmark & \checkmark & \checkmark & \checkmark &  $\hdots$ \\
 
 G351.77 & \checkmark & \checkmark & \checkmark & \checkmark & \checkmark & \checkmark & \checkmark &  $\hdots$ \\
 
 G351.58 & \checkmark & \checkmark & \checkmark & \checkmark & \checkmark & \checkmark & \checkmark &  $\hdots$ \\
 
 G350.18 & \checkmark & \checkmark & \checkmark & $\hdots$ & \checkmark & \checkmark & \checkmark & $\hdots$ \\
 
 G350.10 & \checkmark & \checkmark & \checkmark & \checkmark & \checkmark & \checkmark & \checkmark & $\hdots$ \\
 \hline

\end{tabular}
\caption{Targeted lines detected in each cloud. We consider a detection to be at the 4$\sigma_\text{rms}$ level, the same as the cutoffs for the integrated intensity maps. \textsuperscript{12}CO $J=2\to 1$, \textsuperscript{13}CO $J=2\to 1$, C\textsuperscript{18}O $J=2\to 1$, NH\textsubscript{3} $(1,1)$, and NH\textsubscript{3} $(3,3)$ were detected in all clouds.}
\label{table:detections}
\end{table*}

We used the default ALMA Pipeline Reduction, which utilized the Common Astronomy Software Application (CASA) versions 6.2.1.7-6.4.1.12 \citep{the_casa_team_casa_2022}. We visually inspected and selected the velocity and channel ranges without any spectral features for the presence of continuum emission. If found, the continuum was subtracted.

The native full width at half-maximum (FWHM) beam sizes of the observations are between approximately 28\arcsec\ and 30\arcsec. We smooth the cubes to all have the same beam size of 31\arcsec. For each cloud, we also regrid the cubes to match the \textsuperscript{13}CO $J=2\to 1$ pixel size of 2.935\arcsec and velocity resolution of 0.332 km\ s$^{-1}$. In this channel width of 0.332 km\ s$^{-1}$, the 1$\sigma_\text{rms}$ root-mean-square noise (rms) over the line-free channels varies from about 0.22 to 0.28 Jy\ beam$^{-1}$. We also convert from intensity $I_\nu$ in units of Jy\ beam$^{-1}$ to a brightness temperature (kelvins), in which case the rms noise ranges from about 5.2 to 7.7 mK.

\subsection{ALMA Maps}
\label{subsec:maps}
To make integrated intensity (``moment 0'') maps, we set a cutoff at 4$\sigma_\text{rms}$, then integrate over the velocity range of the cloud in \textsuperscript{13}CO $J=2\to 1$, which is observed for all clouds and is less contaminated with foreground and background emission than the more common isotopologue \textsuperscript{12}CO $J = 2 \to 1$. The velocity range of each cloud is given in Table \ref{table:cloudLocations}. We also generate intensity-weighted velocity (``moment 1''), intensity-weighted velocity dispersion (``moment 2''), and peak intensity maps for each cube. For these images, we again use a 4$\sigma_\text{rms}$ pixel cutoff, as well as a 5$\sigma$ threshold on the moment 0 map, where $\sigma$ is estimated as
\begin{equation}
\label{eq:noise}
\sigma \approx \sqrt{N} \sigma_\text{rms} \Delta v.
\end{equation}
Here, $N$ is the number of channels integrated over for the moment 0 map, and $\Delta v = 0.332$ km\ s$^{-1}$ is the channel velocity width. Although $N$ varies across a single map, since we also use a 4$\sigma_\text{rms}$ cutoff when creating integrated intensity maps, we set $N$ to be constant at the number of channels corresponding to the full integration velocity range, which is likely more than the actual number of channels integrated over for most pixels. This effectively corresponds to the error assuming no masking in the integrated intensity map, which represents a maximum possible $N$, and thus a constant ``maximum" $\sigma$. We then calculate ratio maps between each of the lines and both \textsuperscript{12}CO $J=2\to 1$ and \textsuperscript{13}CO $J=2\to 1$ using the respective moment 0 maps. For each moment 0 ratio map, $R = \frac{M_0^i}{M_0^j}$, between lines $i$ and $j$, we calculate a corresponding error map, where the error $\delta R$ is
\begin{equation}
\label{eq:ratioError}
\delta R = R\sqrt{\left(\frac{\sigma_i}{M_0^i} \right)^2 + \left( \frac{\sigma_j}{M_0^j} \right)^2},
\end{equation}
where $M_0 = \int I_v dv$ is the integrated intensity of a line, and $\sigma$ is the error of the moment 0 map (Eq.\ \ref{eq:noise}). Although this equation does not hold under all circumstances, we use it as an approximation of the error for the integrated intensity ratios. We note that although $\sigma$ is assumed to be constant across each moment 0 map, as discussed above, the ratio error maps are not constant since they are also functions of the individual moment 0 maps. The same holds for the temperature error maps described in Section \ref{subsec:gastemp}.

We also create position-velocity (PV) diagrams for each cube using the packages \texttt{pvextractor}\footnote{\url{https://github.com/radio-astro-tools/pvextractor}} and \texttt{spectral-cube}\footnote{\url{https://github.com/radio-astro-tools/spectral-cube}} \citep{ginsburg_radio-astro-toolsspectral-cube_2019}. We select a path going through the main regions of each cloud, then calculate the PV diagram using the cloud velocity range, with an extra 20 km\ s$^{-1}$ on each side.

\begin{figure*}[t]
\centering
\includegraphics[width=7in]{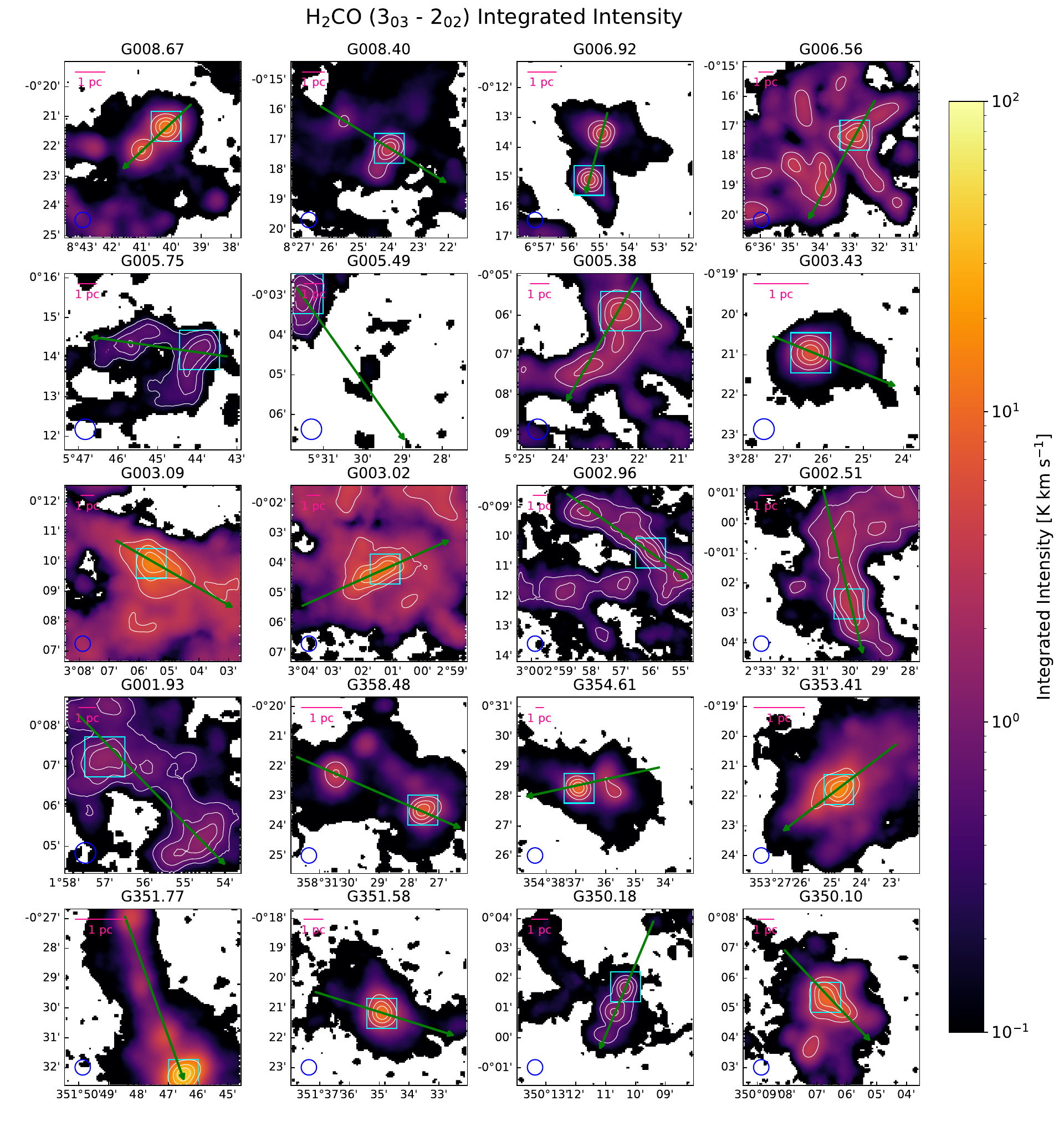}
\caption{Integrated intensity maps of  H\textsubscript{2}CO $(3_{03}-2_{02})$ for each cloud on the same color scale, plotted in Galactic coordinates. Clouds are ordered by Galactic longitude. The light blue boxes are 1' boxed centered on the pixel with peak H\textsubscript{2}CO $(3_{03}-2_{02})$ integrated intensity. The green arrows indicates the slices along which the PV diagrams are taken. The blue circles in the bottom left corner of each map indicate the 31" ALMA beam size. The pink lines in the top left corner of each map indicate a physical size of 1 pc. The uncertainties on these physical sizes may be large as they are proportional to the distance uncertainties. The complete figure set (9 images), which contains integrated intensity maps for all targeted molecular lines, is available in the online journal. We show H\textsubscript{2}CO $(3_{03}-2_{02})$ integrated intensity here because it gives a good representation of the cloud structure of the denser gas.
}
\label{fig:integratedIntensities}
\end{figure*}

\begin{figure*}[t]
\centering
\includegraphics[width=7in]{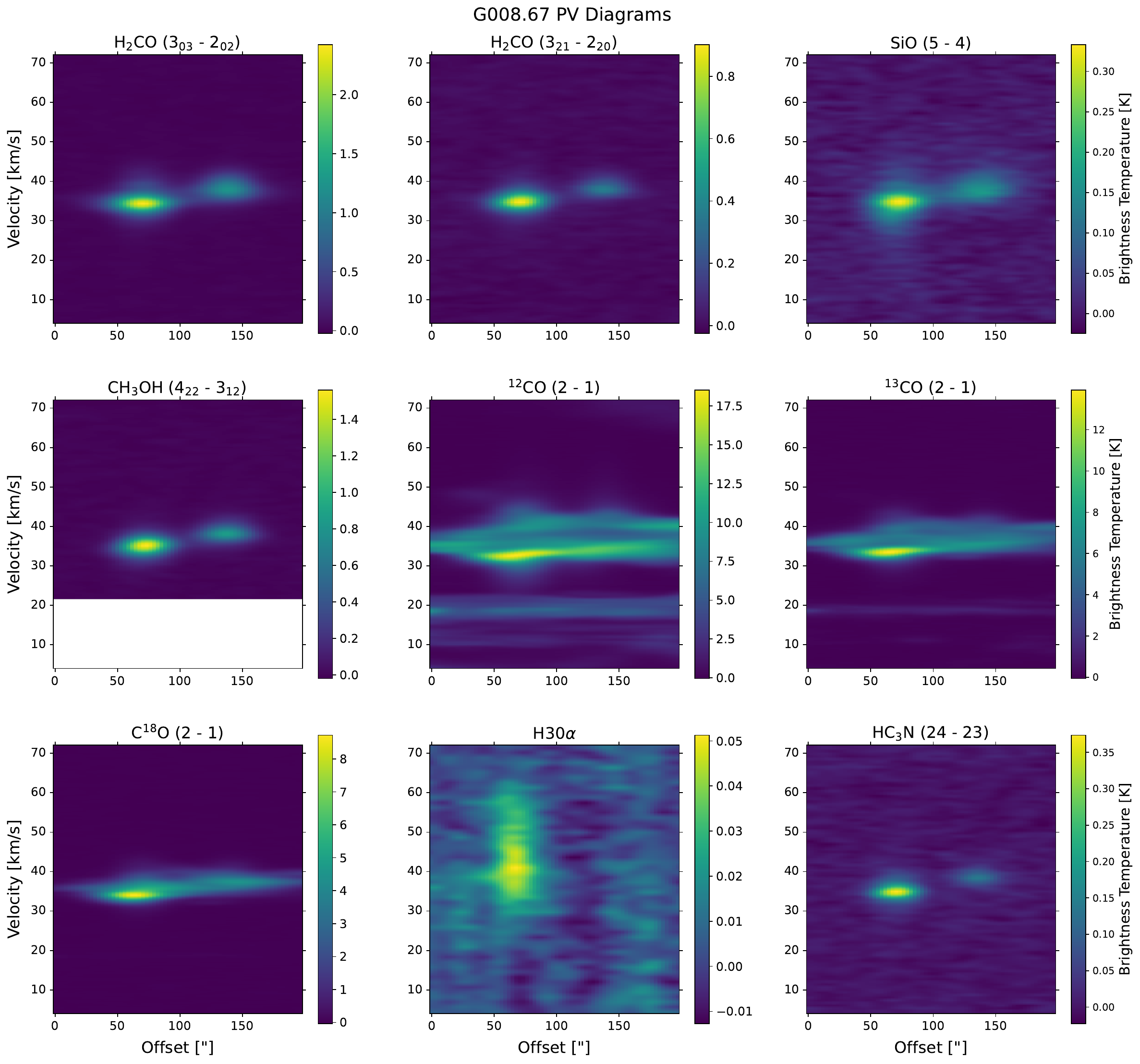}
\caption{PV diagrams of all lines for G008.67 along the slice depicted in Figure \ref{fig:integratedIntensities}. The velocity axis extends 20 km\ s$^{-1}$ beyond the range of the cloud determined with \textsuperscript{13}CO (given in Table \ref{table:cloudLocations}) in both directions. The complete figure set (20 images), which contains PV diagrams for all clouds in our sample, is available in the online journal.
}
\label{fig:pvDiagrams}
\end{figure*}

All clouds display clear emission for all three carbon monoxide isotopologues, although some other lines do not have a significant detection. The detected lines for each cloud are shown in Table \ref{table:detections}. We show the integrated intensity maps of all clouds for each line in Figure \ref{fig:integratedIntensities}, and the PV diagrams in Figure \ref{fig:pvDiagrams}. The full figure sets for these figures are available in the online journal.

\subsection{Ancilllary Data}

\subsubsection{HOPS}

We use HOPS for measurements of metastable ammonia inversion transitions with $(J, K) = (1,1),\ (2,2),\ (3,3),$ and $(6,6)$, for which the rotational temperature is similar to the gas kinetic temperature \citep{huettemeister_multilevel_1995, ott_temperature_2005}. HOPS used the 22 m Mopra radio telescope, which has a main-beam FWHM of 2' at 12 mm \citep{urquhart_characterisation_2010}, and covered a velocity range of -200 to 200 km\ s$^{-1}$, which may exclude parts of the bar dust lanes that are at even higher velocities. Further observation details about the HOPS ammonia catalog can be found in \citet{purcell_h2o_2012}.

The clouds were initially selected to be bright in NH\textsubscript{3} $(3,3)$ emission, and all clouds are observed in the NH\textsubscript{3} $(1,1)$ line, although not all clouds have significant NH\textsubscript{3} $(2,2)$ or $(6,6)$ emission. The ammonia line detections are also shown in Table \ref{table:detections}. For each cloud, we select a region that covers the full extent of the cloud, which in general is larger than the ALMA field of view (FOV) of the cloud. We then make a moment 0 map of each region for all four ammonia lines using the same velocity range as used for the ALMA moment maps.

\subsubsection{Spitzer}

We additionally use data from two Spitzer Galactic plane surveys, the Galactic Legacy Infrared Midplane Survey Extraordinaire (GLIMPSE; \citealt{churchwell_spitzerglimpse_2009}, \citealt{glimpse_team_galactic_2020}) and the Multiband Imaging Photometer Galactic Plane Survey (MIPSGAL; \citealt{rieke_multiband_2004}, \citealt{carey_mipsgal_2009}, \citealt{mipsgal_team_24_2020}). We use the cutout service of the Infrared Science Archive (IRSA) to take 10\arcmin\ wide images containing the central location of each cloud in the 4.5 and 8 $\mu$m bands of the Infrared Array Camera \citep{fazio_infrared_2004} from GLIMPSE. We also take 30\arcmin\ wide images of the same locations in the 24 $\mu$m band of MIPS, then regrid to match the pixel spacing of the GLIMPSE data using the \texttt{reproject}\footnote{\url{https://github.com/astropy/reproject}} package.


\section{Cloud Locations}
\label{sec:loc}

\begin{figure*}[t]
\centering
\includegraphics[width=6in]{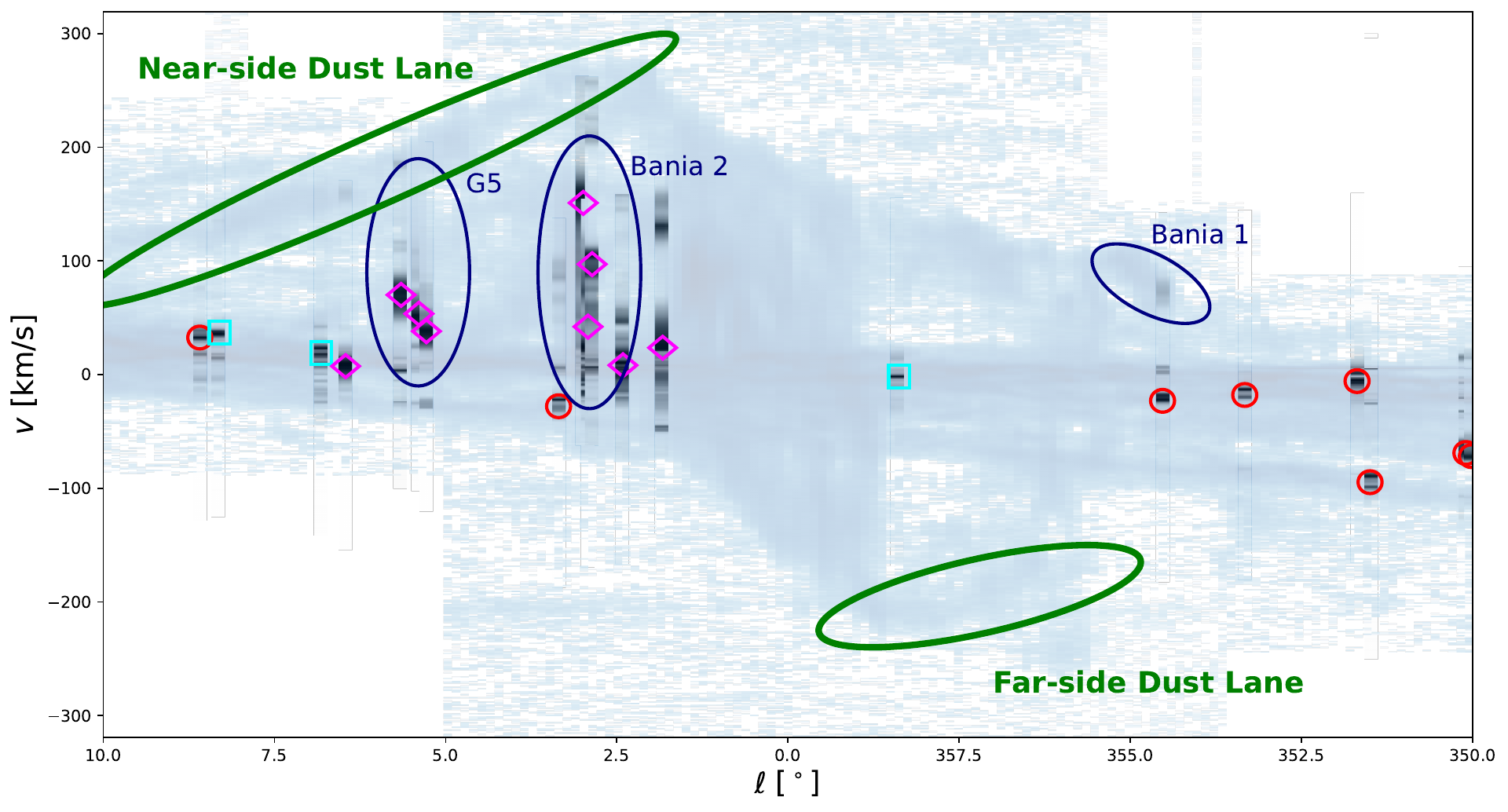}
\caption{The longitude-velocity ($\ell$-v) distribution of the cloud sample overlaid on CO $J = 1 \to 0$ emission from \citet{bitran_large_1997}, in blue. The black vertical stripes are the velocity spectra of each of the clouds, located horizontally at the cloud's longitude. The colored shapes and symbols along each of the velocity spectra are centered on the central velocity of each cloud, and they correspond to the same categories as in Figure \ref{fig:hopsOverlay}. The pink diamonds indicate the clouds that are highly turbulent and have no detected star formation; these are the most likely to be in the bar region. The red circles indicate the clouds with detected star formation but low turbulence; these are likely to be typical star-forming regions in the Galactic disk. Lastly, the blue squares indicate the clouds with low turbulence and no detected star formation; these are likely to be typical non-star-forming clouds in the Galactic disk. The approximate locations of several important features are also circled.}
\label{fig:lvdiagram}
\end{figure*}

As mentioned in Section \ref{subsec:sampselec}, the clouds in the sample are candidates for being in the Galactic bar region due to their bright and broad NH\textsubscript{3} (3,3) emission. We also note that we find fewer clouds with similar properties at $\lvert \ell \rvert > 10^\circ$ in the HOPS data. Although the locations are uncertain, we take the clouds with a NH\textsubscript{3} (1,1) line width greater than 10 km\ s$^{-1}$ to lie along the Galactic bar. We note that, for all 11 clouds in this selection, the line widths are in fact all above 20 km\ s$^{-1}$, and for all nine clouds not in this selection, the line widths are all below 5 km\ s$^{-1}$, indicating two distinct populations rather than a continuum of line widths. The broad-lined clouds are likely to be in the Galactic bar; however, we note below that two of the clouds in this selection may have Galactic longitudes beyond the extent of the bar. This is also further discussed in Section \ref{subsec:loc2}. We thus consider a total of nine clouds to be in the Galactic bar region, and we assume that they lie along the Galactic bar's major axis. We use a Galactic Center distance of $8178 \pm 26$ pc \citep{abuter_geometric_2019} and an angle between the Galactic bar and the line of sight of $30^\circ \pm 2^\circ$ \citep{wegg_structure_2015} to calculate the distance and Galactocentric radius of each cloud. We also include a 1 kpc uncertainty along the line of sight to account for finite bar thickness. We adopt previous distance measurements using Galactic kinematics or associations with nearby cloud complexes for the clouds not in the bar, noting that these distances may be highly uncertain. 

The longitude-velocity ($\ell$-v) diagram in Figure \ref{fig:lvdiagram} can also provide information about the location of the clouds. We see that many of the clouds are consistent with being outside the bar region in the Galactic disk, which is the generally flat feature at all longitudes between velocities of about -50 and 50 km\ s$^{-1}$, and the dust lane features are at higher velocities than all of the clouds in the sample. This is expected given that the HOPS data extends only to $\pm 200$ km s$^{-1}$, which cuts out much of the dust lane emission. However, we note that processes such as cloud-cloud collisions in the Galactic bar region can create features in the $\ell$-v diagram that extend over a wide range of velocities \citep{sormani_geometry_2019}. For example, \citet{gramze_evidence_2023} find that the G5 region around $\ell = 5.4^\circ$ is likely undergoing a cloud-cloud collision and covers a velocity range from approximately 0 to 200 km\ s$^{-1}$. We therefore continue to consider the aforementioned nine broad-lined clouds to be in the Galactic bar. We discuss these bar clouds in more detail in Appendix \ref{appendix:barClouds}, and we discuss the clouds not in the bar below.

\subsection{Clouds Not in the Galactic Bar}

\textit{G008.67}, also known as IRAS 18032-2137, is included in the ALMA Three-millimeter Observations of Massive Star-forming regions (ATOMS) survey \citep{liu_atoms_2020}, which uses kinematic distances calculated from the APEX Telescope Large Area Survey of the Galaxy (ATLASGAL; \citealt{urquhart_atlasgal_2018}). G008.67 is reported to have a distance of 5.2 kpc in the ATOMS survey. It is also included in the ALMA-IMF Large Program \citep{motte_alma-imf_2022}, which also uses ATLASGAL kinematic distances but includes revisions made possible by other surveys such as the Bar and Spiral Structure Legacy Survey \citep{reid_trigonometric_2014}. We adopt the ALMA-IMF kinematic distance of 3.4 $\pm$ 0.3 kpc.

\textit{G003.43}. \citet{longmore_h2o_2017} note that the narrow velocity dispersion of G003.43 is consistent with being outside the Galactic center. Although they adopt a kinematic distance of 21.1 kpc, they find that the probability of the cloud being at the far kinematic distance is low. They conclude that it likely lies on the nearside of the Galactic bar but are unable to calculate a kinematic distance given its proximity to the Galactic Center. The cloud central velocity indicates that it is likely local to the Sun, on the order of 1 kpc, so we adopt a distance of 2.5 kpc but note this is highly uncertain.

\textit{G358.48} does not have a previously reported distance. We assume that its near kinematic distance is more likely than its far kinematic distance, since HOPS has a 3.2 kpc sensitivity limit for typical 400 M$_\odot$ clouds with the NH\textsubscript{3} (1,1) transition \citep{purcell_h2o_2012}, and its central velocity also indicates it as being local to the Sun, so we adopt the same distance of 2.5 kpc as for G003.43.

\textit{G354.61}. Although G354.61 appears to be associated with the clump B1, its central velocity does not match, as can be seen in Figure \ref{fig:lvdiagram}. \citet{jones_h_2013} find its kinematic distance to be 12.3 $\pm$ 2.2 kpc.

\textit{G353.41}, also known as IRAS 17271-3439, is reported to have a kinematic distance of 3.1 kpc in the ATOMS survey, but we instead adopt the revised kinematic distance of 2.0 $\pm$ 0.7 kpc from ALMA-IMF.

\textit{G351.77}, also known as IRAS 17233-3606, is reported to have a distance of 1.34 kpc in the ATOMS survey. This is consistent with the ALMA-IMF distance of 2.0 $\pm$ 0.7 kpc. \citet{reyes-reyes_benchmarking_2024} find a consistent distance of 2.0 $\pm$ 0.1 kpc using astrometric data from Gaia Data Release 3 \citep{gaia_collaboration_gaia_2016, gaia_collaboration_gaia_2023} for sources associated with the cloud, which we adopt.

\textit{G351.58}, also known as IRAS 17220-3609, is reported to have a distance of 8.0 kpc in the ATOMS survey, although \citet{green_distances_2011} place it at a distance of 5.2 kpc using kinematic estimates from maser emission. We adopt the latter since it agrees with the kinematic assessment of the cloud as belonging to the near 3 kpc arm, which is evident in Figure \ref{fig:lvdiagram}.

\begin{figure}[t]
\centering
\includegraphics[width=3in]{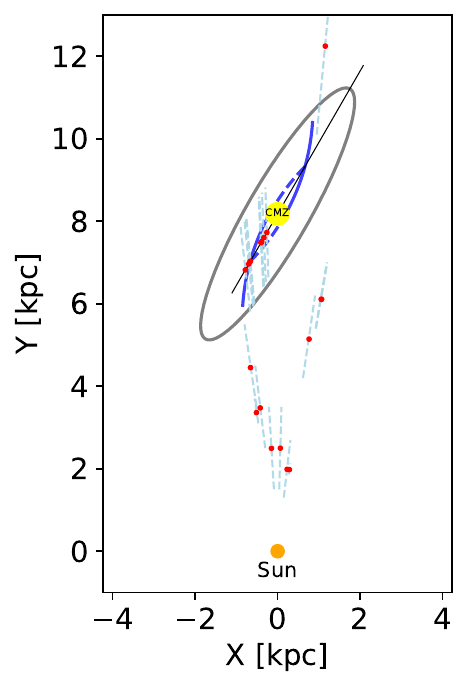}
\caption{Top-down projection view of our cloud distribution. In this cartoon, the bar is at an angle of $30^\circ$ with respect to the Sun, has a semimajor axis of 3.5 kpc, and a semiminor axis length of 0.75 kpc. The black line along the bar axis extends to a Galactic longitude of $\pm 10^\circ$ in both directions. The blue curves within the bar and the circle at the center of the bar are cartoon depictions of the bar inflows and the CMZ. The clouds, denoted by the red dots, are placed at their respective distances with uncertainties along the lines of sight, denoted by the light blue dashed lines, as per Table \ref{table:cloudLocations}. These uncertainties may be underestimates, particularly for kinematic distance measurements, which are highly uncertain at small Galactic longitude; see Section \ref{sec:loc} for further discussion. Several clouds, particularly those in the bar, overlap in the diagram and are indistinguishable from other nearby clouds.}
\label{fig:geometry}
\end{figure}

\textit{G350.18 and  G350.10}. Although the two clouds at the most negative Galactic longitude, G350.18 and G350.10, have line widths above our cutoff of 10 km\ s$^{-1}$, we find that their Galactic longitudes are likely beyond the extent of the Galactic bar. If they do lie along the major axis of the bar, the bar would have to extend to over 4 kpc in the negative Galactic longitude (see Figure \ref{fig:geometry}). Indeed, it has been suggested that the bar may extend out to 5 kpc, or possibly even longer \citep{wegg_structure_2015, sormani_stellar_2022}, so these clouds may lie near or around the tip of the bar, but we cannot conclude with certainty that these clouds belong to the Galactic bar region. We discuss the locations of these clouds further in Section \ref{subsec:loc2}. We refer to previous kinematic measurements of their distances, of which there are several.

\textit{G350.10}, also known as IRAS 17160-3707, is reported to have a distance of 10.53 kpc in the ATOMS survey, although \citet{nandakumar_star-forming_2016} adopt a distance of 6.2 kpc using the kinematics determined by \citet{quireza_radio_2006}. We adopt a value of 6.2 $\pm$ 0.8 kpc, as it matches other kinematic distances to this cloud, which range from 5.7 to 7.3 kpc \citep{caswell_southern_1987, peeters_iso_2002, faundez_simba_2004}. Although the cores of G350.10 and G350.18 are separate as seen in the Spitzer data (see Section \ref{subsec:sf}), the two appear to be linked in the ammonia and CO maps. \citet{longmore_h2o_2017} also consider the two clouds to be two components of the same young massive cluster progenitor gas cloud candidate. We therefore adopt the same distance to G350.18 as that of G350.10.

For these clouds, we also calculate a Galactocentric radius using their distance estimates and Galactic longitudes and latitudes. The locations of all clouds are summarized in Table \ref{table:cloudLocations}, as well as the references for the distance estimates. Our reclassification of G350.18 and G350.10 leaves nine clouds as most likely to lie in the Galactic bar region. We show a top-down projection of the distribution in Figure \ref{fig:geometry}.

\begin{figure*}[t]
\centering
\includegraphics[width=7in]{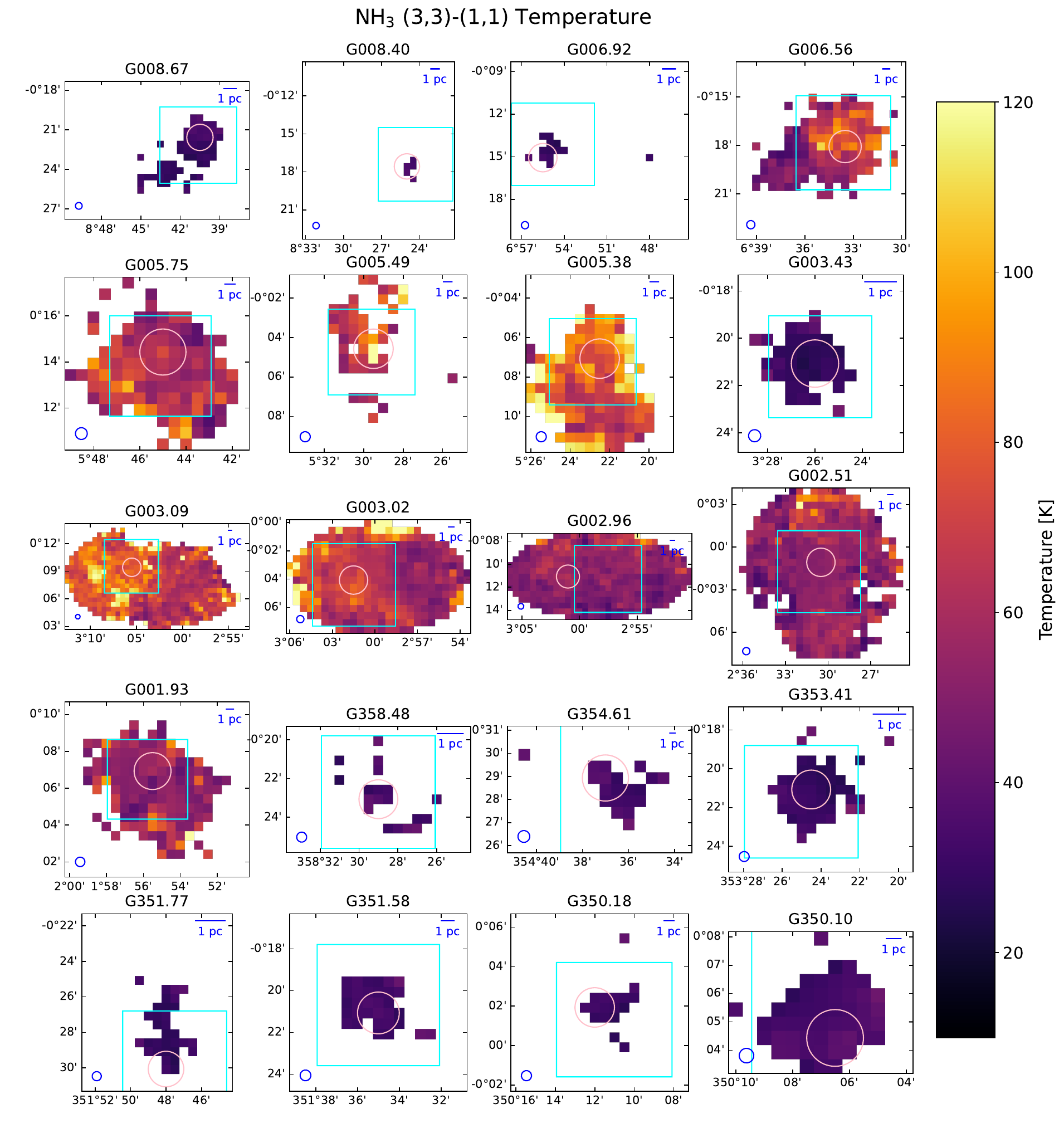}
\caption{Ammonia $(3,3) - (1,1)$ rotational temperature ($T_{13}$) maps for each cloud, on the same color scale, plotted in Galactic coordinates. Clouds are ordered by Galactic longitude. The light blue boxes indicate the ALMA FOV for the cloud. The pink circles are the 2' Mopra HOPS beam, each centered on the pixel with peak NH\textsubscript{3} (3,3) integrated intensity. The blue circles in the bottom left corner of each map indicate the 31" ALMA beam size for comparison. The blue lines in the top right corner of each map indicate a physical size of 1 pc. The uncertainties on these physical sizes may be large as they are proportional to the distance uncertainties.}
\label{fig:NH3temp}
\end{figure*}

\section{Results}
\label{sec:results}

\subsection{Gas Temperature}
\label{subsec:gastemp}

We calculate gas temperatures for all of the clouds using emission from ammonia and formaldehyde.

\subsubsection{Ammonia Temperature}
\label{subsubsec:ammoniaTemp}

We first calculate the rotational temperature using ammonia lines for each cloud following the same procedure outlined in \citet{ott_temperature_2005}. We find that the NH\textsubscript{3} $(1, 1)$ hyperfine structure lines are generally narrow, so we assume that the ammonia emission is optically thin, in which case the column density of an ammonia inversion doublet can be calculated as 
\begin{equation}
\label{eq:NH3col}
N(J, K) = \frac{7.77 \times 10^{13}}{\nu} \frac{J(J+1)}{K^2} \int T_\text{B} dv
\end{equation}
\citep{henkel_dense_2000}, where the column density $N$, rest frequency $\nu$, and integrated main-beam brightness temperature have units of per square centimeter, GHz, and K km\ s$^{-1}$, respectively. We have only metastable ($J = K$) inversions, and the rotational temperature between two such states can be found from the equation
\begin{equation}
\frac{N(J', J')}{N(J, J)} = \frac{g_\text{op}(J')}{g_\text{op}(J)} \frac{2J'+1}{2J+1} \text{exp}\left(\frac{-\Delta E}{T_{JJ'}}\right),
\end{equation}
where $\Delta E$ is the energy level difference between the NH\textsubscript{3} $(J', J')$ and NH\textsubscript{3} $(J, J)$ transitions in K, and $g_\text{op} = 1$ for para-ammonia (i.e., NH\textsubscript{3} $(1, 1)$ and $(2,2)$) and $g_\text{op} = 2$ for ortho-ammonia (i.e., NH\textsubscript{3} $(3, 3)$ and $(6,6)$). We note that the abundance ratio of para- and ortho-ammonia also introduces some additional uncertainty that propagates into the column density and ammonia temperatures calculations.

Solving for the rotational temperature gives
\begin{equation}
\label{eq:NH3temp}
T_{JJ'} = \frac{-\Delta E}{\text{ln}\left(\frac{N(J', J')}{g_\text{op}(J')(2J' + 1)}\right) - \text{ln}\left(\frac{N(J, J)}{g_\text{op}(J)(2J + 1)}\right)}.
\end{equation}
For each cloud, we calculate a temperature map of $T_{12}$, $T_{13}$, and $T_{36}$, or the largest possible subset of the three given that some clouds do not have significant NH\textsubscript{3} $(2,2)$ or $(6,6)$ emission, by applying Eqs. \ref{eq:NH3col} - \ref{eq:NH3temp} to the pairs of ammonia moment 0 maps, as well as a 5$\sigma$ cutoff (Eq.\ \ref{eq:noise}) on both moment 0 maps. We also calculate estimated ammonia temperature error maps using the equation
\begin{equation}
\label{eq:NH3tempErr}
\delta T_{JJ'} = \frac{T^2}{\Delta E}\sqrt{\left(\frac{\delta N(J', J')}{N(J', J')}\right)^2 + \left(\frac{\delta N(J, J)}{N(J, J)}\right)^2},
\end{equation}
where
\[\delta N(J, K) = \frac{7.77 \times 10^{13}}{\nu} \frac{J(J+1)}{K^2} \sigma,\]
where $\sigma$ is the error of the integrated intensity map (Eq.\ \ref{eq:noise}). This equation assumes $\delta N \ll N$, or equivalently $\sigma \ll \int T_B dv$. The ammonia $(3,3) - (1,1)$ ($T_{13}$) temperature maps for all clouds are shown in Figure \ref{fig:NH3temp}. The corresponding error maps are shown in Figure \ref{fig:NH3tempError} in Appendix \ref{appendix:tempErr}.

Although it does not factor into our analysis below, we note that the only four clouds with detections of NH\textsubscript{3} $(6,6)$ are G003.09, G003.02, G002.96, and G002.51, which are the clouds associated with B2, highlighting the extreme nature of this large cloud complex that is the closest region of bright NH\textsubscript{3} $(3,3)$ emission to the CMZ (Figure \ref{fig:hopsOverlay}).

\begin{figure*}[t]
\centering
\includegraphics[width=7in]{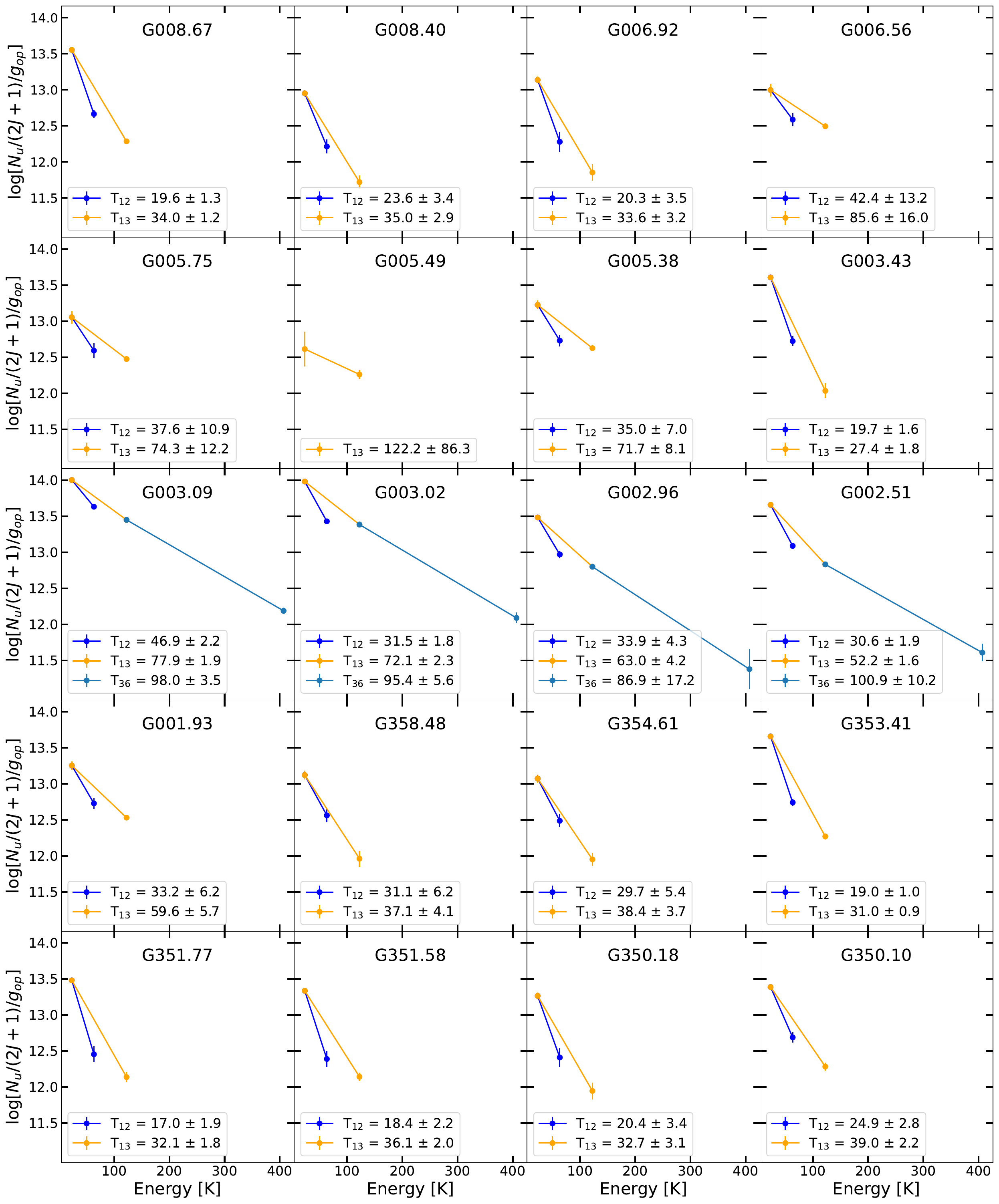}
\caption{Ammonia emission Boltzmann diagrams for all clouds.}
\label{fig:NH3BD}
\end{figure*}

To get a single temperature value for each pair of ammonia lines, we take the temperature and error at the brightest pixel in the NH\textsubscript{3} $(3,3)$ moment 0 map. We show rotation diagrams (or Boltzmann diagrams) of the ammonia lines and the resultant temperature values in Figure \ref{fig:NH3BD}.

\subsubsection{Formaldehyde Temperature}
\label{subsubsec:h2coTemp}

\begin{figure*}[t]
\centering
\includegraphics[width=7in]{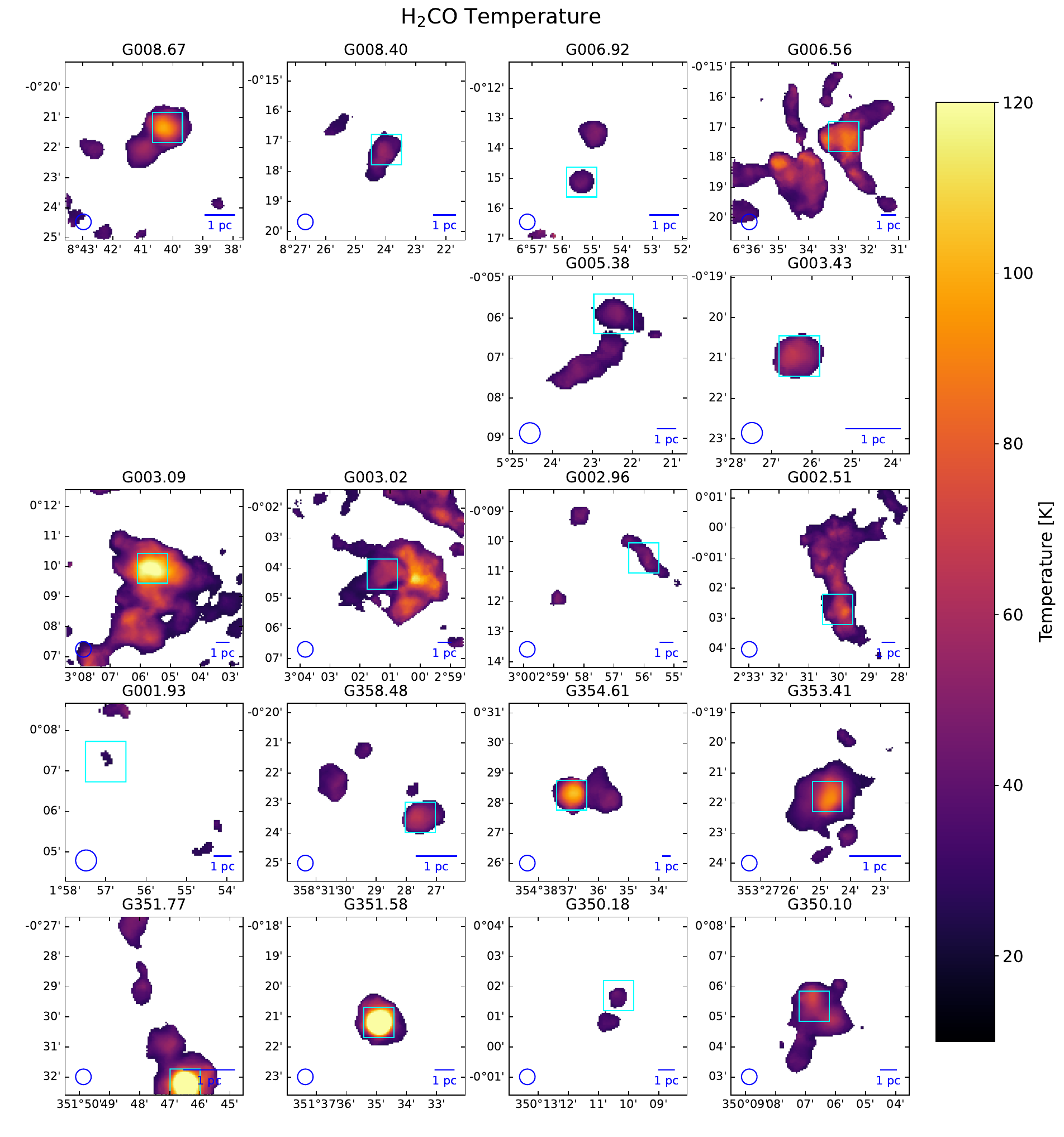}
\caption{Formaldehyde temperature maps for each cloud, on the same color scale, plotted in Galactic coordinates. Clouds are ordered by Galactic longitude. The light blue boxes are 1' boxes centered on the pixel with peak H\textsubscript{2}CO $J=3_{03} \to 2_{02}$ integrated intensity. The blue circles in the bottom left corner of each map indicate the 31" beam size. The blue lines in the bottom right corner of each map indicate a physical size of 1 pc. The uncertainties on these physical sizes may be large as they are proportional to the distance uncertainties. Clouds G005.75 and G005.49 do not show any H\textsubscript{2}CO $J=3_{21} \to 2_{20}$ emission and thus do not have formaldehyde temperature maps.}
\label{fig:H2COtemp}
\end{figure*}

\begin{figure*}[t]
\centering
\includegraphics[width=6.5in]{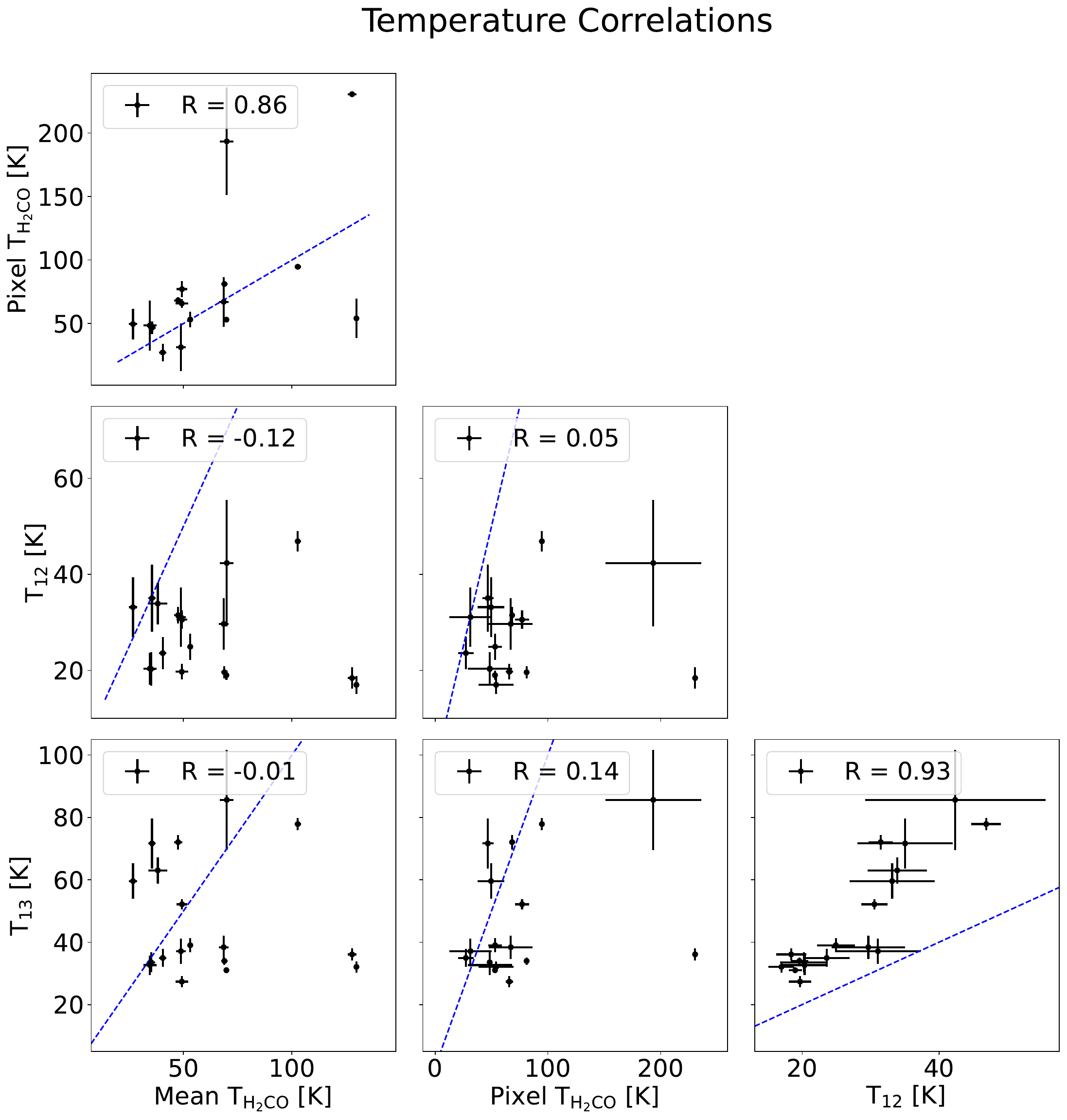}
\caption{Correlations between different temperature measures. Mean T\textsubscript{H$_2$CO} denotes the average formaldehyde temperature within a 1' box centered on the pixel with peak H\textsubscript{2}CO $J=3_{03} \to 2_{02}$ integrated intensity, whereas Pixel T\textsubscript{H$_2$CO} denotes the formaldehyde temperature at the pixel with maximum NH\textsubscript{3} (3,3) integrated intensity once regridded to the ammonia resolution. T\textsubscript{12} represents the ammonia temperature calculated using NH\textsubscript{3} (1,1) and (2,2), taken at the pixel with peak NH\textsubscript{3} (3,3) integrated intensity, and T\textsubscript{13} is the ammonia temperature calculated using NH\textsubscript{3} (1,1) and (3,3) at the same pixel. The dashed blue line in each plot indicates identical temperature estimates.}
\label{fig:tCorr}
\end{figure*}

Next, we calculate the gas temperature based on the line ratio of H\textsubscript{2}CO $(3_{21}-2_{20})$ to H\textsubscript{2}CO $(3_{03}-2_{02})$. We take the ratio between the integrated intensity maps of the two lines, described in Section \ref{subsec:maps}, enforcing a 3$\sigma$ cutoff (Eq.\ \ref{eq:noise}) on both moment 0 maps. We then use the formula
\begin{equation}
\label{eq:formaldehydeTemp}
T_G [K] = 590 R_{\rm{H}_2 \rm{CO}}^2 + 2.88 R_{\rm{H}_2 \rm{CO}} + 23.4,
\end{equation}
which is a polynomial fit to the gas temperature and formaldehyde line ratio $R_{\rm{H}_2 \rm{CO}} = \frac{\int I_v (3_{21} - 2_{20}) dv}{\int I_v (3_{03} \to 2_{02}) dv}$ \citep{ginsburg_dense_2016} derived from the radiative transfer code \textsc{RADEX} \citep{van_der_tak_computer_2007}. This fit uses an assumed gas density of $n(\text{H}_2) = 10^4 \text{ cm}^{-3}$. Unlike ammonia, formaldehyde is an asymmetric top molecule, so the excitation of different transitions, and in turn the line ratio, depend on the H\textsubscript{2} density, although the gas temperature has only a weak density dependence for most of the line ratios we observe ($R_{\rm{H}_2 \rm{CO}} \lesssim 0.3$; \citealt{ginsburg_dense_2016}). We discuss this further in Section \ref{subsec:temperature}. We also calculate temperature error maps, 
\begin{equation}
\label{eq:formaldehydeTempErr}
\delta T_G [K] = \frac{dT_G}{dR} \delta R = 1080R \delta R + 2.88\delta R,
\end{equation}
where $\delta R$ is the error on the formaldehyde ratio (Eq.\ \ref{eq:ratioError}). These errors are likely underestimates for regions where the gas temperature is greater than about 80 K. The resultant formaldehyde temperature maps are shown in Figure \ref{fig:H2COtemp}. We also show the error maps in Figure \ref{fig:H2COtempError} in Appendix \ref{appendix:tempErr}.

The formaldehyde and ammonia temperature depend differently on physical conditions, so a comparison between the two may reveal some insight into the cloud structure. However, their observations are at different resolutions, so to compensate, we return to the formaldehyde data cubes and regrid them to the same spatial and velocity resolution as the HOPS ammonia cubes. We create a regridded temperature map using the same procedure as above, then take the temperature (Eq.\ \ref{eq:formaldehydeTemp}) and error (Eq.\ \ref{eq:formaldehydeTempErr}) at the pixel with peak NH\textsubscript{3} $(3,3)$ integrated intensity. 

We find that the peak NH\textsubscript{3} $(3,3)$ pixel often lies far from the peak emission of the formaldehyde lines, and in one case lies outside the ALMA FOV of the corresponding cloud. The average angular offset between the two peaks is 1.26', and the average physical offset is 2.23 pc, although the uncertainty on the latter may be large due to the individual distance uncertainties. We thus also calculate a potentially more representative formaldehyde temperature value by taking the mean of the temperature map at its original resolution within a 1' box centered on the pixel with maximum H\textsubscript{2}CO $J=3_{03} \to 2_{02}$ integrated intensity (light blue boxes in Figure \ref{fig:H2COtemp}). We take the error on this temperature to be the mean of the temperature error map within the same box.

We use this same box to calculate representative values for other properties of the ALMA data, such as the line widths (Section \ref{subsec:turb}) and ratio maps.

\subsubsection{Ammonia versus Formaldehyde Temperature}
\label{subsubsec:ammFormTemp}

A direct, quantitative comparison between the ammonia and formaldehyde temperature measurements is difficult because of the factor of four difference in beam size between the HOPS and ALMA data and the spatial distance between the peak NH\textsubscript{3} (3,3) and H\textsubscript{2}CO $J = 3_{03} \to 2_{02}$ emission. We still plot the correlations between them in Figure \ref{fig:tCorr}, but we find no significant correlation between the ammonia and formaldehyde temperature. However, there is a clear correlation between the ammonia temperatures calculated using the NH\textsubscript{3} (1,1) and (2,2) emission versus that using the NH\textsubscript{3} (1,1) and (3,3) emission, as well as between the formaldehyde temperature taken at the pixel with peak NH\textsubscript{3} (3,3) emission versus that averaged over a 1' box centered at the peak H\textsubscript{2}CO $J=3_{03} \to 2_{02}$ emission, both of which are expected. 

The lack of any correlation between the ammonia and formaldehyde temperatures indicates they may be tracing gas components at different densities and optical depths. As discussed in more detail in Section \ref{subsec:temperature}, the properties derived from these two molecules, and indeed from many of the molecular lines we observe, are inconsistent with a single gas component.

\subsection{Turbulence}
\label{subsec:turb}

\begin{figure}[t]
\centering
\includegraphics[width=3.3in]{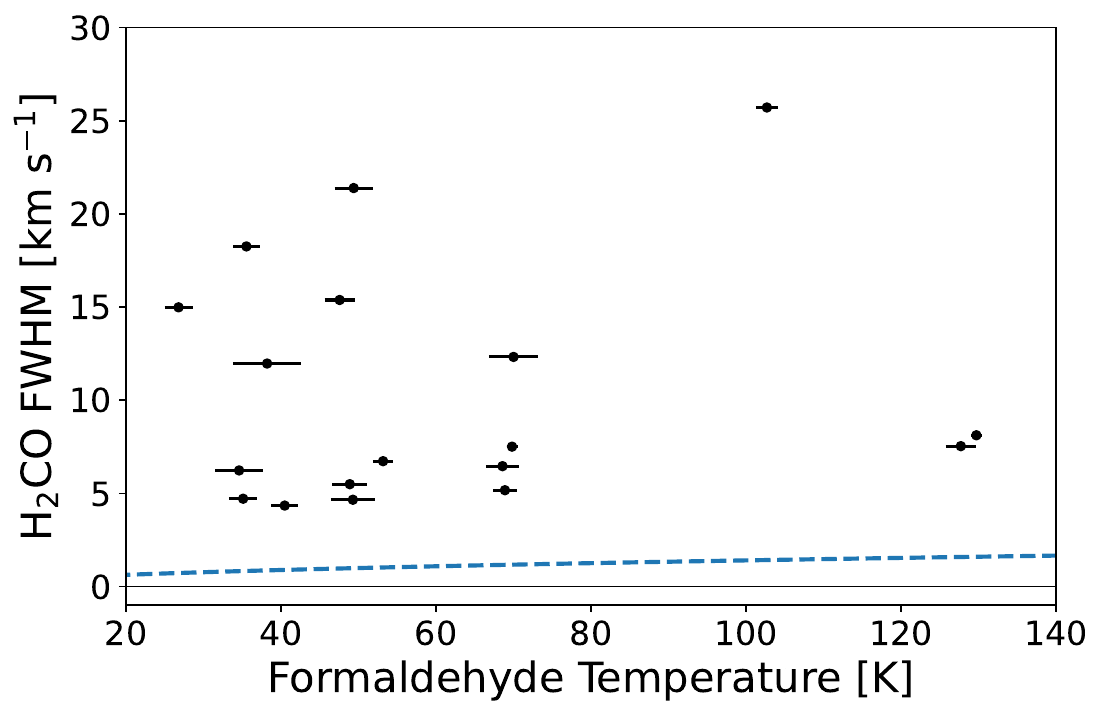}
\caption{H\textsubscript{2}CO FWHM line width against formaldehyde temperature. Both are calculated over a 1' box centered on the pixel with peak H\textsubscript{2}CO $J = 3_{03} \to 2_{02}$ emission. The dashed blue line indicates the thermal line width as a function of temperature, Eq.\ \ref{eq:vThermal}.}
\label{fig:H2COtempLW}
\end{figure}

We calculate a line width for each ALMA line by fitting a Gaussian to the mean spectrum of the 1' box centered on the peak pixel in the H\textsubscript{2}CO $J=3_{03} \to 2_{02}$ integrated intensity map. One cloud exhibits two velocity components in its spectrum, so we fit a double Gaussian and take the wider of the two. Figure \ref{fig:H2COtempLW} shows that there is no correlation between the formaldehyde temperatures and line widths of the clouds, which is expected if the line broadening is due to processes such as turbulence rather than thermal broadening. 

We can calculate the expected degree of thermal broadening for each cloud. Assuming local thermal equilibrium (LTE), the thermal one-dimensional velocities of molecules should follow a Maxwell-Boltzmann distribution. The thermal FWHM is
\begin{equation}
\label{eq:vThermal}
\Delta v_\text{th} = \left( 8\ln2 \frac{k_B T_k}{\mu m_H} \right)^{1/2},
\end{equation}
where $k_B$ is the Boltzmann constant, $T_k$ is the kinetic temperature of the cloud, $\mu = 2.34$ is the mean molecular weight of a molecular cloud \citep{syed_atomic_2020}, and $m_H$ is the mass of a hydrogen atom. Taking the calculated formaldehyde temperatures of each cloud to be representative of the kinetic temperature, we can calculate a thermal line width, then calculate a nonthermal line width corresponding to large- and small-scale velocity variations. Both turbulent and bulk ordered motion contribute to the nonthermal line width, so to isolate the turbulence component, we take
\begin{equation}
\label{eq:vnon-thermal}
\Delta v_\text{turb} = \sqrt{\Delta v^2 - \Delta v_\text{th}^2 - \Delta v_\text{bulk}^2},
\end{equation}
where $\Delta v$ is the observed total line width. We can estimate the contribution from ordered motion through the velocity gradients of the clouds, which can be measured by fitting a linear velocity gradient to the intensity-weighted velocity maps (e.g.,\ \citealt{federrath_link_2016, balser_discovery_2021, wang_role_2024}). An example of this calculation is shown in Appendix \ref{appendix:vg}. Although the velocity gradient may not accurately represent the bulk motion, particularly for clouds with multiple complex velocity components \citep{henshaw_brick_2019}, for the clouds in our sample, which are not very kinematically complex, it likely suffices as an approximate measure of the ordered motion contribution to the line width. As can be seen in Figure \ref{fig:H2COtempLW}, the turbulent line widths are significantly greater than the thermal line widths, which range from $\sim 0.4$ to 1.6 km\ s$^{-1}$. The line width contribution from bulk motion is also small, on the same order as the thermal contribution, indicating that the gas is dominated by turbulence. The one exception is G005.49, which has an estimated bulk motion line width contribution nearly equal to the total line width ($\sim 10 \text{ km s}^{-1}$); however, this is a cloud for which there is no detection of H\textsubscript{2}CO $J=3_{21} \to 2_{20},$ and thus no formaldehyde temperature. For this cloud, we can expect that its line width is dominated by some ordered motion.

The turbulent FWHM line widths, which largely dominate the total FWHM line widths, of formaldehyde for the clouds ranges between $\sim 4$ and 25 km\ s$^{-1}$. The molecular clouds in the Galactic disk studied by \citet{larson_turbulence_1981} have a three-dimensional rms velocity dispersion between $\sim 0.4$ and 9 km\ s$^{-1}$. The one-dimensional FWHM and three-dimensional rms velocity dispersions are related by
\begin{equation}
\label{eq:rmsVSfwhm}
\Delta v = \sqrt{\left(8 \ln 2\right)/3} \ v_\text{rms}.
\end{equation}
The rms velocity dispersion range of the clouds is $\sim 3-20$ km\ s$^{-1}$, which lies between the values expected of clouds in the Galactic disk and clouds in the CMZ, the latter of which have FWHM line widths on the order of $10-50$\ km\ s$^{-1}$ (e.g.,\ \citealt{tsuboi_central_2015}). This suggests that some of the clouds in the sample may indeed be exposed to some of the extreme conditions of the CMZ.

We can approximate the sound speed of the cloud by assuming an isothermal gas with equation of state $P = \rho k_B T/(\mu \text{m}_\text{H})$, in which case the sound speed is
\begin{equation}
\label{eq:soundSpeed}
\text{c}_s = \sqrt{k_B T_k/(\mu \text{m}_\text{H})}.
\end{equation}
The three-dimensional Mach number, assuming isotropic turbulence, is then $\mathcal{M} = v_\text{rms}/\text{c}_s$, where $v_\text{rms}$ is the turbulent three-dimensional rms velocity dispersion of Eq.\ \ref{eq:rmsVSfwhm}. The calculated Mach numbers range from 7 to 45. These values also lie between those typical of molecular clouds in the Galactic disk ($\lesssim 10$; e.g.,\ \citealt{tang_kinetic_2018}, \citealt{syed_atomic_2020}) and those observed in the CMZ ($\gtrsim 25$; e.g.,\ \citealt{kauffmann_galactic_2017}, \citealt{henshaw_molecular_2016}).

We take the errors in the total line widths to be the errors on the Gaussian fits, then propagate accordingly to get errors on the Mach numbers. 

\subsection{Star Formation}
\label{subsec:sf}

\begin{table*}[t]
\centering
\setlength{\tabcolsep}{0.75em}

\begin{tabular}{c  c  c  c  c  c  c} 
 Cloud& $\Delta v$ [km\ s$^{-1}$] & $T_e$ [K] & $D_{\text{H30}\alpha}$ [pc] & $n_e$ [cm\textsuperscript{-3}] & $\int F_{v} dv$ [Jy km\ s$^{-1}$] & $\text{log}_{10} Q$ [s$^{-1}$]\\ [0.5ex] 
 \hline\hline
 G008.67 & $27.21 \pm 1.04$ & $7170.24 \pm 512.06$ & $0.73 \pm 0.09$ & $860.28 \pm 100.02$ & $28.94 \pm 0.07$ & $47.76 \pm 0.09$ \\ 
G354.61 & $38.99 \pm 2.09$ & $6993.59 \pm 665.36$ & $4.34 \pm 0.79$ & $145.23 \pm 26.39$ & $5.50 \pm 0.05$ & $48.12 \pm 0.17$ \\ 
G353.41 & $26.62 \pm 0.22$ & $7558.10 \pm 1588.27$ & $0.79 \pm 0.28$ & $798.87 \pm 279.99$ & $147.07 \pm 0.08$ & $48.00 \pm 0.34$ \\ 
G351.77 & $10.18 \pm 0.81$ & $7560.89 \pm 545.06$ & $0.29 \pm 0.04$ & $2174.67 \pm 308.14$ & $13.31 \pm 0.17$ & $46.98 \pm 0.24$ \\ 
G351.58 & $25.41 \pm 1.07$ & $6677.77 \pm 566.97$ & $1.25 \pm 0.24$ & $504.76 \pm 97.14$ & $61.63 \pm 0.06$ & $48.42 \pm 0.18$ \\ 
G350.18 & $26.62 \pm 1.78$ & $6446.82 \pm 422.08$ & $1.98 \pm 0.29$ & $318.40 \pm 45.86$ & $1.95 \pm 0.05$ & $47.06 \pm 0.12$ \\ 
G350.10 & $28.15 \pm 0.24$ & $6448.32 \pm 422.37$ & $1.62 \pm 0.21$ & $390.11 \pm 51.41$ & $135.03 \pm 0.10$ & $48.90 \pm 0.12$ \\ 

\hline

\end{tabular}
\caption{H30$\alpha$ derived quantities.}
\label{table:SFparameters}
\end{table*}

\begin{figure*}[t]
\centering
\includegraphics[width=6.3in]{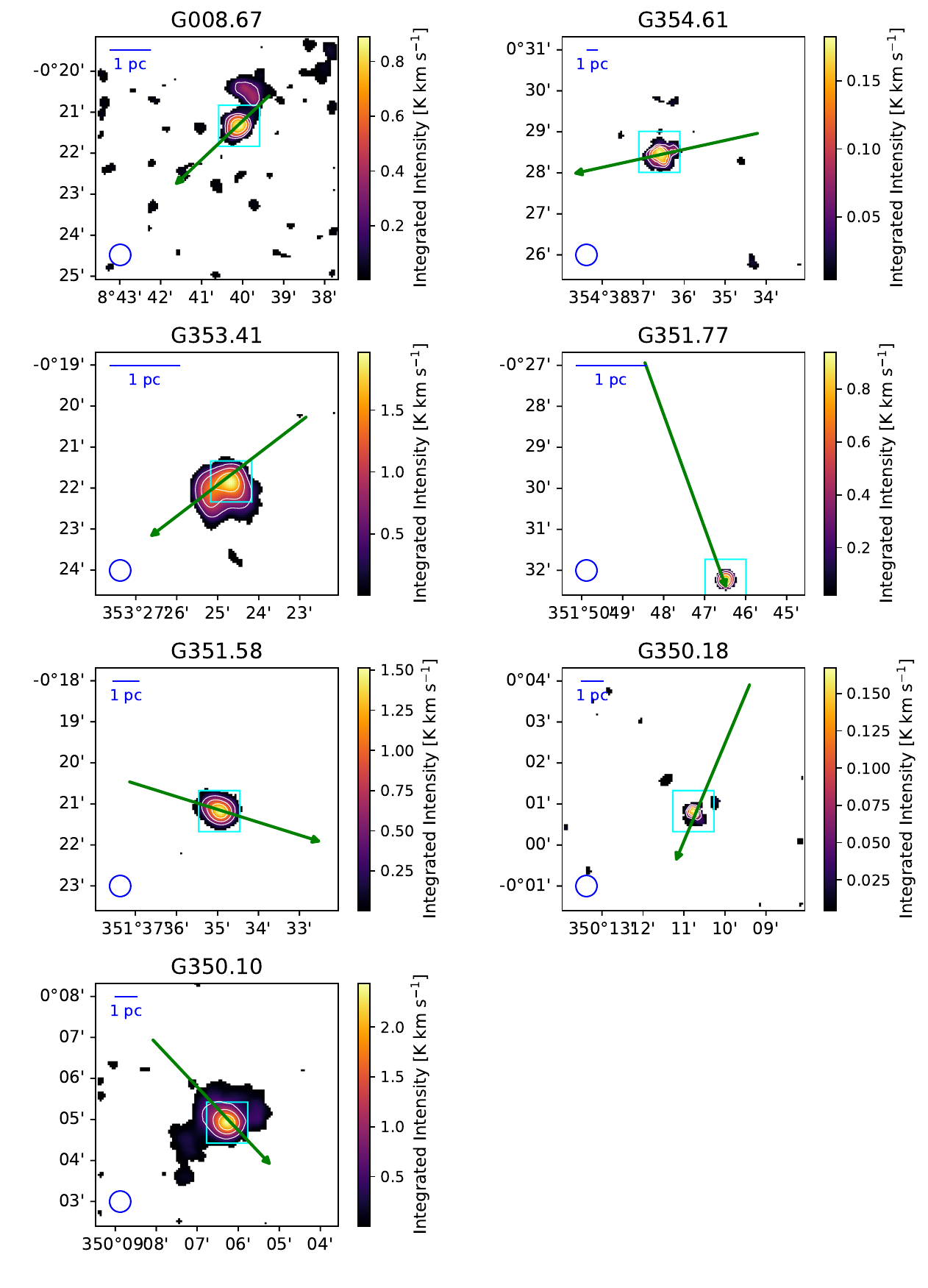}
\caption{H30$\alpha$ integrated intensity maps with white contours, for clouds with detections, calculated as described in Section \ref{subsec:maps}. The light blue boxes are 1' boxes centered on the pixel with peak H30$\alpha$ integrated intensity. The blue circles in the bottom left corner of each map indicate the 31" beam size. The blue lines in the top left corner of each map indicate a physical size of 1 pc. The uncertainties on these physical sizes may be large as they are proportional to the distance uncertainties. The green arrows indicate the paths along which the PV diagram is calculated.}
\label{fig:h30moment0}
\end{figure*}

\begin{figure*}[t]
\centering
\includegraphics[width=6.3in]{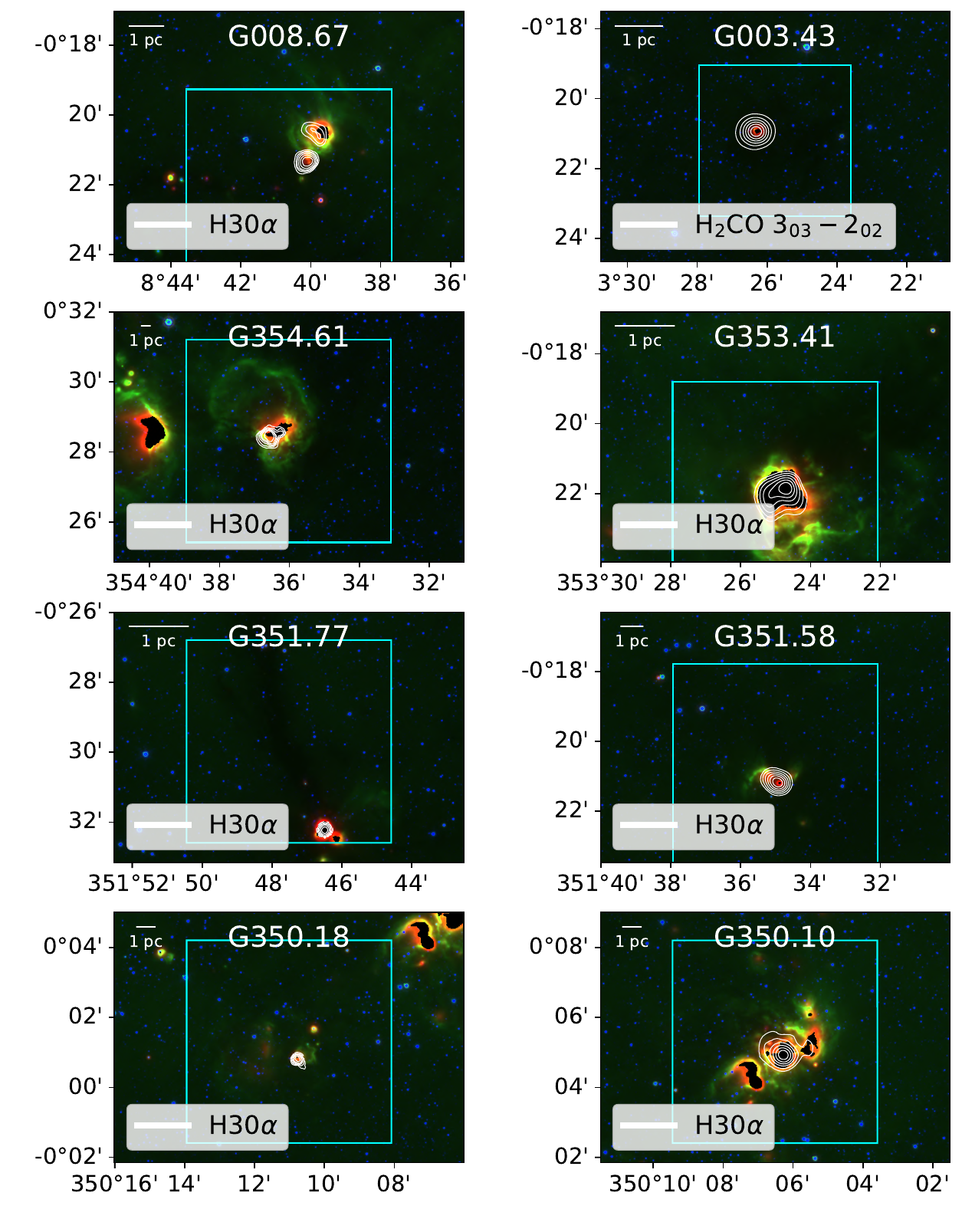}
\caption{Three-color Spitzer images of the clouds with Spitzer detections, with GLIMPSE 4.5 $\mu$m, GLIMPSE 8 $\mu$m, and MIPSGAL 24 $\mu$m in blue, green, and red, respectively, plotted in Galactic coordinates. If the clouds also have H30$\alpha$ detections, the contours show H30$\alpha$ integrated intensity; otherwise, the contours show H\textsubscript{2}CO $J=3_{03} \to 2_{02}$ integrated intensity. The light blue boxes in each image depict the ALMA FOV. The white lines in the top left corner of each map indicate a physical size of 1 pc. The uncertainties on these physical sizes may be large as they are proportional to the distance uncertainties. The 24 $\mu$m data contains artifacts at bright spots, so they are masked by the MIPSGAL processing pipeline.}
\label{fig:spitzer}
\end{figure*}

\begin{figure*}[t]
\centering
\includegraphics[width=6.3in]{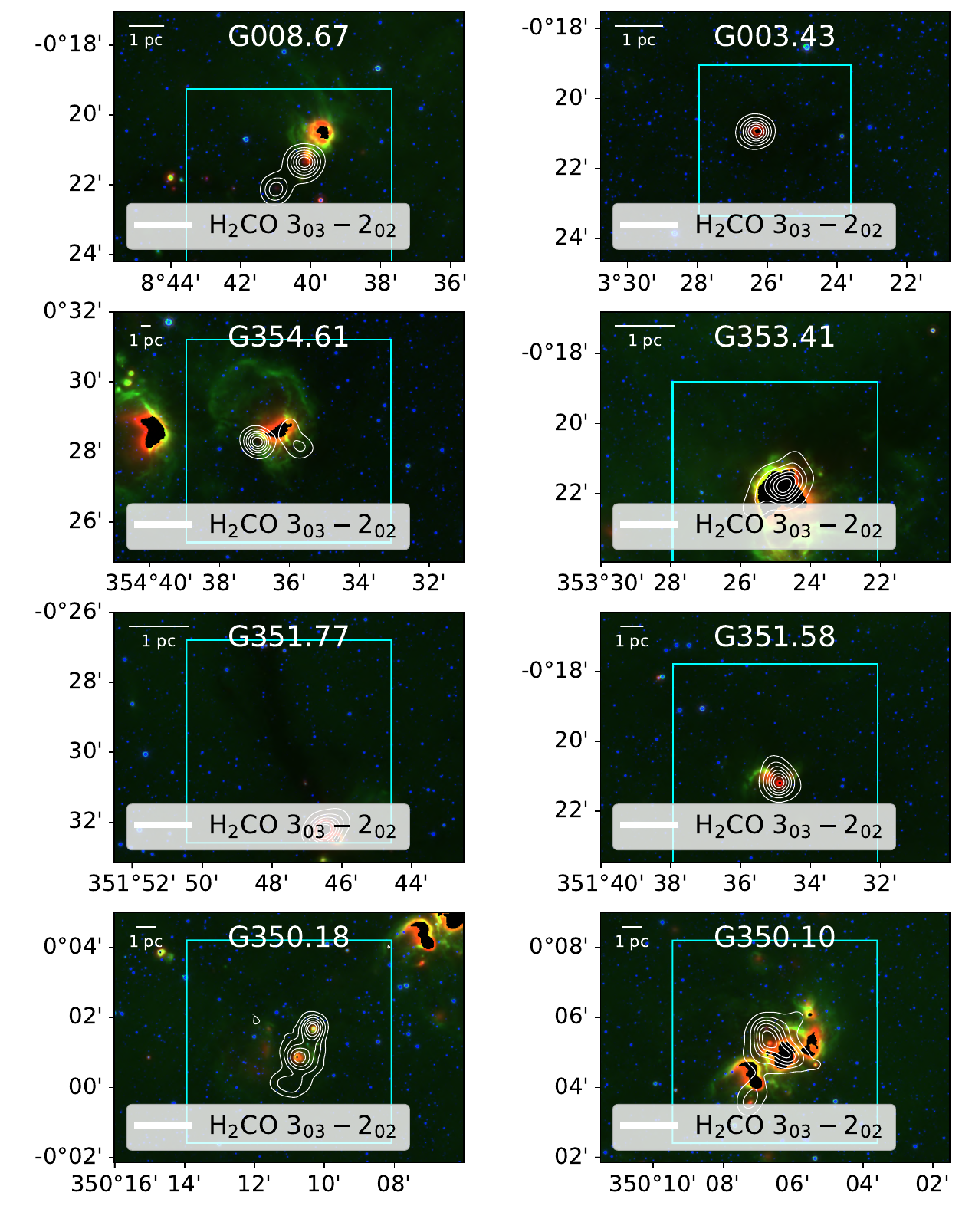}
\caption{Same as Figure \ref{fig:spitzer}, but with H\textsubscript{2}CO $J=3_{03} \to 2_{02}$ integrated intensity contours for all clouds.}
\label{fig:spitzerH2CO}
\end{figure*}

We can calculate the ionizing photon production rate, $Q$, from measurements of H30$\alpha$, via
\begin{equation}
\label{eq:H30a}
\begin{split}
\frac{Q(\text{H}n\alpha)}{\text{s}^{-1}} = & \ 3.99 \times 10^{24} \left(\frac{\alpha_B}{\text{cm}^3 \text{ s}^{-1}}\right)\left(\frac{\epsilon_\nu}{\text{erg s}^{-1}\text{ cm}^{-3}}\right)^{-1} \\ 
& \ \times \left(\frac{\nu}{\text{GHz}}\right)\left(\frac{D}{\text{kpc}}\right)^2 \left(\frac{\int F_v dv}{\text{Jy km\ s}^{-1}}\right)
\end{split}
\end{equation}
\citep{scoville_submillimeter_2013, bendo_tests_2017, kim_new_2018}, where $\alpha_B$ is the case B effective recombination coefficient, and $\epsilon_\nu$ is the emissivity, both of which are functions of electron density and temperature.

To estimate the electron temperature, we use the Galactic disk electron temperature gradient from \citet{quireza_electron_2006},
\begin{equation}
\label{eq:eTempGradient}
T_e [\text{K}] = (5780 \pm 350) + (287 \pm 46) R_\text{Gal} [\text{kpc}],
\end{equation}
which is an empirical fit to \HII{} regions with electron temperatures derived from radio recombination line and continuum measurements, and Galactocentric distances calculated via radial velocity measurements. We can also confirm that these fall under the upper limit on the electron temperature given by the line width of our H30$\alpha$ observations. The thermal contribution to the line width is Gaussian and has a FWHM of
\begin{equation}
\label{eq:thermalFWHM}
\Delta v_\text{th} = \left(8 \text{ln} 2 \frac{k_\text{B} T_e}{m_\text{H}} \right)^{1/2}
\end{equation}
\citep{rivera-soto_recombination_2020}, which is the same form as Eq.\ \ref{eq:vThermal} but replacing the kinetic temperature with electron temperature. Pressure broadening is proportional to $\nu^{-4}$, so it is negligible at the H30$\alpha$ frequency \citep{keto_early_2008}. Thus, the total H30$\alpha$ line should be Gaussian and have a FWHM of
\begin{equation}
\label{eq:totalFWHM}
\Delta v = \sqrt{\Delta v_\text{th}^2 + \Delta v_\text{dy}^2},
\end{equation}
where $\Delta v_\text{dy}$ is the dynamical contribution from unresolved bulk motions. We find that electron temperature calculated from Eq.\ \ref{eq:eTempGradient} is indeed below the upper limit set by Eq.\ \ref{eq:thermalFWHM} (using the Gaussian line width calculated in Section \ref{subsec:turb}) for all the clouds with H30$\alpha$ detections.

Whereas the value of the electron temperature can affect the ionizing photon production rate $Q$ by a factor of up to $\sim$2.5 over a temperature range from 3000 to 15,000 K, the electron density has a relatively small effect on $Q$, with less than a 15\% variation in both the recombination coefficient and emissivity over a density range from $10^2$ and $10^5$ cm$^{-3}$ \citep{bendo_tests_2017}. We estimate the electron density using the fits of $n_e$ against diameter of Galactic \HII{} regions by \citet{hunt_size-density_2009}. To calculate sizes for the clouds, we collapse the H30$\alpha$ PV diagram along the velocity axis to create a one-dimensional cloud profile, then fit a Gaussian. We take the angular diameter of the cloud to be 4 times the standard deviation of the fitted Gaussian, then calculate a physical size using the distance to the cloud. Although this is an inaccurate estimate for elongated clouds, we find that the integrated intensity maps of H30$\alpha$ are roughly circular, shown in Figure \ref{fig:h30moment0} (which also includes the overlain PV diagram path). The resulting cloud sizes all lie within the range of 0.3 and 5 pc, which is covered by the \citet{kim_radio_2001} sample of compact Galactic \HII{} regions; \citeauthor{hunt_size-density_2009} finds a best-fit regression of
\begin{equation}
\label{eq:neDiameter}
\text{log}\ n_e \ [\text{cm}^{-3}] = 2.8 - \text{log}\ D \ [\text{pc}] \,.
\end{equation}

Using these calculated electron densities and temperatures, we interpolate the tables of $\alpha_B$ and $\epsilon_\nu$ values published by \citet{storey_recombination_1995}. We calculate errors on these values using a simple Monte Carlo simulation, assuming a normal distribution for $T_e$ and $n_e$. To calculate the integrated flux density of H30$\alpha$ from our moment 0 maps, we use a 1' box centered on the pixel with peak H30$\alpha$ integrated intensity, then take
\begin{equation}
\label{eq:fluxDensity}
\int F_v\ dv\ [\text{Jy km\ s}^{-1}] = \frac{2k_B \nu^2}{c^2} \iint T_B\ dv\ d\Omega,
\end{equation}
where $\int T_B\ dv$ is the moment 0 map. 

The quantities derived from the H30$\alpha$ line, including the ionizing photon production rate $Q$, are shown for all the clouds with H30$\alpha$ detection in Table \ref{table:SFparameters}. The resultant $Q$ values are mostly consistent with that of a main sequence O star, which has a $\text{log}_{10} Q$ value between 47.88 for an O9.5 star and 49.64 for a an O3 star \citep{martins_new_2005}.

Using the derived $Q$, $n_e$, and $\alpha_B$ values, we can estimate the radius of a Str\"{o}mgren sphere, which represents a region ionized by a single O-type star, using the equation
\begin{equation}
r_S = \left(\frac{3Q}{4\pi \alpha_B n_e^2}\right)^{1/3}.
\end{equation}
We find that the radii derived from this calculation agree with the measured radii of the H30$\alpha$ emission within a factor of 4, with an average $r_{\text{H30}\alpha}/r_S$ ratio of 2.2. Although $n_e$, and thus $\alpha_B$ and $Q$, is calculated using the diameter of the cloud in H30$\alpha$, the relationship between them is non-linear, and so, this provides a check of internal consistency within the calculations.

We also calculate upper limits on $Q$ values for the clouds without significant H30$\alpha$ detections. We take the upper limit on the integrated flux density to be
\begin{equation}
\label{eq:fluxDensityErr}
\int F_v\ dv\ [\text{Jy km\ s}^{-1}] \leq \frac{2k_B \nu^2}{c^2} \sigma \sqrt{N},
\end{equation}
where $\sigma$ is the error on the H30$\alpha$ moment 0 map (Eq. \ref{eq:noise}), and $N$ is the number of pixels in a 1' box on the map. We use the electron temperature gradient (Eq. \ref{eq:eTempGradient}) to estimate electron temperatures for each cloud, and choose the mean of the electron densities of clouds with H30$\alpha$ detections to be an estimate of the electron density of clouds with no detection. We interpolate the electron density and temperature to get $\alpha_B$ and $\epsilon_\nu$ values, then calculate a photoionizing photon production rate with Eq.\ \ref{eq:H30a}; we take a 2$\sigma$ upper limit of twice this $Q$ value. The same method is also applied to calculate errors on the $Q$ values for clouds with H30$\alpha$ emission.

We plot Spitzer images for clouds with detections in Figure \ref{fig:spitzer}, with H30$\alpha$ contours, and in Figure \ref{fig:spitzerH2CO}, with H\textsubscript{2}CO $J=3_{03} \to 2_{02}$ contours. There is generally a correlation between the presence and location of H30$\alpha$ and both 8 and 24 $\mu$m emission, although G003.43 displays 24 $\mu$m emission consistent with the location of formaldehyde emission but does not have an H30$\alpha$ detection. We can also estimate the SFR in the clouds from 24 $\mu$m emission using the relationship from \citet{calzetti_calibration_2007},
\begin{equation}
\label{eq:24muSFR}
\text{SFR} \ [\text{M}_\odot \text{ yr}^{-1}] = 1.27 \times 10^{-38} (L_{24 \mu\text{m}} \ [\text{ergs s}^{-1}])^{0.8850},
\end{equation}
where $L_{24 \mu \text{m}} = \nu L(\nu)$. This relation is derived from extragalactic star-forming regions, so it may not be fully applicable to the clouds in our sample, but it can be used as an approximation. For comparison, we also calculate a star formation rate from the ionizing photon production rate using a conversion from $Q$ to SFR of $7.29 \times 10^{-54} \text{ M}_\odot \text{yr}^{-1}/\text{s}^{-1}$ \citep{murphy_calibrating_2011}, which is calculated using \texttt{STARBURST99} \citep{leitherer_starburst99_1999} and also applies primarily to the galactic scale. It also assumes solar metallicity and a constant SFR over about 100 Myr.

The 24 $\mu$m MIPSGAL processing pipeline masks artifacts in the data, which are particularly prevalent around bright sources \citep{mizuno_processing_2008}. Most of the clouds with H30$\alpha$ emission, which are also bright in the mid-IR, are thus masked, as can be seen in Figures \ref{fig:spitzer} and \ref{fig:spitzerH2CO}. To mitigate this effect, we interpolate the missing values, although this is imperfect as the masked pixels are generally around the brightest locations. We thus treat the resultant SFR values as lower limits.

We integrate the 24 $\mu$m emission, which is given in units of MJy sr\textsuperscript{-1}, over a 2' box centered on the location with peak H30$\alpha$ integrated intensity. We choose a larger box size than for calculating the H30$\alpha$ flux density because we expect the emission to be more extended, as the 24 $\mu$m emission comes from dust around the star-forming region whereas H30$\alpha$ emission comes from, or closer to, the ionized region. We convert these fluxes into luminosities using their distances, then calculate the SFR with Eq.\ \ref{eq:24muSFR}.

To calculate errors on the SFR of these clouds, as well as upper limits on the clouds with no detection, we use the uncertainty maps provided by MIPSGAL. For the clouds with detections, we add the pixel errors in quadrature over the integration box and multiply by the pixel spacing to get a flux density error; for the other clouds, we take the mean over the uncertainty map and multiply by the square root of the number of pixels in a 2' box and the pixel spacing. We then propagate the distance error and the errors in the parameters of Eq.\ \ref{eq:24muSFR} appropriately to get an error. We again use a 2$\sigma$ detection as an upper limit. We note that these errors, as well as the errors for the SFR calculated with H30$\alpha$, are likely underestimates, as there is additional uncertainty from the stochastic distribution around the initial mass function (IMF) for small clusters.

To remove scatter due to cloud size, we calculate an SFR density ($\Sigma_\text{SFR}$) by dividing over the physical area of integration in units of squared parsecs. For the 24 $\mu$m data, this area is a 2' box taken at the cloud distance. For the H30$\alpha$ detections, some of the integrated intensity data is masked by our 4$\sigma_\text{rms}$ cut, as seen in Figure \ref{fig:h30moment0}, so the area is slightly less than a 1' box at the cloud distance. For the H30$\alpha$ upper limits, the area is the full 1' box at the cloud distance. We note that, since $Q$ is proportional to $D^2$, the $\Sigma_\text{SFR}$ calculated with H30$\alpha$ is not directly dependent on distance, but there is still a weak dependence on distance given Eqs. \ref{eq:eTempGradient} and \ref{eq:neDiameter}. Additionally, since $L_{24\mu \text{m}}$ is also proportional to $D^2$, and the SFR calculated with 24 $\mu$m emission is proportional to $L_{24\mu \text{m}}^{0.8850}$, the $\Sigma_\text{SFR}$ calculated with 24 $\mu$m emission has a weak inversely proportional dependence on distance.

\begin{figure}[t]
\centering
\includegraphics[width=3.3in]{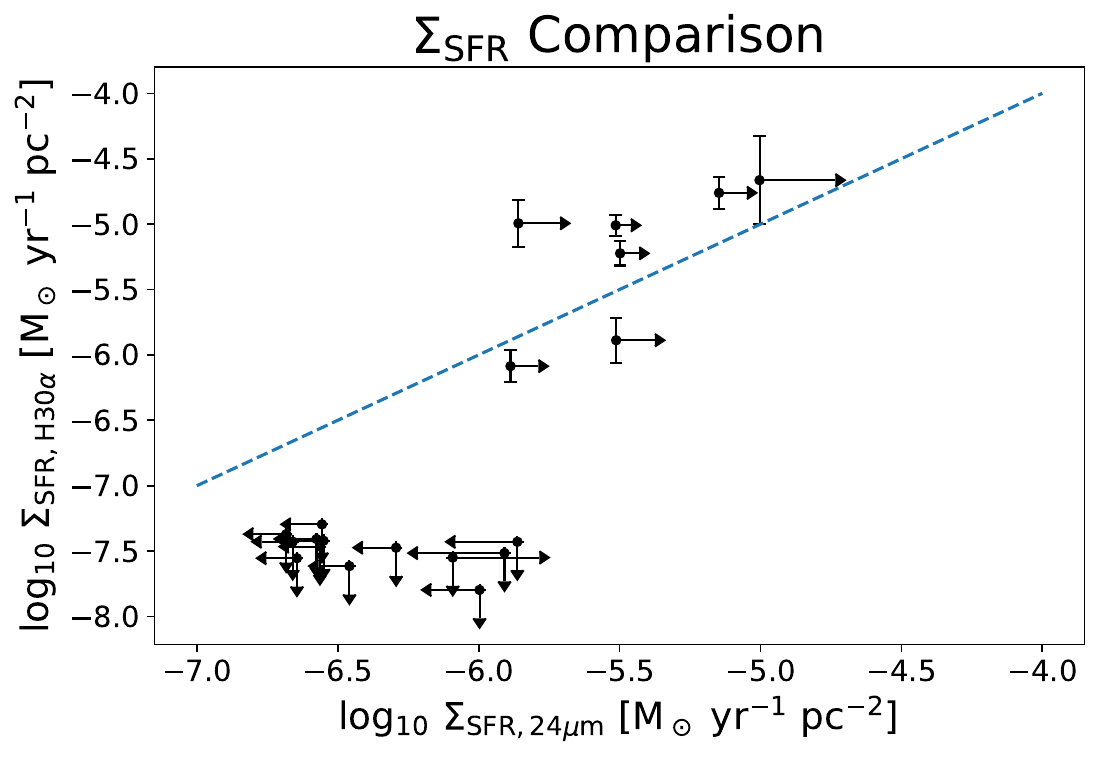}
\caption{Comparison of $\Sigma_\text{SFR}$ calculated using H30$\alpha$ emission vs. that calculated using 24 $\mu$m emission from Spitzer. The blue line assumes the two are equal.}
\label{fig:sfr}
\end{figure}

A comparison between the two $\Sigma_\text{SFR}$ measures is shown in Figure \ref{fig:sfr}. The values generally agree for the clouds that are detected in both. We are also able to place stricter upper limits using H30$\alpha$ due to the higher sensitivity of ALMA.

\citet{calzetti_calibration_2007} notes that emission at 8 $\mu$m is correlated with star formation, but it also depends strongly on metallicity and size, so we do not calculate an SFR using the 8 $\mu$m emission from GLIMPSE. We note, however, that 8 $\mu$m emission is present for all clouds that exhibit 24 $\mu$m emission.


\subsection{Shocks}
\label{subsec:shocks}

Both methanol (CH\textsubscript{3}OH) and SiO are associated with the presence of shocks \citep{meier_spatially_2005, schilke_sio_1997}, although they trace different shock velocities. CH\textsubscript{3}OH can be formed on grain mantles then evaporated into the gas phase through weak shocks with $v_s \lesssim 10$ km\ s$^{-1}$ \citep{bergin_postshock_1998}. On the other hand, evaporation of SiO into the gas phase requires more energetic grain processing through grain core or mantle erosion, which can occur in shocks with $v_s \gtrsim 25$ km\ s$^{-1}$ \citep{garay_silicon_2000}. 

SiO is detected in all clouds except G005.75 and G005.49, and CH\textsubscript{3}OH is detected in all clouds except G005.49.

\subsubsection{SiO}
\label{subsubsec:sio}

We can calculate the column density of SiO, with several assumptions. We assume optically thin emission, LTE, a Rayleigh-Jeans approximation, and negligible background. Then, the column density is
\begin{equation}
\label{eq:sioColDens}
N_\text{tot} = \left( \frac{3k_B}{8\pi^3 \nu S \mu^2}\right) \left( \frac{Q_\text{rot}}{g_J g_K g_I} \right) \text{exp}\left( \frac{E_u}{k_B T_\text{ex}} \right) \int \frac{T_B dv}{f}
\end{equation}
\citep{mangum_how_2015}, where $Q_\text{rot}$ is the partition function, $g_i$ are the degeneracies, $\mu$ is the dipole moment of the molecule, $S$ is the intrinsic line strength, $E_u$ is the upper energy level, $T_\text{ex}$ is the excitation temperature of the gas, and $f$ is the beam filling factor. For a linear molecule like SiO, we have $g_J = 2J +1$, $g_K = g_I = 1$, $Q_\text{rot} = \sum_{J=0}^\infty (2J+1) \text{exp}(-\frac{E_J}{kT})$, and $S = \frac{J}{2J+1}$. For a diatomic molecule, the rotational energy levels are $E_J \approx hBJ(J+1)$, where $B = \frac{\hbar}{4\pi I}$ is the rotational constant of the molecule. For SiO in particular, the dipole moment is $\mu = 3.0982$ Debye \citep{raymonda_electric_1970} and the rotational constant is $B = 21787.5$ MHz \citep{lowry_manson_millimeter_1977}. The column density is proportional to $Q_\text{rot}(T_\text{ex})\text{exp}\left(\frac{1}{T_\text{ex}}\right)$, which has a global minimum at $T_\text{ex} = 31.72$ K. We find that the difference between the column density assuming $T_\text{ex} = 31.72$ K and at the LTE case of $T_\text{ex} = T_\text{k}$ (where we take to be the formaldehyde gas temperature to be the gas kinetic temperature sampled by SiO since the formaldehyde temperature is more likely to sample denser gas than the ammonia temperature) is less than the uncertainty in the integrated intensity of SiO $J = 5 \to 4$, which we take to be within the same 1' box that the formaldehyde temperature is calculated in. However, at a gas density of $n(\text{H}_2) = 10^{4} \text{ cm}^{-3}$, which we assumed when calculating the formaldehyde temperature, the SiO $J=5\to 4$ line is subthermally excited ($T_\text{ex} \ll T_\text{k}$). With \textsc{RADEX}, we find that the SiO column density derived from the LTE assumption is insufficient to explain the observed line intensity. Indeed, for nearly all of the clouds in the sample, either $n(\text{H}_2)$ or $N(\text{SiO})$ must be 1 to 3 orders of magnitude higher, or the kinetic temperature of the gas phase traced by SiO must be significantly higher than that traced by formaldehyde. As we are unable to further constrain the H\textsubscript{2} number density or the gas kinetic temperature sampled by SiO, we treat the SiO column densities calculated under the assumption of LTE as lower limits on the true column density. We also assume the beam filling factor is 1, although it may be lower, in which case the calculated column densities will be lower limits; however, if the beam filling factors for the SiO and CO observations are similar, then the SiO abundances N(SiO)/N(H\textsubscript{2}) are independent of the beam filling factor.

We also calculate column densities of \textsuperscript{13}CO and C\textsuperscript{18}O $J = 2 \to 1$ to obtain SiO abundances. Since it is likely these lines are optically thin, compared to their isotopologue \textsuperscript{12}CO, we can again use Eq.\ \ref{eq:sioColDens}.\footnote{Values for $\mu$ and $B$ for CO taken from \url{https://spec.jpl.nasa.gov/}}. These lines are likely to be closer to LTE than the subthermally excited SiO $J = 5 \to 4$ line.

We use the \textsuperscript{12}C/\textsuperscript{13}C and \textsuperscript{16}O/\textsuperscript{18}O isotope abundance ratios, respectively, which increase with Galactocentric radius \citep{langer_12c_1990}. We use the equations
\begin{equation}
\label{eq:cAbundance}
^{12}\mathrm{C}/^{13}\mathrm{C} = (7.5 \pm 1.9) R_\text{Gal} + (7.6 \pm 12.9)
\end{equation}
and
\begin{equation}
\label{eq:oAbundance}
^{16}\mathrm{O}/^{18}\mathrm{O} = (58.8 \pm 11.8) R_\text{Gal} + (37.1 \pm 82.6)
\end{equation}
from \citet{wilson_abundances_1994}. The \textsuperscript{12}CO column densities derived from the two isotopologues agree well. The SiO abundances derived from both, where we have assumed a typical \textsuperscript{12}CO-to-H\textsubscript{2} abundance ratio of $10^{-4}$ for dense clouds (e.g.,\ \citealt{lacy_detection_1994, burgh_direct_2007}), largely agree within error bars.

Shocks can enhance the abundance of SiO to values greater than 10$^{-10}$ compared to ambient values of 10$^{-12}$--10$^{-11}$ \citep{schilke_sio_1997, garay_silicon_2000}, and they have been shown to enhance abundances to values as high as 10$^{-6}$ at extreme velocities \citep{martin-pintado_sio_1992}. Many of the clouds in our sample exhibit SiO abundances above 10$^{-10}$, with some exceeding 10$^{-9}$, indicating that the gas in these clouds are likely undergoing shocks.

\subsubsection{Methanol}
\label{subsubsec:methanol}

We use the line ratio of CH\textsubscript{3}OH $J = 4_{22} \to 3_{12}$ to \textsuperscript{13}CO $J = 2\to 1$ to assess the weak shocks associated with methanol emission. We take the mean value of the 1' box centered on the pixel with peak H\textsubscript{2}CO $J = 3_{03} \to 2_{02}$ emission, and we take the error to be the mean of the ratio error map within the same box.

The values derived for all properties above are reported in Table \ref{table:cloudprops}.

\subsection{Mass}
\label{subsec:mass}

We calculate masses for each cloud using the \textsuperscript{12}CO, \textsuperscript{13}CO, and C\textsuperscript{18}O integrated intensity maps. As in Section \ref{subsec:shocks}, we calculate the column density maps of \textsuperscript{13}CO and C\textsuperscript{18}O, assuming they are optically thin, then use \textsuperscript{12}C/\textsuperscript{13}C and \textsuperscript{16}O/\textsuperscript{18}O isotope abundance ratios (Eqs. \ref{eq:oAbundance} and \ref{eq:cAbundance}) to convert to column density maps of \textsuperscript{12}CO. From this, we multiply by the typical \textsuperscript{12}CO-to-H\textsubscript{2} abundance ratio of $10^{-4}$ to estimate the H\textsubscript{2} column density. For the two clouds without detections of H\textsubscript{2}CO $(3_{21}-2_{20})$, we estimate the kinetic temperature to be the ammonia (3,3) - (1,1) temperature. In contrast, for the \textsuperscript{12}CO, which is unlikely to be optically thin compared to \textsuperscript{13}CO and C\textsuperscript{18}O, we multiply the integrated intensity by the CO-to-H\textsubscript{2} X-factor of $2.3 \times 10^{20}$ cm$^{-2}$ (K km s$^{-1}$)$^{-1}$ derived by \citet{strong_radial_1988}. We also multiply by the factor 0.8, the typical \textsuperscript{12}CO $J=2 \to 1/J=1\to 0$ line ratio \citep{leroy_heracles_2009}.

For each cloud, we integrate the three derived H\textsubscript{2} column density maps, then multiply by the molecular weight per hydrogen molecule, $\mu_{\text{H}_2} = 2.8$, which accounts for cosmic abundance ratios (e.g.,\ \citealt{kauffmann_mambo_2008}), to calculate an estimated cloud mass. We define the extent of the cloud to integrate over using a 5$\sigma$ (Eq. \ref{eq:noise}) cutoff on the integrated intensity map of H\textsubscript{2}CO $J=3_{03} \to 2_{02}$, because the extent of the cloud is less clear on the CO maps, even for the less abundant isotopologues.

The masses derived from all three CO isotopologues are shown in Table \ref{table:masses}. We also plot several other properties derived against Galactic longitude in Figure \ref{fig:glonProperties}.


\section{Discussion}
\label{sec:discussion}

\subsection{Ammonia and Formaldehyde Thermometers}
\label{subsec:temperature}

As noted in Section \ref{subsubsec:ammFormTemp}, based on our data and temperature determination methods, there appears to be no correlation between the formaldehyde and ammonia temperatures of the clouds, indicating that the two thermometers may be tracing different gas in the clouds or some of the methods to derive the gas temperatures are not applicable to these clouds. The critical densities of the ammonia lines are lower than those of the formaldehyde lines, and so, the ammonia temperature generally traces more diffuse gas than the formaldehyde thermometer \citep{ginsburg_dense_2016}. This may explain why the hot molecular cores seen in the formaldehyde temperature maps are generally not present in the ammonia temperature maps (Figures \ref{fig:NH3temp} and \ref{fig:H2COtemp}), although this may also be in part due to the smaller ALMA beam size of the formaldehyde data compared to the Mopra beam size of the ammonia data.

\begin{figure*}[t]
\centering
\includegraphics[width=6.1in]{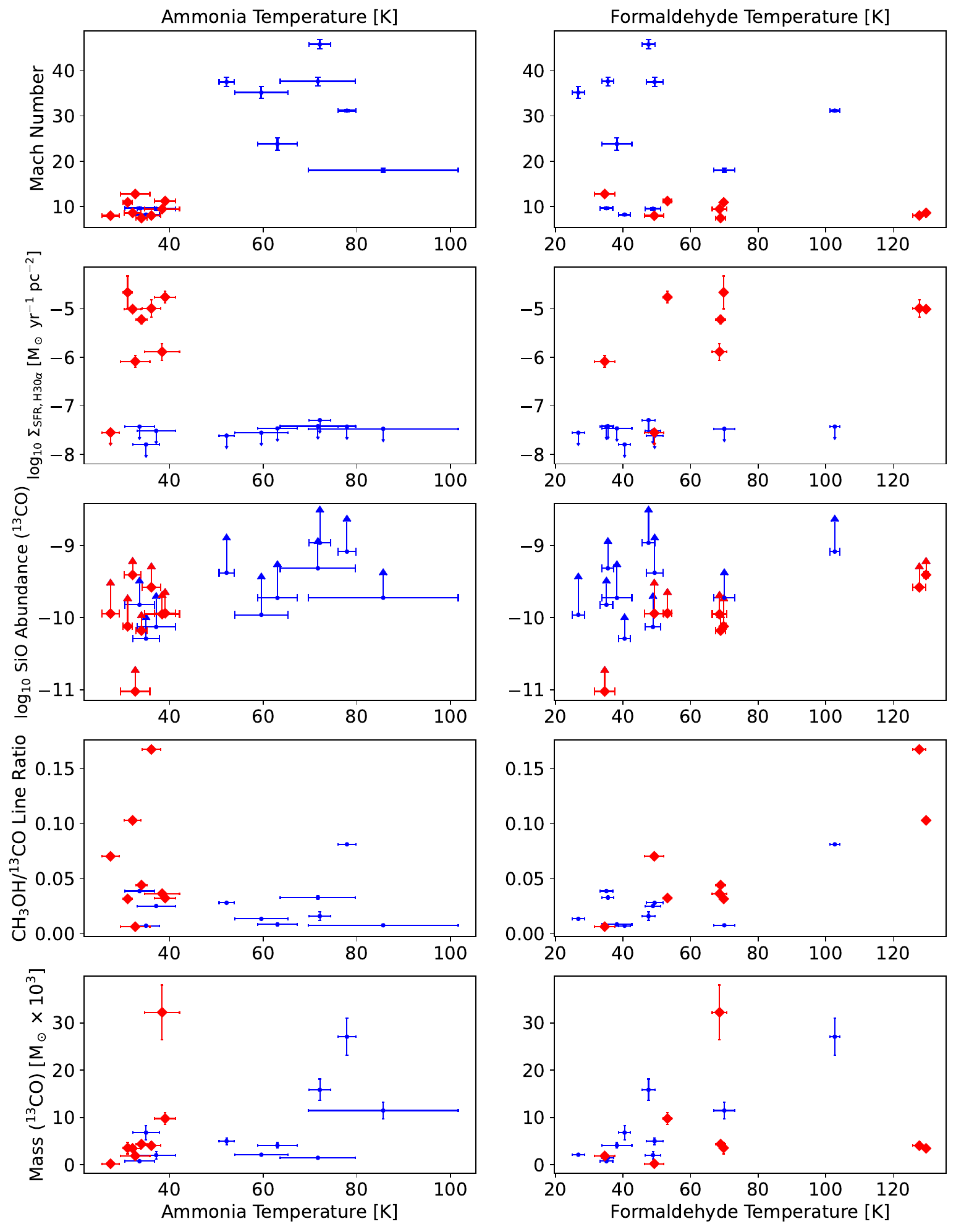}
\caption{Derived cloud properties plotted against both ammonia and formaldehyde temperatures. The red diamonds are clouds with 24 $\mu$m detections, while the blue circles are clouds without. The ammonia temperatures are the value of the ammonia (3,3) - (1,1) temperature maps at the pixel with peak NH\textsubscript{3} (3,3) integrated intensity. The formaldehyde temperatures are the average formaldehyde temperature taken over the 1' box centered on the pixel with peak H\textsubscript{2}CO $J=3_{03} \to 2_{02}$ integrated intensity.}
\label{fig:tempProperties}
\end{figure*}

In Figure \ref{fig:tempProperties}, we plot the Mach number, $\Sigma_\text{SFR}$, SiO abundance, and CH\textsubscript{3}OH/\textsuperscript{13}CO line ratio against the ammonia and formaldehyde temperature for each cloud. The ammonia temperature is well-correlated with the turbulent Mach number, whereas the formaldehyde temperature is not. Although the Mach numbers are calculated using the formaldehyde temperatures and line widths, the result would be similar if the Mach numbers were calculated using ammonia since the thermal line widths (calculated from Eq.\ \ref{eq:vThermal} with the ammonia temperatures) are small compared to the observed ammonia line widths. On the other hand, the opposite is true for correlations with the CH$_3$OH/$^{13}$CO line ratio, which is a weak shock tracer. No strong correlations are apparent for any other properties. Given that the molecules all seem to trace different phases of the gas, each of which likely has different physical properties, strong correlations are not expected.

Another notable feature is that the clouds with the highest ammonia temperatures ($T_{13} > 50$ K) all do not have detected star formation; in other words, all the clouds with detected star formation (with both 24 $\mu$m and H30$\alpha$ emission) are at comparatively lower 

\startlongtable
\movetabledown=2.75in

\begin{rotatetable*}

\begin{deluxetable*}{c  c  c  c  c  c  c  c  c  c  c}
\tabletypesize{\scriptsize}
\tablecaption{Clouds properties derived in Section \ref{sec:results}. The clouds we take to be inside the bar, per the discussion in Sections \ref{sec:loc} and \ref{subsec:loc2}, have their names underlined and in bold. \label{table:cloudprops}}

\tablehead{\colhead{Cloud} & \colhead{T\textsubscript{12}} & \colhead{T\textsubscript{13}} & \colhead{T\textsubscript{H$_2$CO}} & \colhead{$\Delta v_{\text{H}_2\text{CO} 3_{03}-2_{02}}$} & \colhead{$\mathcal{M}$} & \colhead{$\text{log}_{10} \Sigma_{\text{SFR, H30}\alpha}$} & \colhead{$\text{log}_{10} \Sigma_{\text{SFR}, 24\mu\text{m}}$} & \colhead{N(SiO)/N(H\textsubscript{2} (\textsuperscript{13}CO)} & \colhead{N(SiO)/N(H\textsubscript{2}) (C\textsuperscript{18}O)} & \colhead{CH\textsubscript{3}OH/\textsuperscript{13}CO} \\  & K & K & K & km\ s$^{-1}$ & & M$_\odot$ yr$^{-1}$ pc$^{-2}$ & M$_\odot$ yr$^{-1}$ pc$^{-2}$ & & } 
\startdata
G008.67 & $19.59 \pm 1.29$ & $33.98 \pm 1.18$ & $68.87 \pm 1.53$ & $5.17 \pm 0.07$ & $7.47 \pm 0.13$ & $-5.22 \pm 0.09$ & $ \geq-5.50$ & $\geq -10.2$ & $\geq -10.7$ & $0.0442 \pm 0.0004$ \\ 
G008.40 & $23.59 \pm 3.38$ & $34.96 \pm 2.86$ & $40.45 \pm 1.76$ & $4.34 \pm 0.06$ & $8.25 \pm 0.22$ & $\leq -7.80$ & $\leq -6.00$ & $\geq -10.3$ & $\geq -10.6$ & $0.0072 \pm 0.0003$ \\ 
G006.92 & $20.28 \pm 3.47$ & $33.56 \pm 3.16$ & $35.09 \pm 1.85$ & $4.71 \pm 0.09$ & $9.65 \pm 0.32$ & $\leq -7.43$ & $\leq -5.86$ & $\geq -9.82$ & $\geq -10.1$ & $0.0387 \pm 0.0016$ \\ 
G006.56 & $42.36 \pm 13.18$ & $85.63 \pm 16.02$ & $69.98 \pm 3.17$ & $12.33 \pm 0.10$ & $18.05 \pm 0.43$ & $\leq -7.47$ & $\leq -6.29$ & $\geq -9.72$ & $\geq -9.53$ & $0.0076 \pm 0.0008$ \\ 
G005.75 & $37.64 \pm 10.92$ & $74.34 \pm 12.19$ & --- & $11.03 \pm 0.24$ & --- & $\leq -7.37$ & $\leq -6.68$ & --- & --- & $0.0073 \pm 0.0025$ \\ 
G005.49 & --- & $122.16 \pm 86.33$ & --- & $9.43 \pm 0.16$ & --- & $\leq -7.41$ & $\leq -6.58$ & --- & --- & --- \\ 
G005.38 & $35.05 \pm 6.98$ & $71.68 \pm 8.05$ & $35.51 \pm 1.70$ & $18.26 \pm 0.15$ & $37.65 \pm 0.95$ & $\leq -7.42$ & $\leq -6.55$ & $\geq -9.32$ & $\geq -9.05$ & $0.0328 \pm 0.0015$ \\ 
G003.43 & $19.69 \pm 1.62$ & $27.37 \pm 1.83$ & $49.25 \pm 2.84$ & $4.66 \pm 0.04$ & $8.00 \pm 0.24$ & $\leq -7.55$ & $ \geq-6.09$ & $\geq -9.94$ & $\geq -10.3$ & $0.0703 \pm 0.0016$ \\ 
G003.09 & $46.89 \pm 2.17$ & $77.88 \pm 1.90$ & $102.67 \pm 1.41$ & $25.71 \pm 0.08$ & $31.19 \pm 0.24$ & $\leq -7.43$ & $\leq -6.66$ & $\geq -9.09$ & $\geq -8.98$ & $0.0812 \pm 0.0003$ \\ 
G003.02 & $31.48 \pm 1.75$ & $72.09 \pm 2.33$ & $47.54 \pm 1.94$ & $15.38 \pm 0.25$ & $45.81 \pm 1.04$ & $\leq -7.30$ & $\leq -6.56$ & $\geq -8.96$ & $\geq -8.55$ & $0.0159 \pm 0.0041$ \\ 
G002.96 & $33.89 \pm 4.34$ & $63.03 \pm 4.22$ & $38.18 \pm 4.43$ & $11.97 \pm 0.13$ & $23.83 \pm 1.41$ & $\leq -7.47$ & $\leq -6.56$ & $\geq -9.73$ & $\geq -9.41$ & $0.0084 \pm 0.0015$ \\ 
G002.51 & $30.56 \pm 1.91$ & $52.16 \pm 1.60$ & $49.35 \pm 2.45$ & $21.39 \pm 0.28$ & $37.51 \pm 1.06$ & $\leq -7.61$ & $\leq -6.46$ & $\geq -9.38$ & $\geq -9.34$ & $0.0280 \pm 0.0006$ \\ 
G001.93 & $33.15 \pm 6.19$ & $59.59 \pm 5.68$ & $26.76 \pm 1.81$ & $14.99 \pm 0.23$ & $35.17 \pm 1.32$ & $\leq -7.55$ & $\leq -6.64$ & $\geq 9.96$ & $\geq -9.66$ & $0.0134 \pm 0.0009$ \\ 
G358.48 & $31.08 \pm 6.21$ & $37.13 \pm 4.10$ & $48.85 \pm 2.28$ & $5.49 \pm 0.06$ & $9.53 \pm 0.25$ & $\leq -7.52$ & $\leq -5.91$ & $\geq -10.1$ & $\geq -10.5$ & $0.0251 \pm 0.0006$ \\ 
G354.61 & $29.67 \pm 5.36$ & $38.40 \pm 3.72$ & $68.56 \pm 2.16$ & $6.46 \pm 0.10$ & $9.47 \pm 0.21$ & $-5.89 \pm 0.17$ & $ \geq-5.51$ & $\geq -9.95$ & $\geq -10.3$ & $0.0363 \pm 0.0003$ \\ 
G353.41 & $18.98 \pm 0.97$ & $31.02 \pm 0.94$ & $69.79 \pm 0.71$ & $7.51 \pm 0.07$ & $10.95 \pm 0.12$ & $-4.66 \pm 0.34$ & $ \geq-5.00$ & $\geq -10.1$ & $\geq -10.50$ & $0.0316 \pm 0.0002$ \\ 
G351.77 & $16.97 \pm 1.89$ & $32.09 \pm 1.78$ & $129.70 \pm 0.69$ & $8.12 \pm 0.08$ & $8.62 \pm 0.09$ & $-5.01 \pm 0.08$ & $ \geq-5.51$ & $\geq -9.41$ & $\geq -9.84$ & $0.1030 \pm 0.0002$ \\ 
G351.58 & $18.39 \pm 2.21$ & $36.10 \pm 1.99$ & $127.70 \pm 1.97$ & $7.54 \pm 0.12$ & $8.04 \pm 0.15$ & $-4.99 \pm 0.18$ & $ \geq-5.86$ & $\geq -9.58$ & $\geq -10.23$ & $0.1674 \pm 0.0007$ \\ 
G350.18 & $20.36 \pm 3.36$ & $32.65 \pm 3.11$ & $34.57 \pm 3.07$ & $6.23 \pm 0.07$ & $12.79 \pm 0.59$ & $-6.09 \pm 0.12$ & $ \geq-5.89$ & $\geq -11.0$ & $\geq -11.3$ & $0.0063 \pm 0.0010$ \\ 
G350.10 & $24.92 \pm 2.76$ & $39.04 \pm 2.24$ & $53.15 \pm 1.32$ & $6.73 \pm 0.06$ & $11.24 \pm 0.18$ & $-4.76 \pm 0.12$ & $ \geq-5.15$ & $\geq -9.94$ & $\geq -10.2$ & $0.0323 \pm 0.0003$ \\ 
\\
\enddata
 
\end{deluxetable*}
\end{rotatetable*}

\noindent ammonia temperatures. This is not observed with the formaldehyde temperature.

There are several possible reasons for these differences. It may be that formaldehyde is more sensitive to heating from star formation, giving the distinct core structure in the formaldehyde temperature maps, and that ammonia is more sensitive to turbulent heating, resulting in a strong correlation with Mach number. Although neither formaldehyde temperatures nor ammonia temperatures appear to correlate with the SiO abundance, the stronger correlation with the methanol line ratio may indicate that formaldehyde is more sensitive to weak shocks that do not necessarily cause turbulence in the gas. Methanol emission, although associated with shocks, has also been found to be suppressed at sites of star formation due to direct photodissociation or inefficient production from CO on dust grains (e.g.,\ \citealt{meier_spatially_2014, saito_merger-induced_2017}), so star formation is unlikely to explain the correlation between formaldehyde and methanol. A simpler explanation may be that formaldehyde and methanol are chemically similar, whereas the nitrogen atom in ammonia may make it chemically different from methanol. Indeed, astrochemical theory and experiments have shown that both formaldehyde and methanol can be, and likely are, produced by successive hydrogenation of CO on ice and dust grains in molecular clouds (e.g.,\ \citealt{hiraoka_formation_1994, watanabe_efficient_2002, potapov_formation_2017}).

\begin{table*}[t]
\centering
\setlength{\tabcolsep}{0.75em}

\begin{tabular}{c  c  c  c} 
 Cloud & Mass Estimate (\textsuperscript{12}CO) & Mass Estimate (\textsuperscript{13}CO) & Mass Estimate (C\textsuperscript{18}O) \\ & M$_\odot \times 10^3$ & M$_\odot \times 10^3$ & M$_\odot \times 10^3$ \\ [0.5ex] 
 \hline\hline
 G008.67 & $5.3 \pm 0.5$ & $4.3 \pm 0.4$ & $9.1 \pm 0.8$ \\ 
G008.40 & $13.6 \pm 3.0$ & $6.8 \pm 1.5$ & $13.0 \pm 2.9$ \\ 
G006.92 & $1.8 \pm 0.5$ & $0.7 \pm 0.2$ & $1.3 \pm 0.4$ \\ 
\textbf{\underline{G006.56}} & $116.5 \pm 17.8$ & $11.5 \pm 1.8$ & $7.8 \pm 1.2$ \\ 
\textbf{\underline{G005.75}} & $14.2 \pm 2.1$ & $0.9 \pm 0.1$ & $0.3 \pm 0.0$ \\ 
\textbf{\underline{G005.49}} & $4.5 \pm 0.7$ & $0.4 \pm 0.1$ & $0.020 \pm 0.003$ \\ 
\textbf{\underline{G005.38}} & $34.9 \pm 5.2$ & $1.5 \pm 0.2$ & $0.6 \pm 0.1$ \\ 
G003.43 & $0.5 \pm 0.2$ & $0.2 \pm 0.1$ & $0.3 \pm 0.1$ \\ 
\textbf{\underline{G003.09}} & $276.7 \pm 39.8$ & $27.1 \pm 3.9$ & $17.8 \pm 2.6$ \\ 
\textbf{\underline{G003.02}} & $311.6 \pm 44.8$ & $15.9 \pm 2.3$ & $6.1 \pm 0.9$ \\ 
\textbf{\underline{G002.96}} & $84.6 \pm 12.1$ & $4.1 \pm 0.6$ & $2.4 \pm 0.3$ \\ 
\textbf{\underline{G002.51}} & $53.7 \pm 7.6$ & $5.0 \pm 0.7$ & $3.3 \pm 0.5$ \\ 
\textbf{\underline{G001.93}} & $48.6 \pm 6.9$ & $2.1 \pm 0.3$ & $0.9 \pm 0.1$ \\ 
G358.48 & $2.7 \pm 1.1$ & $2.0 \pm 0.8$ & $3.9 \pm 1.5$ \\ 
G354.61 & $46.2 \pm 8.3$ & $32.3 \pm 5.8$ & $60.0 \pm 10.8$ \\ 
G353.41 & $3.4 \pm 1.2$ & $3.5 \pm 1.2$ & $7.0 \pm 2.4$ \\ 
G351.77 & $2.0 \pm 0.1$ & $3.5 \pm 0.2$ & $7.1 \pm 0.5$ \\ 
G351.58 & $5.2 \pm 1.0$ & $4.0 \pm 0.8$ & $10.4 \pm 2.0$ \\ 
G350.18 & $7.7 \pm 1.0$ & $1.8 \pm 0.2$ & $2.7 \pm 0.3$ \\ 
G350.10 & $24.8 \pm 3.2$ & $9.8 \pm 1.3$ & $15.5 \pm 2.0$ \\ 

\hline

\end{tabular}
\caption{Mass estimates of each cloud using the three isotopologues of CO. The clouds we take to be inside the bar, per the discussion in Sections \ref{sec:loc} and \ref{subsec:loc2}, have their names underlined and in bold.}
\label{table:masses}
\end{table*}

\begin{figure*}[t]
\centering
\includegraphics[width=5.5in]{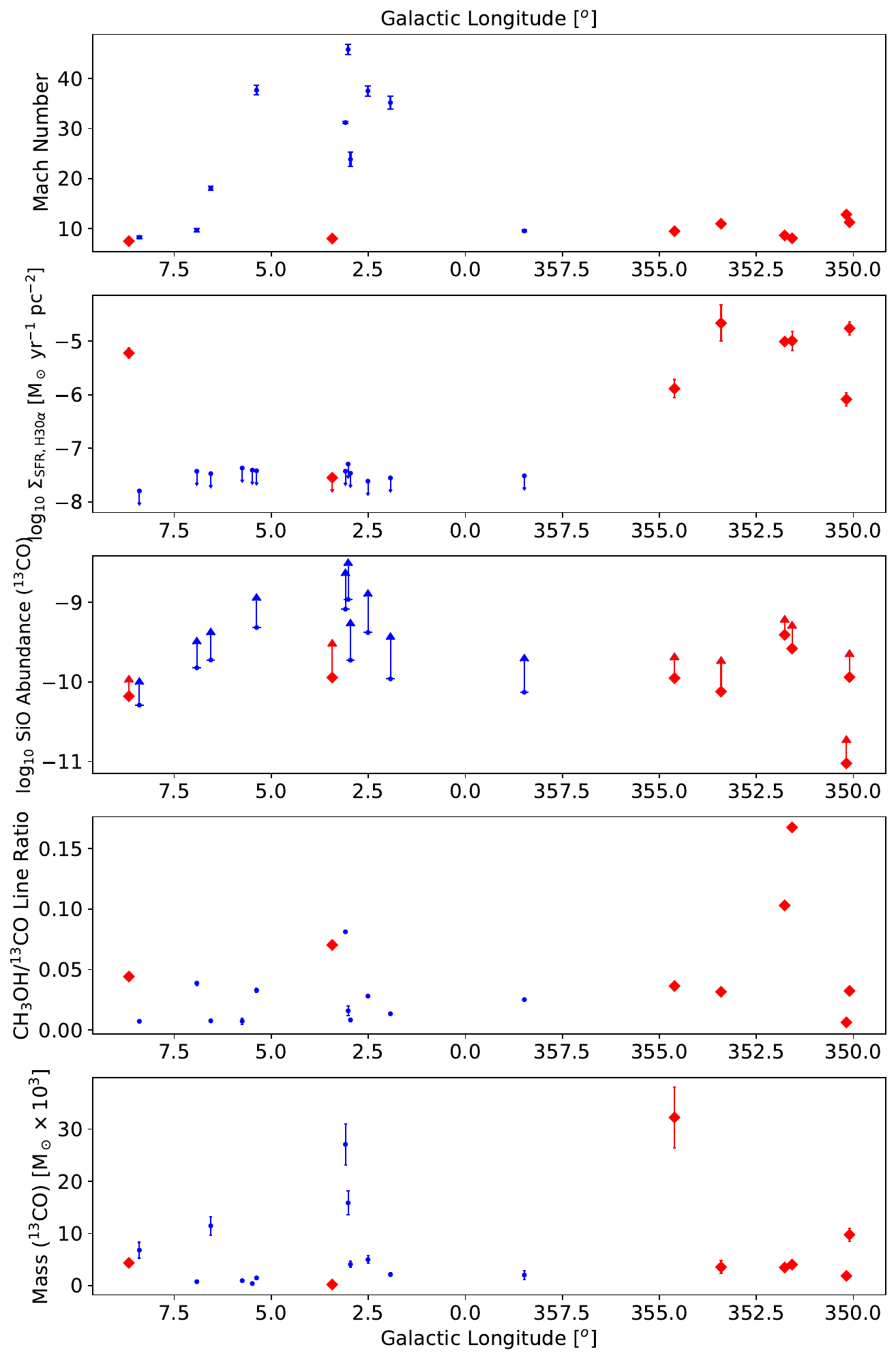}
\caption{Derived cloud properties plotted against Galactic longitude. The red diamonds are clouds with 24 $\mu$m detections, while the blue circles are clouds without. The ammonia temperatures are the value of the ammonia (3,3) - (1,1) temperature maps at the pixel with peak NH\textsubscript{3} (3,3) integrated intensity. The formaldehyde temperatures are the average formaldehyde temperature taken over the 1' box centered on the pixel with peak H\textsubscript{2}CO $J=3_{03} \to 2_{02}$ integrated intensity.}
\label{fig:glonProperties}
\end{figure*}

\begin{figure}[t]
\centering
\includegraphics[width=3in]{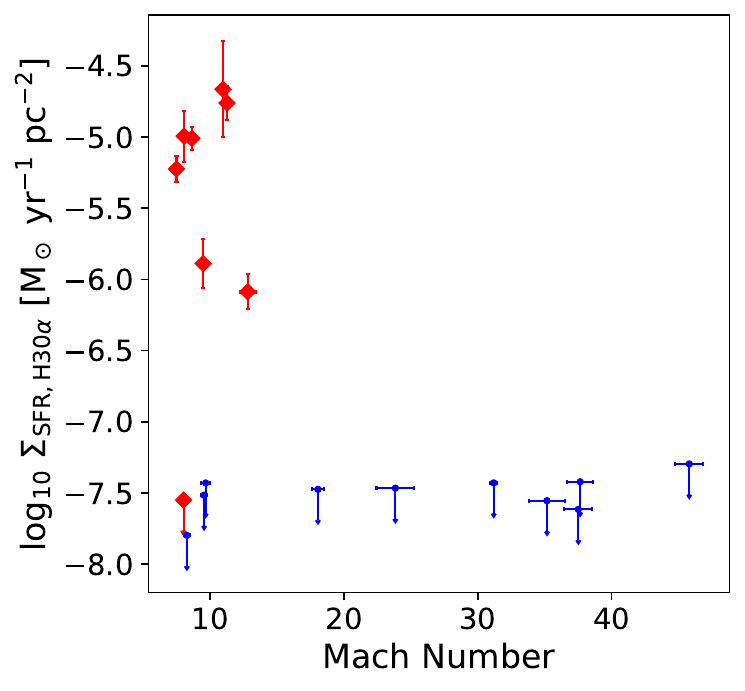}
\caption{$\Sigma_\text{SFR}$, as measured by H30$\alpha$ emission, plotted against Mach number. The red points indicate clouds with a significant H30$\alpha$ detection, and the blue points indicate clouds without one, for which an upper limit is given.}
\label{fig:sfMach}
\end{figure}

Furthermore, while the ammonia temperature is largely independent of H\textsubscript{2} density (e.g.,\ \citealt{ott_temperature_2005}), formaldehyde is an asymmetric top molecule, and thus, the H\textsubscript{2} density affects the excitation of the different energy levels. Although the asymmetry in formaldehyde is very small \citep{mangum_formaldehyde_1993}, making it useful for measuring molecular cloud kinetic temperatures (e.g.,\ \citealt{hogerheijde_millimeter_1995}, \citealt{ao_thermal_2013}, \citealt{ginsburg_dense_2016}, \citealt{tang_kinetic_2018}), most of these studies that survey a large area or multiple clouds typically assume a constant $n(\text{H}_2)$ across the sample. For example, \citet{ginsburg_dense_2016} assumes $n(\text{H}_2) = 10^{4}$ cm\textsuperscript{-3} for most analyses, although they also calculate kinetic temperatures with a constant $n(\text{H}_2) = 10^{3}$ cm\textsuperscript{-3} and $n(\text{H}_2) = 10^{5}$ cm\textsuperscript{-3}. Similarly, \citet{tang_kinetic_2018} assumes a constant $n(\text{H}_2) = 10^{5}$ cm\textsuperscript{-3} across their sample of ATLASGAL clumps. However, given the variety of clouds in our sample, particularly in their location and possible environment, assuming a constant H\textsubscript{2} density is likely an oversimplification. We therefore also explore the H\textsubscript{2} densities required to explain the observed formaldehyde emission using the ammonia temperature as the kinetic temperature of the gas.

For each cloud, we estimate the column density of H\textsubscript{2}CO 
using a $N(\text{H}_2\text{CO})$-to-$N(\text{H}_2)$ abundance ratio of $1.2 \times 10^{-9}$ \citep{wootten_molecular_1978, ginsburg_dense_2016}, where we use the H\textsubscript{2} column densities derived from the LTE calculations of \textsuperscript{13}CO and C\textsuperscript{18}O in Section \ref{subsubsec:sio}. In Sections \ref{subsec:mass}, we also calculate the H\textsubscript{2} density directly using the \textsuperscript{12}CO emission and a CO-to-H\textsubscript{2} X-factor; however, as discussed in \ref{subsec:xfactor}, we find that this is likely less accurate, particularly for clouds in the bar.

We use \textsc{RADEX} to calculate the $R_{\rm{H}_2 \rm{CO}}$ line ratio at the cloud's ammonia temperature (T\textsubscript{13}), H\textsubscript{2}CO $J=3_{03} \to 2_{02}$ line width, the two derived H\textsubscript{2}CO column densities, and a range of H\textsubscript{2} volume densities ($n = 10^{2.5} - 10^{7}$ cm$^{-3}$). We then interpolate and attempt to calculate the H\textsubscript{2} volume density given the observed line ratio for both column densities. The line width, line ratio, and column densities are taken to be the average within within the same 1' box that the formaldehyde temperature is calculated in.

We find that, for all but four of the clouds, there is no H\textsubscript{2} density for which the expected line ratio from \textsc{RADEX} matches the observed line ratio. The four clouds for which there is an H\textsubscript{2} density at which the two agree are the four clouds where the formaldehyde and ammonia temperatures from Section \ref{subsec:temperature} agree within uncertainties. In the cases where T\textsubscript{H$_2$CO} $>$ T\textsubscript{13}, the formaldehyde column density must be several orders of magnitude higher for there to be a H\textsubscript{2} density at which the colder T\textsubscript{13} can explain the observed line ratio. On the other hand, for the clouds where T\textsubscript{H$_2$CO} $<$ T\textsubscript{13}, there is no combination of formaldehyde column density and H\textsubscript{2} density at which the warmer T\textsubscript{13} can explain the observed line ratio.

Given that the critical densities of the ammonia lines are orders of magnitude lower than those of the formaldehyde lines \citep{shirley_critical_2015, ginsburg_dense_2016}, we expect that the two temperatures trace different density components of the gas. This is further complicated by the significant difference in beam size of the two observations. We therefore do not necessarily expect the two to agree, although we do not conclude that one traces the overall properties of the clouds better than the other. More data at a comparable resolution is needed to better understand the relationships between the ammonia and formaldehyde thermometers.

\subsection{Star Formation and Turbulence}
\label{subsec:sfTurb}

We note in Section \ref{subsec:temperature} that the clouds with higher ammonia temperatures all do not have 24 $\mu$m or H30$\alpha$ detections. At the same time, we see in Figure \ref{fig:tempProperties} that these clouds have the highest Mach numbers ($\mathcal{M} > 15$) in the sample. This indicates that the turbulence in these clouds may be inhibiting star formation to some degree.

We plot the $\Sigma_\text{SFR}$ measured by H30$\alpha$ against the Mach number in Figure \ref{fig:sfMach}. Although there is no direct inverse correlation between the two, we find that all the clouds with detected star formation are at relatively low Mach numbers; in other words, the highly turbulent clouds are not star forming. 

It is well known that the SFR in the CMZ is significantly lower than expected given the high gas densities and that this discrepancy may be due to turbulent pressure \citep{longmore_variations_2013, kruijssen_what_2014, barnes_star_2017}. The relationship between SFR and turbulence is less understood on the scale of individual molecular clouds, as turbulence can both create density fluctuations needed for local collapse (e.g.,\ \citealt{mac_low_control_2004}) and also prevent global cloud collapse (e.g.,\ \citealt{federrath_link_2016}). Meanwhile, star formation itself can inject energy into the surrounding gas, enhancing turbulence \citep{yun_turbulent_2021}; we see that, although the clouds with detected star formation are comparatively less turbulent, their Mach numbers still indicate supersonic motions, likely caused by dynamical activities related to stellar feedback. We conclude from our observations that the turbulent pressure is suppressing star formation in the clouds with high Mach numbers.  

\subsection{What is Heating the Clouds?}

The gas temperatures of most of the clouds in our sample are well above the typical values in the Galactic disk of $\sim 10-20$ K, and some are well above the CMZ average of $\sim$60 K \citep{ginsburg_dense_2016}.

There are likely four main heating mechanisms for molecular clouds here: radiative photoelectric heating, X-ray heating, cosmic ray heating, and turbulent heating (e.g.,\ \citealt{ao_thermal_2013}).

Although the Galactic center and CMZ regions have an elevated X-ray luminosity \citep{yuasa_suzaku_2008, muno_diffuse_2004} and cosmic ray ionization rate \citep{yusef-zadeh_cosmic-ray_2007, indriolo_herschel_2015}, we do not expect this to be the case for the Galactic bar and disk regions in which the clouds in our sample lie. \citet{ginsburg_dense_2016} finds that, although in some regions of the CMZ, heating from cosmic rays may be important, the energetics of the CMZ are largely dominated by turbulent heating. X-ray heating and cosmic ray heating can thus be disfavored as heating mechanisms in these clouds.

In the clouds with no detections of H30$\alpha$ or Spitzer 24 $\mu$m emission, it is highly unlikely that radiation plays a major role in heating since the nondetection of these indicates the lack of high-mass stars. Thus, for these clouds, the most likely heating mechanism is turbulence. We can estimate the temperature due to turbulent heating:
\small
\begin{equation}
\label{eq:turbTemp}
\begin{split}
T_\text{turb} = & \ 62 \left( \frac{n}{10^{4.5} \text{ cm}^{-3}} \right)^{1/6} \left( \frac{L}{5 \text{ pc}} \right)^{-1/3} \\
& \ \times \left(\frac{ \text{d}v/\text{d}r}{5 \text{ km\ s}^{-1} \text{ pc}^{-1}}\right)^{-1/3} \left( \frac{V_\text{FWHM}}{20 \text{ km\ s}^{-1}} \right) \text{ K},
\end{split}
\end{equation}
\normalsize
which is a simplified analytic solution to the turbulence-dominated thermal equilibrium equation, assuming gas densities $n \leq 10^{5} \text{ cm}^{-3}$, where $L$ is the size of the cloud, d$v/$d$r$ is the velocity gradient of the cloud, and $V_\text{FWHM}$ is the observed FWHM line width  \citep{ao_thermal_2013}.

\begin{figure}[t]
\centering
\includegraphics[width=3in]{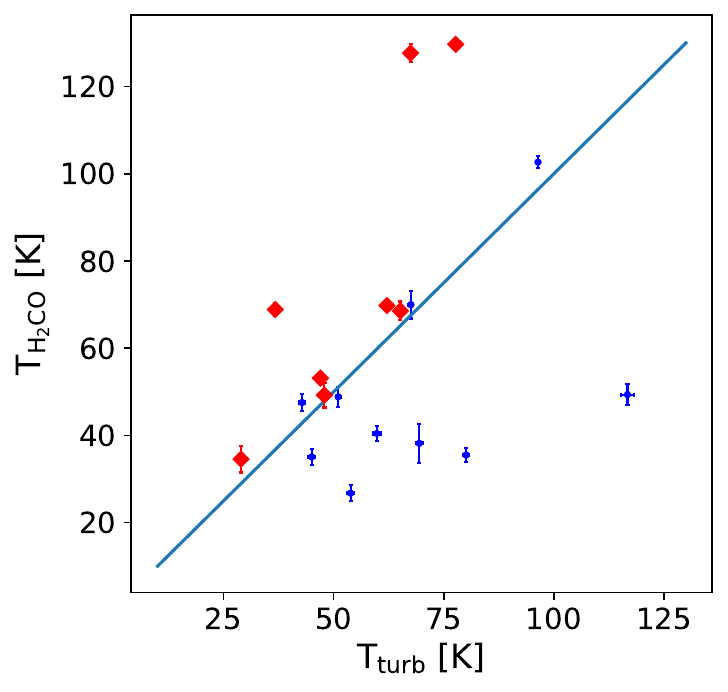}
\caption{Observed formaldehyde temperature plotted against the estimated temperature from turbulent heating (Eq.\ \ref{eq:turbTemp}). The red diamonds are clouds with 24 $\mu$m detections, while the blue circles are clouds without. The blue line indicates where the two temperatures are equal. For clouds corresponding to points below the blue line, the turbulence-driven estimated temperature is greater than the observed temperature, so turbulence is (more than) sufficient to explain the observed temperature. For clouds corresponding to points above the blue line, turbulence is insufficient to explain the observed temperature.}
\label{fig:turbTemp}
\end{figure}

Since the structure of the formaldehyde emission is generally more complex than the H30$\alpha$ emission, we fit a circular two-dimensional Gaussian around the peak H\textsubscript{2}CO $(3_{21}-2_{20})$ emission to calculate the cloud angular diameter, rather than fit a 1-dimensional Gaussian to the flattened PV diagram as in Section \ref{subsec:sf}. We again take the angular diameter to be four times the standard deviation of the Gaussian fit, then calculate the physical size of the cloud using its distance. 

We use the cloud velocity gradients measured in Section \ref{subsec:turb} and assume a gas density of $10^4 \text{ cm}^{-3}$, the same as used to derive Eq.\ \ref{eq:formaldehydeTemp}, although we note that the estimated turbulent kinetic temperature depends relatively weakly on the gas density, cloud size, and velocity gradient compared to the line width. We calculate the estimated turbulence-driven kinetic temperatures of all our clouds and plot the results against the observed formaldehyde temperatures in Figure \ref{fig:turbTemp}, noting the clouds that are detected with Spitzer.

We see that, for the clouds with no detected Spitzer emission, or in other words, with no detected star formation, turbulent heating is generally sufficient to explain the observed temperatures of the clouds, but for the clouds with detected Spitzer emission, it is generally insufficient. We can then expect that turbulence is the dominant heating mechanism in the clouds without star formation, whereas radiation is dominant in the clouds with star formation.

Although turbulence is likely heating the non-star-forming clouds, it is uncertain what exactly is causing the turbulence. High-velocity shocks could be one potential mechanism driving turbulence. Such shocks would be traced by SiO \citep{martin-pintado_sio_1992, schilke_sio_1997, garay_silicon_2000}, but we find no correlation between the SiO abundances and the kinetic temperature (see Figure \ref{fig:tempProperties}). Our analysis of the SiO column densities in Section \ref{subsubsec:sio} suggests that the kinetic temperature of the gas phase traced by SiO is possibly significantly higher than that traced by formaldehyde, and given such a multiphase scenario, a correlation between properties of phases traced by different molecules is not expected. We also note that, due to the likely subthermal emission of the SiO $J = 5\to 4$ line, the SiO abundances calculated in Section \ref{subsubsec:sio} are lower limits. However, the values do appear to be, on average, higher for clouds in the bar (see Figure \ref{fig:glonProperties}), and in general are elevated above ambient values of $10^{-12}$ to $10^{-11}$ \citep{schilke_sio_1997, garay_silicon_2000}. On the other hand, shocks at lower velocity, as traced by methanol, may also be contributing to turbulence and heating. We do see a weak correlation between the methanol line ratio and the formaldehyde temperature, as noted in Section \ref{subsec:temperature}, although this is complicated by the fact that the clouds with star formation also appear to show elevated methanol line ratios. We can see in the integrated intensity maps in Figure Set \ref{fig:integratedIntensities} that the methanol emission is slightly more extended than the SiO emission for many clouds, indicating that weak shocks may be prevalent on a wider scale, potentially heating the gas, generating turbulence, and mildly elevating the SiO abundance as well. As such, shocks may be a reasonable source for the observed turbulence, particularly for the clouds in the bar, although the exact source is uncertain. The uncertainty is compounded by the assumption of gas in a single phase in our calculations, whereas the clouds are certainly much more complex.

\subsection{Where Are the Clouds Located?}
\label{subsec:loc2}

In Section \ref{sec:loc}, we discussed the locations of the clouds based primarily on their line widths and prior kinematic distance determinations. With the greater context of the various properties we have derived, we provide further evidence that some of the clouds in the sample lie in the Galactic bar region.

We initially considered all clouds with an NH\textsubscript{3} (1,1) line width greater than 10 km\ s$^{-1}$ to be on the Galactic bar, since such broad lines are atypical for clouds in the Galactic disk. However, we noted that the two clouds at the most negative Galactic longitude, G350.18 and G350.11, are at Galactocentric radii around 4 kpc, which may be beyond the extent of the Galactic bar, and so preliminarily concluded that they do not lie in the bar region, despite having broad lines. However, given that the bar may extend to 5 kpc or greater \citep{wegg_structure_2015}, it is possible that these two clouds are in the region around the tip of the bar, where gas accumulates at the apocenter of x\textsubscript{1} orbits.  With our ALMA observations, we find that these two clouds have relatively narrow formaldehyde line widths, and note that their Mach numbers and ammonia temperatures, both of which are calculated independent of distance, are lower than all of the other clouds we assumed to be in the bar. We thus still consider these two clouds to be part of a separate population from the nine clouds in the bar region.

In fact, we can broadly categorize the clouds in our sample into three groups. The first is highly turbulent clouds with no detected star formation. The Mach numbers of these clouds are all greater than 15, and many match or exceed values expected in the CMZ ($\gtrsim 25$). Although we can not calculate Mach numbers for G005.49 and G005.75 since they are not detected with the H\textsubscript{2}CO $J=3_{21} \to 2_{20}$ line, we include them in this group because they are associated with the same cloud complex G5 as G003.58; have very high ammonia temperatures, as do the other clouds in this group ($T_{13} > 50$ K); and also do not exhibit any star formation. Furthermore, G005.49 is the only cloud for which the bulk motion, as measured by the velocity gradient in Section \ref{subsec:turb}, appears to dominate the line width, indicating that some gradient-inducing process, such as shear, rotation, or a cloud-cloud collision, is affecting the cloud. In total, there are nine clouds in this group, and these nine are the same as the clouds we considered to be in the Galactic bar in Section \ref{sec:loc}. As previously noted, seven of these clouds are associated with either G5 or B2, both of which have been suggested to be high-velocity collisions in the Galactic bar region, possibly between the dust lanes and overshooting gas \citep{fux_3d_1999, sormani_geometry_2019, gramze_evidence_2023}. The other two clouds in this group, G006.56 and G001.93, have not previously been studied in detail, but their similar features to the clouds in G5 and B2 indicate that they may also be undergoing collisions, perhaps as they cross the nearside dust lane, or subject to other extreme physical processes in the Galactic bar region. In particular, G001.93 appears to be located spatially near the CMZ, so it may belong to a collision between the dust lane and CMZ itself, which is a second type of extended velocity feature observed in the dynamical simulations of \citet{sormani_geometry_2019}, or it may be subject to conditions closer to those in the CMZ. We conclude that, regardless of whether these clouds are undergoing high-velocity collisions, the extreme processes in the Galactic bar region are inhibiting gravitational collapse and suppressing star formation.

The second group contains stars with detected star formation but low turbulence. As noted in Section \ref{subsec:sfTurb}, the eight clouds with 24 $\mu$m detections, seven of which are also detected with H30$\alpha$, exhibit Mach numbers comparable to typical molecular clouds in the Galactic disk ($\lesssim$ 10), so these are likely normal Galactic disk star-forming regions. G350.18 and G350.11, which may lie around the tip of the Galactic bar, belong to this group. If these two clouds are indeed near the end of the bar, this may indicate that the processes causing the observed turbulence and inhibited star formation in the Galactic bar nearer to the Galactic Center, such as collisions and shears, are less prominent around the edge of the Galactic bar, because these two clouds do not exhibit such turbulence or suppressed star formation. This may agree with previous observations of nearby barred galaxies that show inhibited star formation along the length of the bar but strong star formation at the center and ends of the bar (e.g.,\  \citealt{reynaud_kinematics_1998, james_stellar_2016}).

The remaining three clouds, G008.40, G006.92, and G358.48, exhibit neither turbulence nor star formation. These are also likely typical Galactic disk molecular clouds with unusually bright NH\textsubscript{3} (3,3) emission, and they may become star-forming in the future. 

We note here that while the calculation of $\Sigma_\textrm{SFR}$ is weakly dependent on distance, as noted in Section \ref{subsec:sf}, the detection or non-detection of star formation is enough to distinguish between our three groups. The only other properties calculated in Section \ref{sec:results} that are dependent on distance are the SiO abundance (Section \ref{subsubsec:sio}) and mass (Section \ref{subsec:mass}). However, we do not use these properties when distinguishing between our three groups. We do note that the SiO abundance may be generally elevated for clouds in the bar (see Figure \ref{fig:glonProperties}), consistent with the increased presence of strong shocks in the bar. We also discuss a relationship between the calculated mass and cloud location in Section \ref{subsec:xfactor}.

We find that all of the clouds likely to be in the bar are located on the nearside of the bar, at positive Galactic longitudes. This may be due to projection effects of the bar. At a bar angle of 30$^\circ$ \citep{wegg_structure_2015}, a Galactic longitude of $-10^\circ$ along the bar major axis corresponds to a Galactocentric radius of nearly twice that of a Galactic longitude of $10^\circ$. In other words, the farside of the bar is contained within a much smaller longitude range: assuming a semimajor axis of 3.5 kpc, the bar extends 8.4$^\circ$ in the negative longitude direction but 16$^\circ$ in the positive longitude direction. Furthermore, the distance from the Sun to the farside of the bar is greater than that to the nearside of the bar: a cloud on the bar major axis at -5$^\circ$ would be about 2.5 kpc further than a cloud on the bar major axis at 5$^\circ$. The combination of these two projection effects may make observing clouds on the farside of the bar more difficult than clouds on the nearside of the bar.

\subsection{X-Factor in the Bar}
\label{subsec:xfactor}

\begin{figure}[t]
\centering
\includegraphics[width=3in]{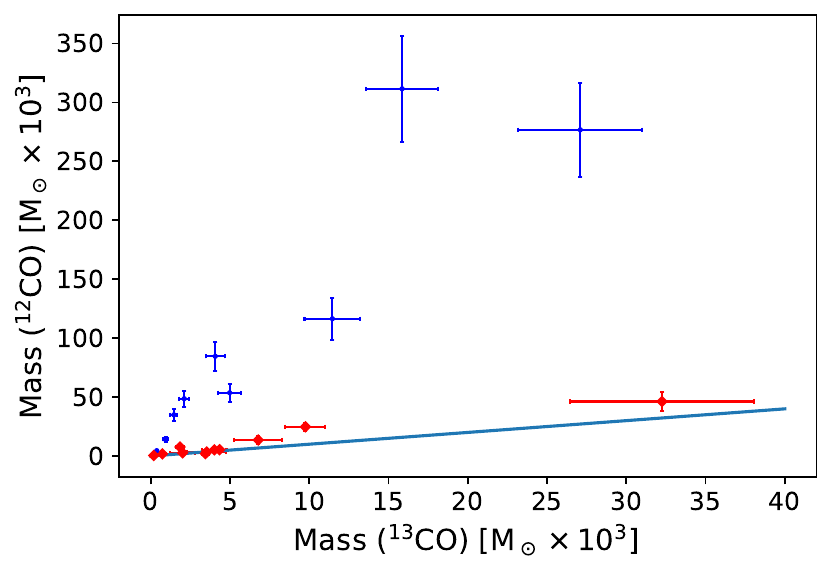}
\includegraphics[width=3in]{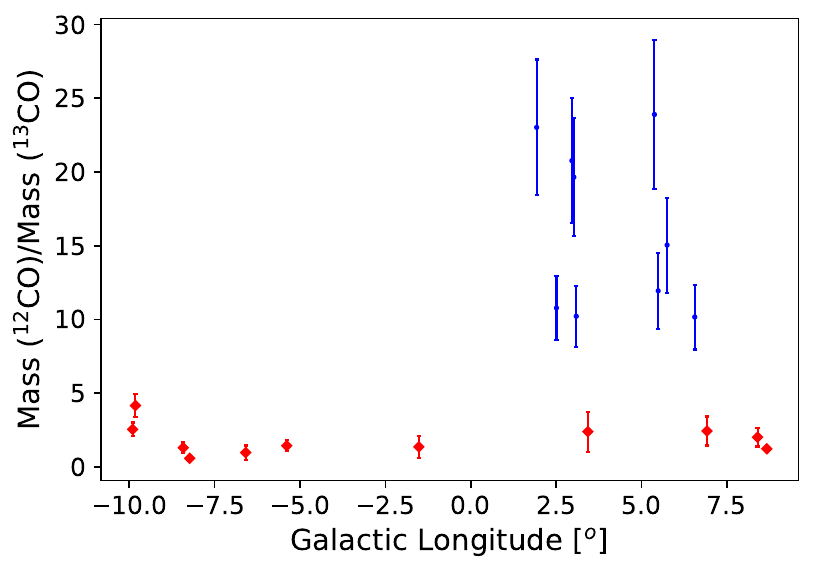}

\caption{\textit{Top:} mass estimate calculated using the integrated intensity of \textsuperscript{12}CO and an X-factor of $2.3 \times 10^{20}$ cm$^{-2}$ (K km s$^{-1}$)$^{-1}$ against the mass estimated calculated using the integrated intensity of \textsuperscript{13}CO using the equation of local thermal equilibrium. The blue circles are the clouds likely to be in the bar region as per the discussion in Section \ref{subsec:loc2}, and the red diamonds are the clouds likely not in the bar region. The blue line indicates where the two mass estimates are equal. \textit{Bottom:} ratio of the two mass estimates against Galactic longitude. The colors and symbols are the same as above.}
\label{fig:massComp}
\end{figure}

The mass estimates calculated using the equation of LTE (Eq. \ref{eq:sioColDens}) and the integrated intensities of \textsuperscript{13}CO and C\textsuperscript{18}O agree to within a factor of about 2-3, although there is one outlier for which it seems most of the C\textsuperscript{18}O lies outside the extent of the cloud as defined by a 5$\sigma$ cutoff on the H\textsubscript{2}CO $J=3_{03} \to 2_{02}$. On the other hand, the mass calculated using the integrated intensity of \textsuperscript{12}CO and the typical Galactic X-factor of $2.3 \times 10^{20}$ cm$^{-2}$ (K km s$^{-1}$)$^{-1}$ is quite discrepant from the other two estimates, as seen in Figure \ref{fig:massComp}. In particular, for the clouds most likely to be on the bar, the X-factor calculation significantly overestimates the mass compared to the LTE calculations, whereas the difference is smaller for the clouds not on the bar. 

\citet{ferriere_spatial_2007} find the X-factor in the Galactic Center region to be lower than the typical Galactic value, on the order of $10^{19}$ cm$^{-2}$ (K km s$^{-1}$)$^{-1}$. \citet{gramze_evidence_2023} find a similar value of $1.5 \times 10^{19}$ cm$^{-2}$ (K km s$^{-1}$)$^{-1}$ for G5, and conclude that the lower value extends beyond just the Galactic center, although they note that the special conditions of G5 may not be representative of the entire bar lane. Our findings provide further evidence that this reduced X-factor value does indeed extend along the Galactic bar, which is consistent with the dynamically disturbed nature of clouds on the bar. If we take the LTE mass estimate with \textsuperscript{13}CO to be more representative of the true cloud mass, then we can calculate an X-factor for each cloud in the bar, the average of which is $(1.59 \pm 0.5) \times 10^{19}$ cm$^{-2}$ (K km s$^{-1}$)$^{-1}$, where the uncertainty is taken to simply be the standard deviation over the sample. This value is consistent with that calculated by \citet{gramze_evidence_2023} and indicates that the standard Galactic X-factor is an overestimate by 10-20 times in this region.

\citet{gramze_evidence_2023} suggest that this discrepancy is due to low optical depth, since the typical X-factor of $2.3 \times 10^{20}$ cm$^{-2}$ (K km s$^{-1}$)$^{-1}$ assumes optical thickness \citep{strong_radial_1988}. We estimate the optical depth for each of the clouds using a root finding algorithm on the equation

\begin{equation}
\label{eq:opticalDepth}
\frac{I_\nu (^{12}\text{CO})}{I_\nu (^{13}\text{CO})} = \frac{1-e^{-\tau_{12}}}{1-e^{-\tau_{12}/\text{R}_\text{C}}},
\end{equation}
where $I_\nu$ is the integrated intensity of the line, $\tau_{12}$ is the optical depth of \textsuperscript{12}CO, and R\textsubscript{C} is the \textsuperscript{12}C/\textsuperscript{13}C isotope abundance ratio (Eq. \ref{eq:cAbundance}). We use the average integrated intensity over a 1' box centered on the pixel with maximum \textsuperscript{12}CO integrated intensity. The resulting optical depth values are shown in Table \ref{table:opticalDepths}.

\begin{table*}[t]
\centering
\setlength{\tabcolsep}{0.75em}

\begin{tabular}{c  c  c  c} 
 Cloud & $\tau_{12}$ (Eq. \ref{eq:cAbundance}) & $\tau_{12}$ ($\text{R}_\text{C} = 40$) & Location \\ [0.5ex] 
 \hline\hline
G008.67 & $24$ & $22$ & Out \\ 
G008.40 & $14$ & $16$ & Out\\ 
G006.92 & $14$ & $13$ & Out\\ 
G006.56 & $1.0$ & $3.2$ & In\\ 
G005.75 & $0.087$ & $2.1$ & In\\ 
G005.49 & $0.085$ & $1.8$ & In\\ 
G005.38 & $0.47$ & $2.7$ & In\\ 
G003.43 & $11$ & $9.1$ & Out\\ 
G003.09 & $0.78$ & $4.3$ & In\\ 
G003.02 & $0.35$ & $3.5$ & In\\ 
G002.96 & $0.097$ & $3.1$ & In\\ 
G002.51 & $1.9$ & $7.2$ & In\\ 
G001.93 & $0.11$ & $3.7$ & In\\ 
G358.48 & $20$ & $16$ & Out\\ 
G354.61 & $16$ & $16$ & Out\\ 
G353.41 & $32$ & $23$ & Out\\ 
G351.77 & $25$ & $18$ & Out\\ 
G351.58 & $12$ & $16$ & Out\\ 
G350.18 & $8.8$ & $14$ & Out\\ 
G350.10 & $12$ & $19$ & Out\\ 

\hline

\end{tabular}
\caption{Optical depth of \textsuperscript{12}CO for the clouds, calculated using an R\textsubscript{C} value following Eq. \ref{eq:cAbundance} or using $\text{R}_\text{C} = 40$. The location column indicates whether the cloud is inside or outside the bar, as per the discussion in Section \ref{subsec:loc2}}
\label{table:opticalDepths}
\end{table*}

We see that the optical depths of the clouds on the bar are significantly lower than those not on the bar, and are mostly within the optically thin regime. Since the optical depth is strongly dependent on R\textsubscript{C}, we also perform the calculation assuming a constant R\textsubscript{C} = 40 \citep{gramze_evidence_2023} to ensure that this difference is not only due to the large uncertainty of the dependence of R\textsubscript{C} on the Galactocentric radius (Eq. \ref{eq:cAbundance}). These are also shown in Table \ref{table:opticalDepths}. Although taking R\textsubscript{C} = 40 increases the optical depths of the clouds in the bar, they are still all lower than those of the clouds outside the bar, and we note that the values using Eq. \ref{eq:cAbundance} are likely more reflective of the actual optical depth. We thus conclude that the lower optical depth of the clouds on the bar are the likely cause of the reduced X-factor for these clouds, in agreement with \citet{gramze_evidence_2023}. This gives further evidence that the X-factor in the bar region as a whole is lower than the typical Galactic value, and the previously reported values of the mass inflow rate into the CMZ \citep{sormani_mass_2019, hatchfield_dynamically_2021}, which use the Galactic X-factor, are overestimates.

We previously noted that the masses estimated from the integrated intensities of \textsuperscript{13}CO and C\textsuperscript{18}O agree to within a factor of about 2-3, excluding one outlier. Interestingly, we note that there are nine clouds for which the mass estimate from C\textsuperscript{13}O is greater than that from \textsuperscript{18}CO, and these nine clouds are the clouds likely to be on the bar. Over these clouds (excluding the outlier for which $\text{M}_{\text{C}^{13}\text{O}}/\text{M}_{\text{C}^{18}\text{O}} = 19$), the average ratio is $\text{M}_{\text{C}^{13}\text{O}}/\text{M}_{\text{C}^{18}\text{O}} = 2$, whereas for the 11 clouds not on the bar, the average ratio is $\text{M}_{\text{C}^{13}\text{O}}/\text{M}_{\text{C}^{18}\text{O}} = 0.5$. One possible reason for this discrepancy is the uncertainty in location of the clouds, since, although the two mass estimates scale equivalently with distance, they are dependent in different ways on Galactocentric radius, as per the \textsuperscript{12}C/\textsuperscript{13}C and \textsuperscript{16}O/\textsuperscript{18}O isotope abundance ratios (Eqs. \ref{eq:cAbundance} and \ref{eq:oAbundance}). It may also be that these abundance ratios themselves do not apply to the clouds, or the Galactic bar region as a whole, in the same way as the typical Galactic X-factor does not apply.

\section{Conclusion}
\label{sec:conclusion}

We observed the molecular lines SiO $J=5\to 4$, H\textsubscript{2}CO $J=3_{21} \to 2_{20}$, H\textsubscript{2}CO $J=3_{03} \to 2_{02}$, HC\textsubscript{3}N $J = 24 \to 23$, CH\textsubscript{3}OH $J = 4_{22} \to 3_{12}$, C\textsuperscript{18}O $J = 2\to 1$, \textsuperscript{13}CO $J = 2\to 1$, \textsuperscript{12}CO $J = 2\to 1$, and H30$\alpha$ with the ALMA ACA for 20 clouds with Galactic longitudes $\lvert \ell \rvert < 10^\circ$ that are outside the CMZ. These clouds were selected from the HOPS survey, which covers velocities from -200 to 200 km s$^{-1}$ and Galactic latitudes $\lvert b \rvert < 5^\circ$, for their bright and broad NH\textsubscript{3} (3,3) emission. These spectral lines and bands probe several important processes (e.g.,\ \citealt{schilke_sio_1997, meier_spatially_2005, ott_temperature_2005, calzetti_calibration_2007, ginsburg_dense_2016}), and we measure temperatures, shocks, turbulence, $\Sigma_\text{SFR}$, and masses for all the clouds. We also used observations of the metastable ammonia inversion transitions with $(J, K) = (1,1), (2,2), (3,3)$ and $(6,6)$ from HOPS to measure ammonia temperatures, as well as the 4.5, 8, and 24 $\mu$m bands from Spitzer as tracers of star formation. 

We found that although the clouds do not share a velocity with the large-scale dust lanes feeding the CMZ, nine of them have similar properties to clouds that do have those velocities, such as high Mach numbers and ammonia temperatures. It is possible that these clouds are associated with the dust lanes but have been accelerated to velocities closer to zero by collisions. Of these clouds, most are associated with the known clumps G5 and B2, which may be sites of dust lane gas that has overshot the CMZ and is colliding with the dust lane on the opposite side \citep{fux_3d_1999, sormani_geometry_2019, gramze_evidence_2023}, although at least two of the clouds that we expect to lie on the Galactic bar have not been previously studied in detail. Furthermore, we showed that turbulence in the clouds along the Galactic bar, possibly from high-velocity collisions, is inhibiting star formation, as all of the clouds with high turbulence are not star forming. Heating solely from turbulent pressure is sufficient to heat these clouds to their observed temperatures, whereas photoelectric heating is dominant in the star-forming clouds. 

Simple modeling and calculations with \textsc{RADEX}, which assumes a single gas phase, suggests that all transitions are inconsistent with a single gas temperature and density. Different molecules appear to trace different gas phases in the molecular clouds. Ammonia appears to trace colder gas in the clouds, formaldehyde appears to trace warm gas, and SiO may be tracing shocked and hot gas, with each of these gas phases likely at a different density. A more detailed analysis that accounts for the complex, multiphase nature of the clouds can be carried out with further data, including observations of ammonia at comparable resolution.

We found that the typical \citet{strong_radial_1988} CO-to-H\textsubscript{2} X-factor of $2.3 \times 10^{20}$ cm$^{-2}$ (K km s$^{-1}$)$^{-1}$ significantly overestimates the mass of clouds in the bar, and that this overestimation is likely due to the lower optical thickness of the \textsuperscript{12}CO in these clouds compared to those outside the bar. We measured the average X-factor for these clouds to be $(1.59 \pm 0.5) \times 10^{19}$ cm$^{-2}$ (K km s$^{-1}$)$^{-1}$, which is comparable to the value of $1.5 \times 10^{19}$ cm$^{-2}$ (K km s$^{-1}$)$^{-1}$ obtained by \citet{gramze_evidence_2023} from observations of the cloud G5, which they also find to have optically thin, or slightly optically thick, \textsuperscript{12}CO. \citeauthor{gramze_evidence_2023} note that G5 may be not be representative of gas in the bar, but our findings suggest that this reduced X-factor indeed applies to the bar region as a whole, in which case previous measurements of the CMZ gas inflow rate \citep{sormani_mass_2019, hatchfield_dynamically_2021} are overestimates.

Our observations of clouds located in and out of the Galactic bar region highlight the dynamical activity of the Galactic bar outside of just the CMZ. Although some recent statistical surveys of barred galaxies suggest that star formation is environmentally independent and not lower in the bar region (e.g.\ \citealt{muraoka_co_2019, diaz-garcia_molecular_2021, querejeta_stellar_2021}), we have shown that, in the Milky Way, turbulence and other extreme processes along the bar lane inflows, induced by the bar potential, suppress star formation, an observation that has been reported for both individual barred galaxies and statistical studies (e.g.,\ \citealt{tubbs_inhibition_1982, reynaud_kinematics_1998, james_stellar_2016,maeda_statistical_2023}). Our findings further provide evidence that molecular clouds acquire CMZ-like properties as they travel along the bar inflows, prior to reaching the CMZ itself. This transitional region of the Milky Way merits further study and observation as a link between the Galactic disk and the CMZ, and a deeper understanding of its dynamics can provide insight into the inner Galaxy as a whole.

\begin{acknowledgements}

This paper makes use of the following ALMA data: ADS/JAO.ALMA\#2019.2.00068.S, ADS/JAO.ALMA\#2021.2.00001.S, \newline
ADS/JAO.ALMA\#2022.1.00591.S. ALMA is a partnership of ESO (representing its member states), NSF (USA) and NINS (Japan), together with NRC (Canada), MOST and ASIAA (Taiwan), and KASI (Republic of Korea), in cooperation with the Republic of Chile. The Joint ALMA Observatory is operated by ESO, AUI/NRAO and NAOJ. The National Radio Astronomy Observatory is a facility of the National Science Foundation operated under cooperative agreement by Associated Universities, Inc.

This research has made use of the NASA/IPAC Infrared Science Archive, which is funded by the National Aeronautics and Space Administration and operated by the California Institute of Technology.

A.N. is a student at the National Radio Astronomy Observatory and was funded as a participant in the National Radio Astronomy Observatory Research Experience for Undergraduates Site, supported by the National Science Foundation under grant No. 1852401.

M.C.S. acknowledges financial support from the European Research Council under the ERC Starting Grant ``GalFlow'' (grant 101116226).

A.G. acknowledges support from the NSF under grants AAG 2206511 and CAREER 2142300.

R.S.K. acknowledges financial support from the European Research Council via the ERC Synergy Grant ``ECOGAL'' (project ID 855130),  from the German Excellence Strategy via the Heidelberg Cluster of Excellence (EXC 2181 - 390900948) ``STRUCTURES'', and from the German Ministry for Economic Affairs and Climate Action in project ``MAINN'' (funding ID 50OO2206). R.S.K. is grateful for computing resources provided by the Ministry of Science, Research and the Arts (MWK) of the State of Baden-W\"{u}rttemberg through bwHPC and the German Science Foundation (DFG) through grants INST 35/1134-1 FUGG and 35/1597-1 FUGG, and also for data storage at SDS@hd funded through grants INST 35/1314-1 FUGG and INST 35/1503-1 FUGG. R.S.K. also thanks the Harvard-Smithsonian Center for Astrophysics and the Radcliffe Institute for Advanced Studies for their hospitality during his sabbatical, and the 2024/25 Class of Radcliffe Fellows for highly interesting and stimulating discussions.

\end{acknowledgements}

\software{
Python, IPython \citep{perez_ipython_2007}, 
NumPy \citep{harris_array_2020}, 
Matplotlib \citep{hunter_matplotlib_2007}, 
SciPy \citep{virtanen_scipy_2020}, 
Astropy \citep{the_astropy_collaboration_astropy_2022}, \texttt{spectral-cube} \citep{ginsburg_radio-astro-toolsspectral-cube_2019}, CASA \citep{the_casa_team_casa_2022}, CARTA \citep{comrie_carta_2021}, \textsc{RADEX} \citep{van_der_tak_computer_2007}}

\facility{ALMA, IRSA}

\appendix

\section{Clouds in the Galactic Bar}
\label{appendix:barClouds}

As noted in Section \ref{sec:loc}, for the clouds we assume to be in the Galactic bar, we calculate distances using the geometry of the bar and the Galactic longitudes of the clouds, assuming they lie along the Galactic bar's major axis.

\textit{G005.75, G005.49, and G005.38} are spatially associated with the cluster G5, which has observational evidence suggesting that it is the site of a cloud-cloud collision between the dust lane on the nearside of the Galaxy and gas that originated from the dust lane on the farside of the Galaxy and overshot the CMZ \citep{gramze_evidence_2023}. The line-of-sight velocities of these three clouds are all consistent with that of G5, which \citet{gramze_evidence_2023} found to extend from as low as 0-200 km\ s$^{-1}$. We adopt geometric distances of 7.0 $\pm$ 1.0, 7.0 $\pm$ 1.0, and 7.1 $\pm$ 1.0 kpc for these three clouds, respectively.

\textit{G003.09, G003.02, and G002.96} are spatially associated with the cluster B2, which has also been suggested to be a similar site of a collision between the dust lane and overshooting material from the other side, or gas accreting onto x\textsubscript{2} orbits in the bar region \citep{sormani_geometry_2019}. \textit{G002.51} may also be associated with the same clump. Although the central velocities of these four clouds are not similar, B2 itself extends over a velocity range greater than 150 km\ s$^{-1}$ \citep{fux_3d_1999}, so these clouds likely belong to B2 given their aforementioned broad lines. We adopt geometric distances of 7.5 $\pm$ 1.0 kpc for G003.09, G003.02, and G002.96, and 7.6 $\pm$ 1.0 kpc for G002.51.

\textit{G006.56 and G001.93} are not associated with known clusters in the bar but still exhibit broad lines, so we consider them as likely to be in the bar. We adopt geometric distances of 6.9 $\pm$ 1.0 kpc and 7.7 $\pm$ 1.0 kpc for these two clouds, respectively.

\section{Temperature Error Maps}
\label{appendix:tempErr}

\begin{figure*}[t]
\centering
\includegraphics[width=7in]{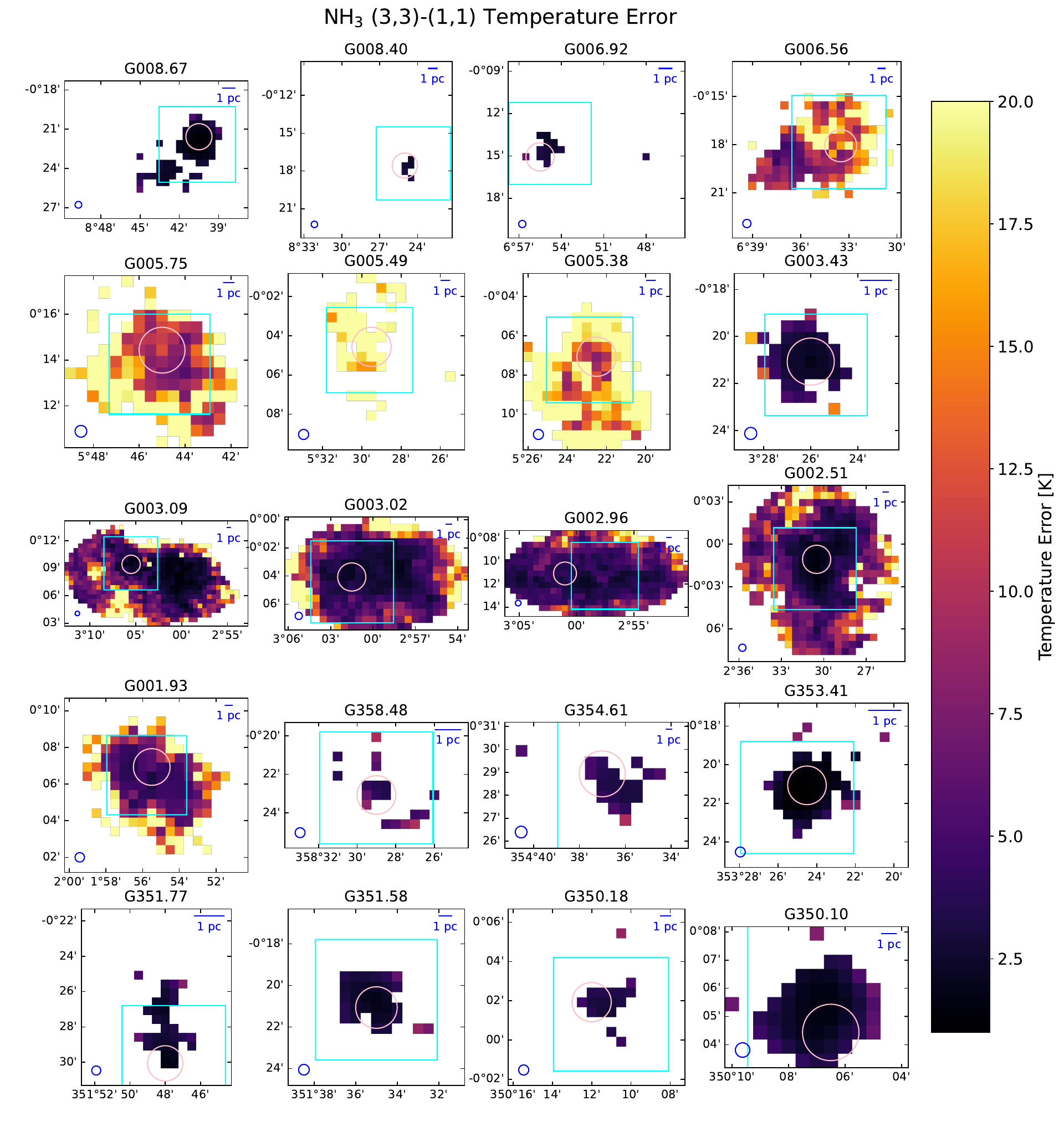}
\caption{Ammonia $(3,3) - (1,1)$ temperature error maps for each cloud, with the same overlays as Figure \ref{fig:NH3temp}}
\label{fig:NH3tempError}
\end{figure*}

\begin{figure*}[t]
\centering
\includegraphics[width=7in]{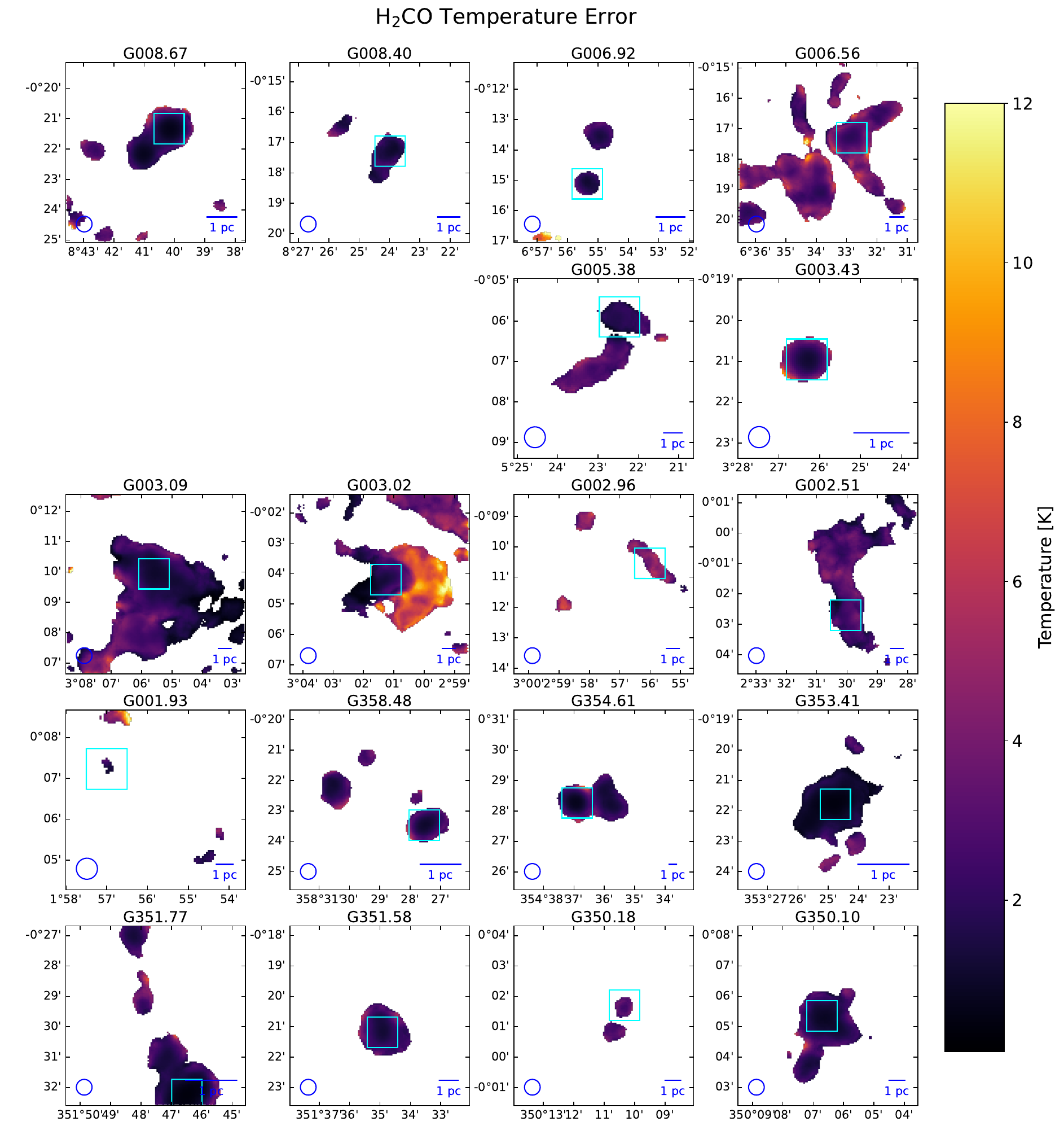}
\caption{Formaldehyde temperature error maps for each cloud, with the same overlays as Figure \ref{fig:H2COtemp}.}
\label{fig:H2COtempError}
\end{figure*}

As mentioned in Section \ref{subsubsec:ammoniaTemp}, we include the ammonia temperature error maps for each cloud in Figure \ref{fig:NH3tempError}. Likewise, as mentioned in Section \ref{subsubsec:h2coTemp}, we include the formaldehyde temperature error maps for each cloud in Figure \ref{fig:H2COtempError}.

\section{Velocity Gradient Fitting}
\label{appendix:vg}

To determine the velocity gradients of the clouds, we fit a plane of form

\begin{equation}
z = ax + by + c
\end{equation}
to the H\textsubscript{2}CO $(3_{03}-2_{02})$ intensity-weighted velocity (``moment 1'') map of each cloud, where $z$ is the velocity, $(x,y)$ are the sky pixel coordinates, and $(a,b,c)$ are fit constants. We take the velocity gradient magnitude to be $\text{d}v/\text{d}r = \sqrt{a^2 + b^2}$, which can be converted to units of kilometers per second per parsec using the distance to the cloud. We multiply this by the beam size in radians to estimate the contribution of bulk motion to the line width.

\begin{figure*}[t]
\centering
\includegraphics[width=7in]{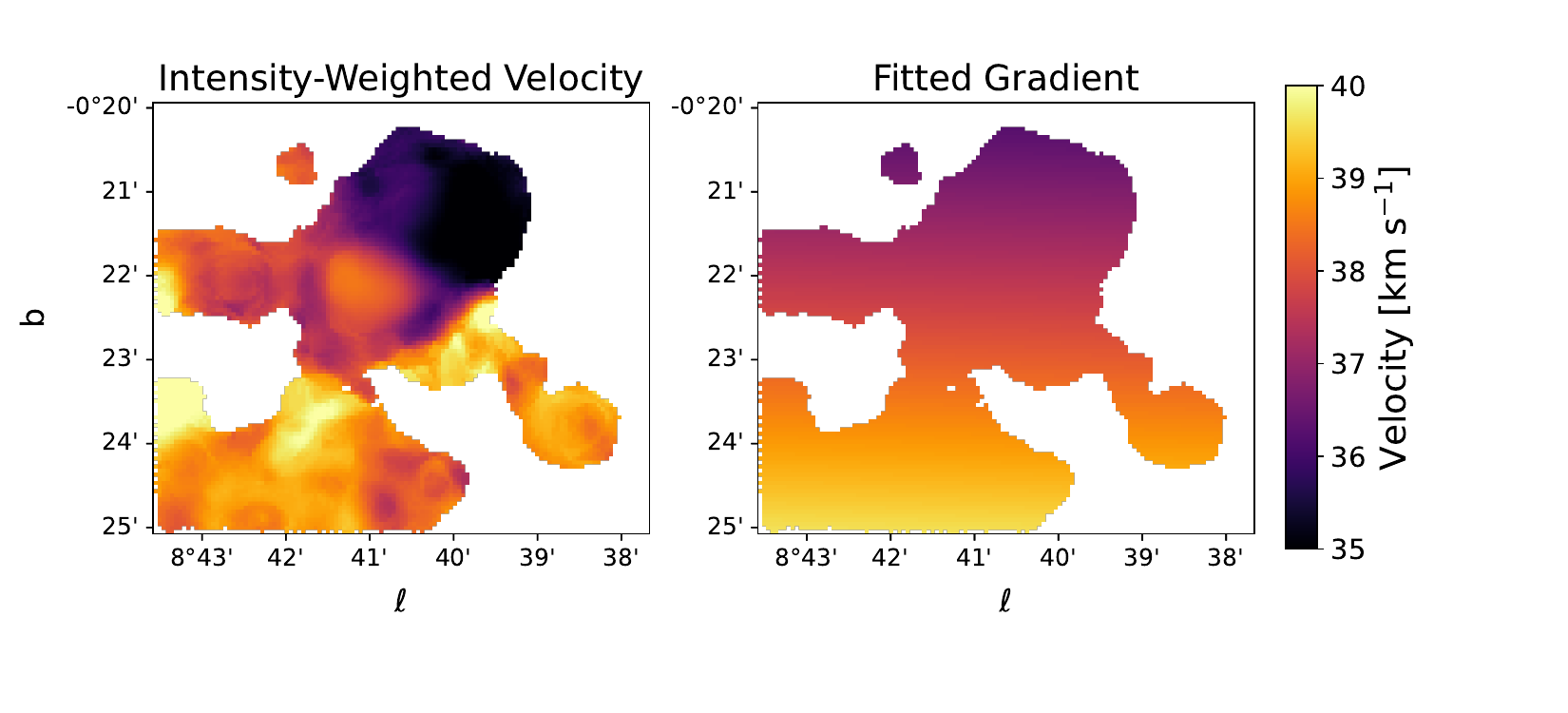}
\includegraphics[width=4in]{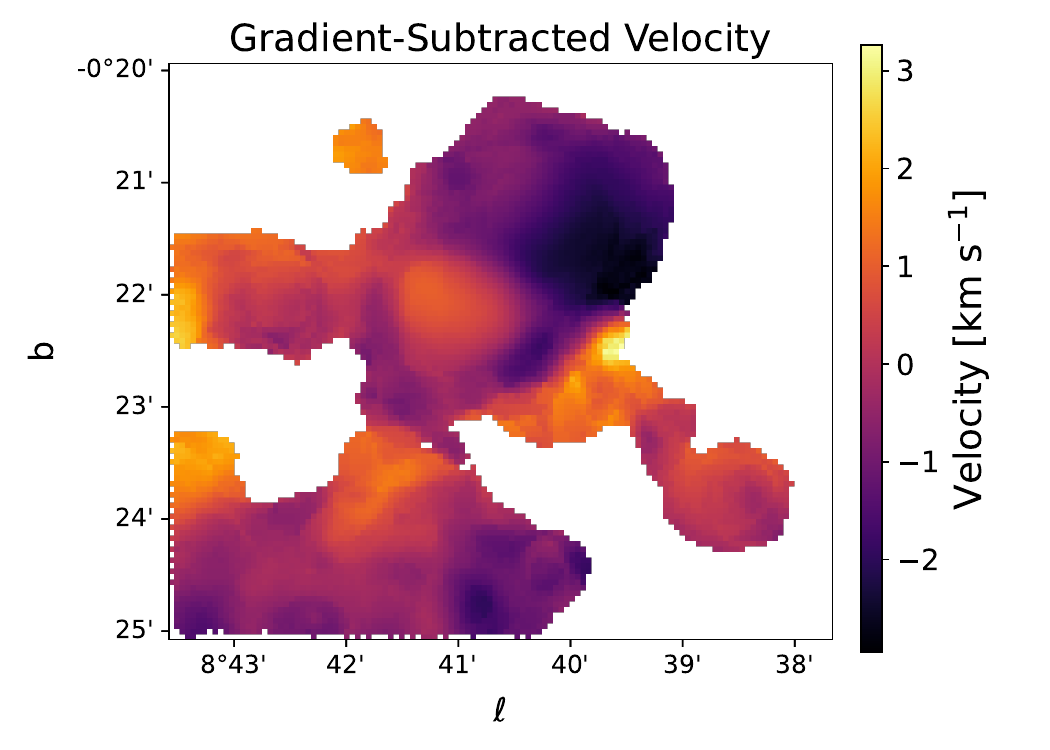}
\caption{Map of the intensity-weighted velocity of H\textsubscript{2}CO $(3_{03}-2_{02})$ for cloud G008.67 (left), the gradient fit to the velocities (right), and the residual velocities after subtracting the gradient (bottom). The gradient likely represents large-scale bulk motions of the cloud, separate from turbulence.}
\label{fig:velGrad}
\end{figure*}

In Figure \ref{fig:velGrad}, we show an example of the gradient fitting procedure for cloud G008.67.

\bibliography{bibliography.bib}
\bibliographystyle{aasjournal}

\end{document}